\documentclass{aa}
\usepackage{graphicx}
\usepackage{txfonts}
\usepackage{natbib}
\bibpunct{(}{)}{;}{a}{}{,}
\usepackage{lscape}
\usepackage{rotating}

\begin{document}

\title{A Sino-German $\lambda$6\ cm polarisation survey of the
  Galactic plane} \subtitle{IX. \ion{H}{II} regions}
\author{X. Y. Gao\inst{1,2}, P. Reich\inst{3}, L. G. Hou\inst{1,2},
  W. Reich\inst{3}, J. L. Han\inst{1,2,4}}

\offprints{X. Y.~Gao and P.~Reich\\ \email{xygao@nao.cas.cn,
    preich@mpifr-bonn.mpg.de}}

\institute{National Astronomical Observatories, CAS, Jia-20 Datun
  Road, Chaoyang District, Beijing 100012, PR China \and CAS Key
  Laboratory of FAST, National Astronomical Observatories, Chinese
  Academy of Sciences \and Max-Planck-Institut f\"{u}r
  Radioastronomie, Auf dem H\"{u}gel 69, 53121 Bonn, Germany \and
  School of Astronomy, University of Chinese Academy of Sciences,
  Beijing 100049, China}

\date{Received; accepted}

\abstract 
{Large-scale radio continuum surveys provide data to get insights into
  the physical properties of radio sources. \ion{H}{II} regions are
  prominent radio sources produced by thermal emission of ionised gas
  around young massive stars.}
{We identify and analyse \ion{H}{II} regions in the Sino-German
  $\lambda$6\ cm polarisation survey of the Galactic plane.}
{Objects with flat radio continuum spectra together with infrared
  and/or H$\alpha$ emission were identified as \ion{H}{II}
  regions. For \ion{H}{II} regions with small apparent sizes, we
  cross-matched the $\lambda$6\ cm small-diameter source catalogue
  with the radio \ion{H}{II} region catalogue compiled by Paladini and
  the infrared \ion{H}{II} region catalogue based on the WISE
  data. Effelsberg $\lambda$21\ cm and $\lambda$11\ cm continuum
  survey data were used to determine source spectra. High angular
  resolution data from the Canadian Galactic Plane Survey and the NRAO
  VLA Sky Survey were used to solve the confusion when low angular
  resolution observations were not sufficient. Extended \ion{H}{II}
  regions were identified by eye by overlaying the Paladini and the
  WISE \ion{H}{II} regions onto the $\lambda$6\ cm survey images for
  coincidences. The TT-plot method was employed for spectral index
  verification.}
{A total of 401 \ion{H}{II} regions were identified and their flux
  densities were determined with the Sino-German $\lambda$6\ cm survey
  data. In the surveyed area, 76 pairs of sources are found to be
  duplicated in the Paladini \ion{H}{II} region catalogue, mainly due
  to the non-distinction of previous observations with different
  angular resolutions, and 78 objects in their catalogue are
  misclassified as \ion{H}{II} regions, being actually planetary
  nebulae, supernova remnants, or extragalactic sources that have
  steep spectra. More than 30 \ion{H}{II} regions and \ion{H}{II}
  region candidates from our $\lambda$6\ cm survey data, especially
  extended ones, do not have counterparts in the WISE \ion{H}{II}
  region catalogue, of which 9 are identified for the first time. Our
  results imply that some more Galactic \ion{H}{II} regions still
  await to be discovered and the combination of multi-domain
  observations is important for \ion{H}{II} region
  identification. Based on the newly derived radio continuum spectra
  and the evidence of infrared emission, the previously identified
  SNRs G11.1$-$1.0, G20.4+0.1, and G16.4$-$0.5 are believed to be
  \ion{H}{II} regions.}
{}


\titlerunning{\ion{H}{II} regions in the $\lambda$6\ cm survey}
\authorrunning{X. Y. Gao et al.}

\maketitle
\section{Introduction}

\ion{H}{II} regions are clouds of ionised gas. They appear as
flat-spectrum radio sources at higher radio frequencies (usually
$>$1~GHz) when optically thin. \ion{H}{II} regions are excellent
tracers for Galactic spiral pattern \citep{Hou09, hh14_8.5}. Their
emission measures and rotation measures, and hence the derived thermal
electron densities and magnetic fields \citep[e.g.][]{Gao10,
  Harvey-Smith11} provide important information of the Galactic
interstellar medium. The determination of free-free absorption by
\ion{H}{II} regions at low radio frequencies helps to derive the
synchrotron emissivity of our Galaxy \citep[e.g.][]{Su17}.

Previous large-scale radio continuum and recombination line surveys
revealed a large number of \ion{H}{II} regions
\citep[e.g.][]{Altenhoff70, Caswell87, Kuchar97}. By selecting from 24
published lists and catalogues, \citet{Paladini03} compiled a widely
used radio catalogue of Galactic \ion{H}{II} regions containing 1\,442
entries. Recently, through large surveys of radio recombination lines
(RRLs) carried out by big dishes such as the Arecibo and Green Bank
telescopes, many more new \ion{H}{II} regions have been discovered
\citep[e.g.][]{Anderson11, Bania12}. A catalogue including over 8\,000
Galactic \ion{H}{II} regions and \ion{H}{II} region candidates from
the Wide-field Infrared Survey Explorer (WISE) data was compiled by
\citet{abb+14}. The updated
version\footnote{http://astro.phys.wvu.edu/wise/} of this catalogue
has been made publicly available after including the follow-up work of
\citet{aaj+15, Makai17}, and \citet{aal+18}.

The Sino-German $\lambda$6\ cm polarisation survey of the Galactic
plane \citep{Sun07, Gao10, Sun11a, Xiao11} surveyed the Galactic disc
from $\ell$ = 10$\degr$ to 230$\degr$ in Galactic longitude and
between $b = \pm5\degr$ in Galactic latitude. It provides a large
field in which to identify Galactic \ion{H}{II}
regions. High-frequency radio data are ideal for detecting very young
\ion{H}{II} regions, which are often optically thick at low
frequencies. Moreover, \ion{H}{II} regions with a flat spectrum for
the thermal bremsstrahlung emission become more pronounced at high
radio frequencies than synchrotron radio sources, which have steep
spectrum and dominate at lower frequencies \citep{Sun11a, Xu13}. From
this survey, several new \ion{H}{II} regions have already been
identified, i.e. G124.0+1.4 and G124.9+0.1 in \citet{Sun07},
G148.8+2.3, G149.5+0.0 (possible part of \ion{H}{II} region SH 2-205)
and G169.9+2.0 in \citet{Shi08b}, and G98.3$-$1.6 and G119.6+0.4 in
\citet{Xiao11}. A large $\lambda$6\ cm \ion{H}{II} region catalogue
was presented by \citet{Kuchar97} based on high angular resolution
survey images obtained by the telescopes at Green Bank \citep[87GB
  survey,][hereafter 87GB]{Condon89} and Parkes \citep[Parkes-MIT-NRAO
  surveys,][hereafter PMN]{Condon93,Tasker94}. It contains 760
Galactic \ion{H}{II} regions with sizes ranging up to 10$\arcmin$. The
Sino-German $\lambda$6\ cm data with a beam size of 9$\farcm$5 can be
used for comparison with the small-diameter \ion{H}{II} regions in
\citet{Kuchar97}, and also to complement new and larger \ion{H}{II}
regions with sizes exceeding 10$\arcmin$.

In this work we identify and analyse \ion{H}{II} regions in the
Sino-German $\lambda$6\ cm survey. We briefly introduce the data used
in Sect.~2. In Sect.~3 we discuss the method for \ion{H}{II} region
identification. We show the results in Sect.~4, and summarise in
Sect.~5.

\section{Data}

\subsection{Urumqi $\lambda$6\ cm data}

The Sino-German $\lambda$6\ cm polarisation survey of the Galactic
plane\footnote{The $\lambda$6\ cm survey data can be downloaded from
  www.mpifr-bonn.mpg.de/survey.html or
  http://zmtt.bao.ac.cn/6cm/surveydata.html} was carried out with the
Urumqi 25m radio telescope of the Xinjiang Astronomical Observatory,
Chinese Academy of Sciences. The telescope was equipped with a
dual-channel $\lambda$6\ cm receiver constructed at the
Max-Planck-Institut f{\"u}r Radioastronomie, Germany, which worked at
a central observing frequency of 4800~MHz with a bandwidth of 600~MHz,
or at 4963~MHz with a reduced bandwidth of 295~MHz to avoid
interference. The angular resolution of the survey is 9$\farcm$5
$\times$ 9$\farcm$5 and the sensitivity is about 1~mK T$_{b}$. The
total intensity calibration for all observations was based on the
calibrators 3C286, 3C48, and 3C138. Descriptions of a detailed system
set-up and data reduction can be found in \citet{Sun06, Sun07} and
\citet{Gao10}. From the $\lambda$6\ cm survey, \citet{Reich14}
identified 3\,832\footnote{The $\lambda$6\ cm sources at
  $\ell=13\fdg377, b=0\fdg136$ and $\ell=13\fdg404, b=0\fdg147$ are
  replaced by $\ell=13\fdg356, b=0\fdg125$ and $\ell=13\fdg178,
  b=0\fdg045$ through refit. Two additional $\lambda$6\ cm sources at
  $\ell=41\fdg224, b=0\fdg354$ and $\ell=63\fdg041, b=-0\fdg344$
  determinded by Gaussian fitting are added in this work.}
small-diameter sources with apparent sizes of less than 16$\arcmin$
via two-dimensional elliptical Gaussian fits. The fitted major and
minor axes were recorded together with the position angles with
respect to the north Galactic pole. The positional accuracy is better
than 1$\arcmin$, found by comparing with point sources of the high
angular resolution NRAO VLA Sky Survey (NVSS) \citep{Condon98}. In
this work the information (e.g.  positions, sizes, and flux densities)
of the $\lambda$6\ cm small-diameter sources are taken from
\citet{Reich14}.

\subsection{Effelsberg $\lambda$11\ cm data}

The Effelsberg $\lambda$11\ cm Galactic plane survey was conducted
with the Effelsberg 100m radio telescope \citep{Reich84, Reich9011,
  Fuerst90a}. The angular resolution of the $\lambda$11\ cm survey is
4$\farcm$3 $\times$ 4$\farcm$3. The sensitivity is about
20~mK\ T$_{b}$. The survey covers the Galactic plane from $358\degr
\leqslant \ell \leqslant 240\degr$ and $|b| \leqslant 5\degr$, and can
therefore be used for a full comparison with the Urumqi $\lambda$6\ cm
survey. A catalogue that includes 6\,483 $\lambda$11\ cm
small-diameter sources with apparent sizes up to 12$\arcmin$ was
compiled by \citet{Fuerst90b}. We compiled an additional version of
the Effelsberg $\lambda$11\ cm source catalogue after convolving the
data to 9$\farcm$4 in order to compare the $\lambda$11\ cm data to the
Effelsberg $\lambda$21\ cm data at the same angular resolution. These
data were used when the $\lambda$11\ cm data at 4$\farcm$3 resolution
did not cover the entire \ion{H}{II} region.

\subsection{Effelsberg $\lambda$21\ cm data}

The Effelsberg 100m radio telescope was also used for a
$\lambda$21\ cm Galactic plane survey \citep{Reich9021, Reich97}. The
angular resolution of the Effelsberg $\lambda$21\ cm data is
9$\farcm$4 $\times$ 9$\farcm$4, similar to that of the Urumqi
$\lambda$6\ cm survey. The sensitivity of the total intensity data is
about 40~mK T$_{b}$. The survey covered the range in Galactic
longitude nearly identical to the Effelsberg $\lambda$11\ cm survey,
being $357\degr \leqslant \ell \leqslant 240\degr$, but the Galactic
latitude range is limited to $|b| \leqslant 4\degr$. A total of 2\,714
point-like sources have been identified in the survey
\citep{Reich9021, Reich97}. Both the Effelsberg $\lambda$11\ cm and
$\lambda$21\ cm data play important roles in determining source
spectra together with the Urumqi $\lambda$6\ cm data.

\begin{table*}[!tbhp]
\caption{Surveys compiled by \citet{Paladini03} that overlap the
  $\lambda$6\ cm survey region. The results of the 10~GHz continuum
  survey from \citet{Handa87} are incorporated for comparison.}
\label{survey}
\scriptsize
\begin{tabular}{lllc}
\hline\hline

\multicolumn{1}{c}{Abbreviation -- Reference}  &\multicolumn{1}{c}{telescopes}  &\multicolumn{1}{c}{observing frequency} &\multicolumn{1}{c}{HPBW}  \\
                                       &                                &\multicolumn{1}{c}{(GHz)}               &\multicolumn{1}{c}{($\arcmin$)}   \\
\hline

F72 -- \citet{Felli72}                               & NRAO 300 feet                             & 1.4, continuum                          & 10 \\
A70 -- \citet{Altenhoff70}                           & NRAO 300 / 140 feet / Ft. Davis 85 feet   & 1.4/2.7/5.0, continuum                  & 9.4 $\times$ 10.4 / 10.9 $\times$ 11.7 / 10.8 $\times$ 10.8  \\
B69 -- \citet{Beard69}                               & Parkes 64m                                & 1.4/2.7, continuum                      & 14.0 / 7.4 \\
D70 -- \citet{Day70}                                 & Parkes 64m                                & 2.7, continuum                          & 8.2 \\
G70 -- \citet{Goss70a}                               & Parkes 64m                                & 2.7, continuum                          & 8 \\
W70 -- \citet{Wendker70}                             & NRAO 140 feet                             & 2.7, continuum                          & 10.6 $\times$ 11.6 \\
R86 -- \citet{Reich86}                               & Effelsberg 100m                           & 2.7, continuum                          & 4.27 \\
F87 -- \citet{Fuerst87}                              & Effelsberg 100m                           & 2.7, continuum                          & 4.27 \\
M67 -- \citet{Mezger67}                              & NRAO 140 feet                             & 5.0, continuum                          & 6.45 $\times$ 6.3 \\
Wil70 -- \citet{Wilson70}$^{\dagger}$                  & Parkes 64m                                & 5.0, H$_{109\alpha}$                      & 4 \\
R70 -- \citet{Reifenstein70}                         & NRAO 140 feet                             & 5.0, H$_{109\alpha}$ and continuum        & 6.5 \\
A78 -- \citet{Altenhoff78}$^{*}$                      & Effelsberg 100m                           & 4.9, continuum                          & 2.6 \\
D80 -- \citet{Downes80}$^{\dagger\dagger}$               & Effelsberg 100m                           & 4.9/4.8, H$_{110\alpha}$ and H$_2$CO      & 2.6 \\
C87 -- \citet{Caswell87}$^{\dagger\dagger\dagger}$        & Parkes 64m                                & 5, H$_{109\alpha}$ and H$_{110\alpha}$      & 4.4 \\
K97 -- \citet{Kuchar97}                              & NRAO 300 feet / Parkes 64m                & 4.9, continuum                          & 3.7 $\times$ 3.3 / 5.0$^{\dagger\dagger\dagger\dagger}$ \\
W82 -- \citet{Wink82}                                & Effelsberg 100m / Tucson 11m              & 4.9/14.8/86, continuum                  & 2.6/1.0/1.3 \\
H87 -- \citet{Handa87}                               & Nobeyama 45m                              & 10.05/10.55, continuum                  & 3 \\
W83 -- \citet{Wink83}                                & Effelsberg 100m                           & 14.7,  H$_{76\alpha}$ and continuum       & 1 \\

\hline
\end{tabular}
\tablefoot{
\scriptsize
$^{*}$ ADS gave Altenhoff et al. (1979), but it should be Altenhoff et al. (1978).\\
$\dagger$ continuum data mainly from \citet{Goss70b}. \\
$\dagger\dagger$ continuum results from A78; \\
$\dagger\dagger\dagger$ continuum results from \citet{Haynes78, Haynes79}; \\
$\dagger\dagger\dagger\dagger$ 3$\farcm$7 $\times$ 3$\farcm$3 for 87GB and 5$\farcm$0 for PMN surveys. \\
}
\end{table*}

\subsection{Canadian Galactic Plane Survey $\lambda$21\ cm data}

By using the synthesis telescopes of the Dominion Radio Astrophysical
Observatory, the Canadian Galactic Plane Survey (CGPS) mapped the
Galactic plane at $\lambda$21\ cm within the range of $66\degr
\leqslant \ell \leqslant 175\degr$ and $-3\degr < b < 5\degr$
\citep{Taylor03, Landecker10}. The missing large-scale total-intensity
emission of the interferometer data was compensated by the data from
the Stockert $\lambda$21\ cm survey \citep{Reich82, Reich86a}. The
angular resolution of the CGPS data is about $\sim$1$\arcmin$, ideal
for resolving Galactic sources, especially in confused areas where the
Urumqi $\lambda$6\ cm beam size is too large.

\subsection{NVSS $\lambda$21\ cm data}

The NRAO VLA Sky Survey is a $\lambda$21\ cm interferometric continuum
survey covering the entire sky of declination $\delta >$ $-$40$\degr$
\citep{Condon98}. The survey provides images with an angular
resolution of 45$\arcsec$ and a catalogue that contains over 1.8
million discrete sources. They were used to extract radio continuum
information where CGPS data are absent in this work. As mentioned by
\citet{Condon98}, the flux density of extended sources with sizes more
than a few times the beamwidth will be missing due to the lack of
shorting spacings. Therefore, the NVSS data can only be used for
point-like sources and as a hint for structures that are not too
extended.

\subsection{WISE \ion{H}{II} region catalogue}

By analysing the WISE mid-infrared images ($\lambda$ = 3.4, 4.6, 12,
and 22 $\mu$m at a resolution of 6$\farcs$1, 6$\farcs$4, 6$\farcs$5,
and 12$\farcs$0, respectively), \citet{abb+14} compiled an infrared
\ion{H}{II} region and \ion{H}{II} region candidate catalogue,
covering all Galactic longitudes with Galactic latitude $|b| \leqslant
8\degr$. Four types were classified in the catalogue by comparison
with the high angular resolution radio interferometric survey data
(e.g. CGPS, NVSS). Known \ion{H}{II} regions are labelled `K',
representing the \ion{H}{II} regions that have either radio
recombination line measurements or measured H$\alpha$ emission; `G'
stands for a group of known \ion{H}{II} regions and several
\ion{H}{II} region candidates where \ion{H}{II} region candidates are
located on or within the photo-dissociation region of the known
\ion{H}{II} region; \ion{H}{II} region candidates labelled `C' are
objects that have characteristic \ion{H}{II} region mid-infrared (MIR)
morphology and a radio emission counterpart, but without RRL or
H$\alpha$ observations; radio-quiet \ion{H}{II} region candidates `Q'
meet all the conditions of `C', but without radio counterparts. After
adding and removing some sources, there are 8\,407 total entries of
the newest WISE \ion{H}{II} region catalogue.

\section{Methodology: identification of \ion{H}{II} regions}

To identify \ion{H}{II} regions in the Sino-German $\lambda$6\ cm
survey, we separated the Urumqi $\lambda$6\ cm sources into two
groups: small-diameter sources from \citet{Reich14} and extended
sources (apparent size $>$ 16$\arcmin$) that have not been catalogued
yet. In this work, we only consider \ion{H}{II} regions whose
boundaries and flux densities can be reliably determined by the Urumqi
data.

To identify small-diameter \ion{H}{II} regions, the intrinsic sizes of
\citet{Reich14} sources were first calculated by deconvolution
following
\begin{equation}
\begin{aligned}
\theta_{maj} = \sqrt{\theta_{fit,maj}^{2} - 9\farcm5^{2}},\\
\theta_{min} = \sqrt{\theta_{fit,min}^{2} - 9\farcm5^{2}},
\end{aligned}
\label{eq:eq1}
\end{equation}  
where $\theta_{fit}$ represent the fitted sizes of the major and minor
axes of the $\lambda$6\ cm source in \citet{Reich14} and 9$\farcm$5 is
the angular resolution of the Urumqi survey. A cross-match was made
between the Urumqi $\lambda$6\ cm small-diameter sources and the
Paladini \ion{H}{II} region catalogue. To obtain a redundant sample, a
loose matching condition was set that the distance for the coincidence
must be less than or equal to $r_{\rm p} + r_{\rm \sigma_{p}} + r_{\rm
  6cm} + r_{\rm \sigma_{6cm}} + r_{\rm error}$, where $r_{\rm p}$ and
$r_{\rm \sigma_{p}}$ are the source radii and errors given by
\citet{Paladini03}, and $r_{\rm 6cm}$, $r_{\rm \sigma_{6cm}}$, and
$r_{\rm error}$ are the deconvolved half width of the major axis
($\theta_{maj}$) of the $\lambda$6\ cm sources, their errors, and the
positional errors given in \citet{Reich14}. We obtained a total of 419
matching pairs for small-diameter \ion{H}{II} regions. The Paladini
\ion{H}{II} region catalogue includes 1\,442 entries, of which 777
objects from 17 catalogues are located in the area of the Urumqi
survey (see Table~\ref{survey}). Unfortunately, the positional
accuracy given in the Paladini catalogue is limited to 0$\fdg$1 in
Galactic coordinates. Hence, we had to extract the source positions
from the original literature. One Paladini source often has several
measurements at various frequencies with different central
coordinates. Preference is given to the observation that is more
recent, where the observing frequency is closer to 4.8~GHz, and/or
where the angular resolution is higher. According to
Table~\ref{survey}, the priority sequence follows \citet[][hereafter
  K97]{Kuchar97}, the 4.9~GHz data of \citet[][hereafter W82]{Wink82},
\citet[][hereafter A78]{Altenhoff78}, \citet[][hereafter
  Wil70]{Wilson70}, \citet[][hereafter C87]{Caswell87},
\citet[][hereafter R70]{Reifenstein70}, \citet[][hereafter
  M67]{Mezger67}, \citet[][hereafter B69]{Beard69}, \citet[][hereafter
  G70]{Goss70a}/\citet[][hereafter D70]{Day70}, and \citet[][hereafter
  A70]{Altenhoff70}. We also have data from other references; for
example, \citet[][hereafter D80]{Downes80} observed H$_{110\alpha}$
and H$_{2}$CO for some \ion{H}{II} regions. Their continuum parameters
were taken from A78 and the flux densities of the sources were
re-estimated from the peak flux densities given in A78. Therefore, the
flux densities reported by D80 are often nearly identical to those of
A78. Similarly, C87 measured H$_{109\alpha}$ and H$_{110\alpha}$, and
W70 measured H$_{109\alpha}$ of some \ion{H}{II} regions and their
continuum results are taken from \citet{Haynes78, Haynes79} and
\citet{Goss70b}, respectively. Although observations above 14~GHz
\citep[][hereafter W82 and W83]{Wink82, Wink83} have the highest
angular resolution, their coordinates are not given priority because
they may only show small compact components. \citet{Reich86} and
\citet[][hereafter R86 and F87, which were interchanged in the
  Paladini \ion{H}{II} region catalogue]{Fuerst87} studied 9 and 11
Galactic objects, respectively, of which 5 and 7 were identified as
\ion{H}{II} regions. Several surveys \citep[e.g.][hereafter F72 and
  W70]{Felli72, Wendker70} have angular resolutions (7$\arcmin$ $\sim$
11$\arcmin$) similar to that of the Urumqi $\lambda$6\ cm survey so
that we expected similar observable radio structures and morphologies.

The WISE \ion{H}{II} region catalogue introduced in Sect.~2.6 is
believed to be the most complete star forming region sample to date. A
second cross-match was then made between the Urumqi $\lambda$6\ cm
small-diameter sources and the 4\,168 WISE \ion{H}{II} region and
\ion{H}{II} region candidates within the range of the Urumqi survey
with the same matching condition as described above.

\citet{Anderson11} found that the mid-infrared and the radio continuum
emission of Galactic \ion{H}{II} regions share similar morphologies
and angular extents. It was emphasised by \citet{abb+14} that the WISE
\ion{H}{II} regions are related to their radio correspondence based on
spatial coincidence. Therefore, we overlaid both the
\citet{Paladini03} objects and the WISE \ion{H}{II} regions and
\ion{H}{II} region candidates whose diameters are larger than
8$\arcmin$ onto our $\lambda$6\ cm images. Coincidences for extended
\ion{H}{II} regions were then searched by eye. We also kept those
extended sources in the $\lambda$6\ cm image that have not been
identified that way. They may be new \ion{H}{II} regions. To eliminate
the projection effect in the radio-infrared association and to verify
the nature of those un-identified extended sources, spectra were
fitted using the Effelsberg $\lambda$21\ cm, $\lambda$11\ cm, and the
Urumqi $\lambda$6\ cm data, and/or were tested by the TT-plot method
\citep{Turtle62} for both the small-diameter and the extended sources
chosen by the above procedure. The sources that show an optically
thin, flat ($\alpha \sim -0.1$, S$_{\nu}$ $\sim$ $\nu^{\alpha}$), or a
slightly positive spectral index were identified as \ion{H}{II}
regions. Additional supporting evidence came from H$\alpha$ data
\citep{Finkbeiner03}, the Digitised Sky Survey
\citep[][DSS]{Lasker90}, and IRIS 60$\mu$m data \citep{Miville05}. To
avoid the inclusion of supernova remnants (SNRs) and planetary nebulae
(PNe), the Green SNR catalogue \citep{Green17}, the Galactic PNe
catalogues from \citet{Condon98p}, \citet{Parker06},
\citet{Miszalski08}, \citet{Frew13}, \citet{Parker16},
\citet{Fragkou18}, and \citet{Irabor18} were consulted.

\section{Results}

\begin{table*}[tbhp]
\caption{Small-diameter \ion{H}{II} regions identified in the Urumqi
  $\lambda$6\ cm survey by comparison with the Paladini \ion{H}{II}
  region catalogue.}
\scriptsize
\label{pala}
\setlength{\tabcolsep}{0.8mm}
\begin{tabular}{cccrccccccrccl}
  \hline
  \hline
\multicolumn{1}{c}{P number} & \multicolumn{1}{c}{G-NAME}   & \multicolumn{1}{c}{GLong}        & \multicolumn{1}{c}{GLat}        & \multicolumn{1}{c}{Ref}    & \multicolumn{1}{c}{$D$}      & \multicolumn{1}{c}{Ref} & \multicolumn{1}{c}{EB S$_{\rm 21cm}$} & \multicolumn{1}{c}{EB S$_{\rm 11cm}$} & \multicolumn{1}{c}{Urumqi S$_{\rm 6cm}$} & \multicolumn{1}{c}{S$_{\rm Ref}$} & \multicolumn{1}{c}{Freq} & \multicolumn{1}{c}{Ref} & \multicolumn{1}{c}{Notes} \\
\multicolumn{1}{c}{}         & \multicolumn{1}{c}{}         & \multicolumn{1}{c}{($\degr$)} & \multicolumn{1}{c}{($\degr$)} & \multicolumn{1}{c}{}       &\multicolumn{1}{c}{(kpc)}     & \multicolumn{1}{c}{}    & \multicolumn{1}{c}{(Jy)}            & \multicolumn{1}{c}{(Jy)}            & \multicolumn{1}{c}{(Jy)}               & \multicolumn{1}{c}{(Jy)}            &\multicolumn{1}{c}{(GHz)} & \multicolumn{1}{c}{}    & \multicolumn{1}{c}{} \\
\hline

 129 &G10.1$-$0.4   & 10.073   &$-$0.412   &A78   & 4.95$\pm$0.51   & srm+14  & 47.01  & 54.819 & 63.835 & 3.00   & 4.9 &A78   & included in the $\lambda$6\ cm source G10.16$-$0.35 \\
     &              &          &           &      &                 &         &        &        &        & 3.0    & 4.9 &D80   & \\
 131 &G10.2$-$0.4   & 10.190   &$-$0.426   &D80   & 4.95$\pm$0.51   & srm+14  & 47.01  & 54.819 & 63.835 & 5.1    & 4.9 &D80   & included in the $\lambda$6\ cm source G10.16$-$0.35 \\
 132 &G10.2$-$0.3   & 10.159   &$-$0.349   &K97   & 4.95$\pm$0.51   & srm+14  & 47.01  & 54.819 & 63.835 & 55     & 2.7 &A70   & included in the $\lambda$6\ cm source G10.16$-$0.35 \\
     &              &          &           &      &                 &         &        &        &        & 65     & 2.7 &G70   & new 14.7~GHz flux density \\
     &              &          &           &      &                 &         &        &        &        & 66.353 & 4.9 &K97   & \\
     &              &          &           &      &                 &         &        &        &        & 47     & 5.0 &A70   & \\
     &              &          &           &      &                 &         &        &        &        & 47.48  & 4.9 &A78   & \\
     &              &          &           &      &                 &         &        &        &        & 47.5   & 4.9 &D80   & \\
     &              &          &           &      &                 &         &        &        &        & 63.2   & 5.0 &M67   & \\
     &              &          &           &      &                 &         &        &        &        & 51.8   & 5.0 &R70   & \\
     &              &          &           &      &                 &         &        &        &        & 48.4   & 5.0 &Wil70 & \\
     &              &          &           &      &                 &         &        &        &        & 31.67  &10.0 &H87   & \\
     &              &          &           &      &                 &         &        &        &        &<36.6   &14.7 &W83   & \\
 134 &G10.3$-$0.2   & 10.315   &$-$0.150   &K97   & 4.95$\pm$0.51   & srm+14  & 14.61  & 13.77  & 15.923 & 22     & 2.7 &G70   & same as P135 \\
     &              &          &           &      &                 &         &        &        &        & 20.472 & 4.9 &K97   & \\
     &              &          &           &      &                 &         &        &        &        & 12.48  & 4.9 &A78   & \\
     &              &          &           &      &                 &         &        &        &        & 12.5   & 4.9 &D80   & \\
     &              &          &           &      &                 &         &        &        &        & 20.2   & 5.0 &M67   & \\
     &              &          &           &      &                 &         &        &        &        &<9.2    &14.7 &W83   & \\
 135 &G10.3$-$0.1   & 10.307   &$-$0.145   &R70   &                 &         & 14.61  & 13.77  & 15.923 & 20     & 2.7 &A70   & same as P134 \\
     &              &          &           &      &                 &         &        &        &        & 20     & 5.0 &A70   & \\
     &              &          &           &      &                 &         &        &        &        & 13.6   & 5.0 &R70   & \\
     &              &          &           &      &                 &         &        &        &        & 10.81  &10.0 &H87   & \\
     &$\cdots$      &          &           &      &                 &         &        &        &        &        &     &      & \\
\hline\hline
\end{tabular}

\tablefoot{\scriptsize Column 1: sequential number in
  \citet{Paladini03}; Column 2: G-name of the \ion{H}{II} region;
  Columns 3, 4, and 5: Galactic longitude, latitude, and their
  reference; Columns 6 and 7: distance for the \ion{H}{II} region and
  reference; Columns 8, 9, and 10: integrated flux densities based on
  the Effelsberg $\lambda$21\ cm, $\lambda$11\ cm, and the Urumqi
  $\lambda$6\ cm data. Some of the $\lambda$11\ cm results (with 2
  digits) are from the data at original resolution and some (with 3
  digits) are from the data which were convolved to an angular
  resolution of 9$\farcm$4; Columns 11, 12, and 13: integrated flux
  densities, the corresponding frequencies, and their references. Most
  of these are based on \citet{Paladini03}, except those of
  \citet{Handa87}. The flux densities from \citet{Paladini03} were all
  examined by comparing the results to the original
  references. Corrections were made if improper values were found;
  Column 14: remarks for the source.\\ Full table is accessible from
  the CDS.}

\tablebib{\scriptsize brm+09: \citet{brm+09}; chr+14: \citet{chr+14};
  fb84: \citet{fb84}; fb15: \citet{fb15}; hbc+07: \citet{hbc+07};
  hh14: \citet{hh14_8.5}; irm+13: \citet{irm+13}; mdf+11:
  \citet{mdf+11}; mrm+09: \citet{mrm+09}; okh+10: \citet{okh+10};
  rus03: \citet{rus03}; rag07: \citet{rag07}; rbr+10: \citet{rbr+10};
  rbs+12: \citet{rbs+12}; srm+14: \citet{srm+14}; wsr+14:
  \citet{wsr+14}; xrzm06: \citet{xrzm06}; xlr+13: \citet{xlr+13};
  zzr+09: \citet{zzr+09}; zrm+13: \citet{zrm+13}; zms+14:
  \citet{zms+14}.}
\end{table*}

\begin{table*}
\caption{$\lambda$6\ cm small-diameter \ion{H}{II} regions that have
  WISE \ion{H}{II} region counterparts but are not in the Paladini
  catalogue.}
\scriptsize
\label{wise6}
\setlength{\tabcolsep}{0.6mm}
\begin{tabular}{cccrrrccrccccl}
  \hline
  \hline
\multicolumn{1}{c}{No.}  &\multicolumn{1}{c}{G-NAME}  &\multicolumn{1}{c}{GLong$_{\rm 6cm}$}  &\multicolumn{1}{c}{GLat$_{\rm 6cm}$}  &\multicolumn{1}{c}{$\theta_{maj}$}  &\multicolumn{1}{c}{$\theta_{min}$}  &\multicolumn{1}{c}{$S_{\rm 6cm}$}  &\multicolumn{1}{c}{GLong$_{\rm wise}$}  &\multicolumn{1}{c}{GLat$_{\rm wise}$}   &\multicolumn{1}{c}{diameter}      &\multicolumn{1}{c}{WISE type} &\multicolumn{1}{c}{$D$}   & \multicolumn{1}{c}{Ref}  & \multicolumn{1}{c}{Notes} \\
                         &                            &\multicolumn{1}{c}{($\degr$)}          &\multicolumn{1}{c}{($\degr$)}         &\multicolumn{1}{c}{($\arcmin$)}    &\multicolumn{1}{c}{($\arcmin$)}    &\multicolumn{1}{c}{(Jy)}           &\multicolumn{1}{c}{($\degr$)}           &\multicolumn{1}{c}{($\degr$)}           &\multicolumn{1}{c}{($\arcmin$)}   &                              & (kpc)                    &                          & \\
\hline
    1 & G12.2$-$0.1  & 12.232 &  $-$0.110   &   8.4  &   5.7  &  5.405      &  12.145  &     0.000  &  13.4   &K   &                 &             & one $\lambda$6\ cm source matches thirteen WISE \ion{H}{II} regions \\
      &              &        &             &        &        &             &  12.184  &  $-$0.113  &   0.7   &K   &                 &             & \\
      &              &        &             &        &        &             &  12.192  &  $-$0.104  &   1.0   &K   &                 &             & \\
      &              &        &             &        &        &             &  12.194  &  $-$0.144  &   0.6   &K   &                 &             & \\
      &              &        &             &        &        &             &  12.194  &  $-$0.112  &   0.6   &K   &                 &             & \\
      &              &        &             &        &        &             &  12.194  &  $-$0.129  &   1.6   &K   &                 &             & \\
      &              &        &             &        &        &             &  12.199  &  $-$0.034  &   1.4   &K   &                 &             & \\
      &              &        &             &        &        &             &  12.202  &  $-$0.110  &   0.8   &K   &                 &             & \\
      &              &        &             &        &        &             &  12.209  &  $-$0.105  &   0.7   &K   &                 &             & \\
      &              &        &             &        &        &             &  12.216  &  $-$0.119  &   0.8   &K   &                 &             & \\
      &              &        &             &        &        &             &  12.217  &  $-$0.138  &   2.3   &K   &                 &             & \\
      &              &        &             &        &        &             &  12.229  &  $-$0.115  &   2.0   &K   &                 &             & \\
      &              &        &             &        &        &             &  12.311  &  $-$0.059  &   6.0   &K   &                 &             & \\
    2 & G19.8+0.3    & 19.775 &    0.285    & $<$3.1 & $<$3.1 &  0.655      &  19.741  &     0.280  &   1.4   &K   & 14.20$\pm$0.61  & hh14        & one $\lambda$6\ cm source matches two WISE \ion{H}{II} regions \\
      &              &        &             &        &        &             &  19.781  &     0.287  &   3.7   &K   & 17.28$\pm$1.22  & hh14        & \\
    3 & G23.5+1.6    & 23.547 &    1.589    & $<$3.1 & $<$3.1 &  0.147      &  23.569  &     1.582  &   3.9   &K   &  2.94$\pm$0.42  & hh14        & \\
    4 & G26.4+1.5    & 26.373 &    1.451    &    8.1 &    5.7 &  1.216      &  26.385  &     1.399  &  19.8   &K   &                 &             & one $\lambda$6\ cm source matches two WISE \ion{H}{II} regions \\
      &              &        &             &        &        &             &  26.385  &     1.401  &   3.8   &K   &  3.03$\pm$0.40  & hh14        & \\
      &$\cdots$      &        &             &        &        &             &          &            &         &    &                 &             & \\
\hline\hline
\end{tabular}

\tablefoot{\scriptsize Columns 1 and 2: sequential number and G-name;
  Columns 3 and 4: Galactic longitude and latitude of the
  $\lambda$6\ cm source; Columns 5 and 6: intrinsic size
  $\theta_{maj}$ and $\theta_{min}$ of the $\lambda$6\ cm source,
  obtained following Eq.~(\ref{eq:eq1}) by assuming the sources to be
  Gaussian. The observed sizes ($\theta_{fit,maj}$ and
  $\theta_{fit,min}$) used for the deconvolution are taken from
  \citet{Reich14}. The Galactic position angle for each source can be
  found in \citet{Reich14}; Column 7: integrated flux density at
  $\lambda$6\ cm; Columns 8 and 9: Galactic longitude and latitude of
  the matched WISE \ion{H}{II} region; Column 10: diameter of the
  matching WISE \ion{H}{II} region; Column 11: group type of the WISE
  \ion{H}{II} region, K$^{*}$ indicates currently known \ion{H}{II}
  regions (original C-type WISE \ion{H}{II} regions) according to
  \citet{Kim18}; Columns 12 and 13: distance information and their
  references; Column 14: remarks for the source. Some other references
  in the 14th column: e.g. \citet{Suad12}. \\ Full table is accessible
  from the CDS.}

\tablebib{\scriptsize abb+14: \citet{abb+14}; aaj+15: \citet{aaj+15};
  aal+18: \citet{aal+18}; cpt06: \citet{cpt06}; fb15: \citet{fb15};
  hh14: \citet{hh14_8.5}; mdf+11: \citet{mdf+11}; rus03:
  \citet{rus03}; rag07: \citet{rag07}; xlr+13: \citet{xlr+13}.  }
\end{table*}

\begin{table*}
\caption{Extended \ion{H}{II} regions in the $\lambda$6\ cm survey.}
\scriptsize
\label{extended}
\setlength{\tabcolsep}{0.7mm}
\begin{tabular}{cccrccrcrcccl}
  \hline
  \hline
\multicolumn{1}{c}{No.}  &\multicolumn{1}{c}{G-NAME}  &\multicolumn{1}{c}{GLong$_{\rm 6cm}$}  &\multicolumn{1}{c}{GLat$_{\rm 6cm}$}  &\multicolumn{1}{c}{Apparent $\theta_{maj}$}  &\multicolumn{1}{c}{Apparent $\theta_{min}$}   &\multicolumn{1}{c}{$S_{\rm 6cm}$}  &\multicolumn{1}{c}{GLong$_{\rm wise}$}     &\multicolumn{1}{c}{GLat$_{\rm wise}$}    &\multicolumn{1}{c}{Diameter}        & $D$      & Ref. &\multicolumn{1}{c}{Notes} \\
                         &                            &\multicolumn{1}{c}{($\degr$)}          &\multicolumn{1}{c}{($\degr$)}         &\multicolumn{1}{c}{($\arcmin$)}    &\multicolumn{1}{c}{($\arcmin$)}    &\multicolumn{1}{c}{(Jy)}           & \multicolumn{1}{c}{($\degr$)}             &\multicolumn{1}{c}{($\degr$)}            &\multicolumn{1}{c}{($\arcmin$)}     &  (kpc)   &      & \\
\hline

      1  &G11.7$-$1.7  &  11.69  &$-$1.74     &19.8  &15.7   &    5.084      &  11.662  &  $-$1.692     &  30.2     &   1.6$\pm$0.2      & rus03      &  SH 2-37, P150 \\
      2  &G13.8$-$0.8  &  13.80  &$-$0.81     &18.0  &11.5   &   12.183      &  13.776  &  $-$0.795     &  18.1     &                    &            &  RCW~156, P176 \\
      3  &G15.1+3.3    &  15.11  &   3.35     &22.0  &20.3   &    7.743      &  15.128  &     3.310     &  37.8     &   2.8$\pm$0.6      & rus03      &  SH 2-46, P202 \\
      4  &G16.9$-$1.1  &  16.87  &$-$1.05     &21.8  &13.9   &    7.447      &  16.857  &  $-$1.155     &  23.9     &                    &            &  SH 2-50, including P216, see \citet{Sun11b} \\
      5  &G16.9+0.8    &  16.94  &   0.77     &  60  &55     &  162.677      &  16.993  &     0.874     &  37.5     &   2.0$\pm$0.25     & rus03      &  SH 2-49, RCW~165, P218,P219,P220,P224  \\
         &$\cdots$     &                      &      &       &               &          &               &           &                    &            &  \\
     75  &G17.0+1.7    &  17.02  &   1.72     &27.8  &15.7   &    6.353      &          &               &           &                    &            &  P221/P222, flat radio spectrum \\
     76  &G20.3$-$0.9  &  20.27  &$-$0.89     &18.5  &13.0   &    3.560      &          &               &           &                    &            &  P266/P267/P268, flat radio spectrum \\
     77  &G63.7$-$0.7  &  63.65  &$-$0.72     &17.0  &13.6   &    0.858      &          &               &           &                    &            &  P622, flat radio spectrum \\
     78  &G66.4$-$1.2  &  66.41  &$-$1.17     & 100  &96     &    7.874      &  66.660  &  $-$1.167     &  62.5     &                    &            &  DU~34, including P631, Q-type WISE \ion{H}{II} region? \\
     79  &G68.4+0.2    &  68.39  &   0.19     &23.6  &17.3   &    1.404      &          &               &           &                    &            &  flat radio spectrum, possible infrared emission associated \\
         &$\cdots$     &         &            &      &       &               &          &               &           &                    &            &  \\     
\hline\hline
\end{tabular}

\tablefoot{\scriptsize Columns 1 and 2: sequential number and G-name;
  Columns 3 and 4: Galactic longitude and latitude of the
  $\lambda$6\ cm source; Columns 5 and 6: apparent size $\theta_{maj}$
  and $\theta_{min}$ of the $\lambda$6\ cm source measured in Galactic
  longitude and latitude directions; Column 7: integrated flux density
  at $\lambda$6\ cm; Columns 8 and 9: Galactic longitude and latitude
  of the matched WISE \ion{H}{II} region; Column 10: diameter of the
  matching WISE \ion{H}{II} region; Column 11 and 12: distance
  information and their references; Column 13: remarks for the
  source. Some other references in the 13th column:
  e.g. \citet{Sieber84}, \citet{Suad12}, \citet{Gao11x},
  \citet{Gao13}, and \citet{Graham82}.  \\ Full table is accessible
  from the CDS.}

\tablebib{\scriptsize abb+14: \citet{abb+14}; aaj+15: \citet{aaj+15};
  aal+18: \citet{aal+18}; rmb+09: \citet{rmb+09}; chr+14:
  \citet{chr+14}; fb84: \citet{fb84}; fb15: \citet{fb15}; hh14:
  \citet{hh14_8.5}; rus03: \citet{rus03}; rag07: \citet{rag07}.}
\end{table*}

After scrutinising the small-diameter and the extended \ion{H}{II}
regions on the $\lambda$6\ cm images, we identified 401 \ion{H}{II}
regions in total from the Sino-German $\lambda$6\ cm polarisation
survey of the Galactic plane. We list the results in Table~\ref{pala},
Table~\ref{wise6}, and Table~\ref{extended}. Full tables can be
accessed via CDS. Table~\ref{pala} contains the $\lambda$6\ cm
small-diameter \ion{H}{II} regions with counterparts in the
\citet{Paladini03} \ion{H}{II} region catalogue. Table~\ref{wise6}
includes the $\lambda$6\ cm small-diameter \ion{H}{II} regions that
have WISE \ion{H}{II} region counterparts but were not included in
\citet{Paladini03}. Table~\ref{extended} lists the extended
\ion{H}{II} regions identified in the Urumqi $\lambda$6\ cm survey.
In all these three tables, we provide the central positions of the
\ion{H}{II} regions, the integrated flux densities especially the new
$\lambda$6\ cm results based on the Urumqi data, and distance
information from new measurements in the literature if available.

For objects in Table~\ref{pala}, the flux densities based on the
Effelsberg $\lambda$21\ cm and $\lambda$11\ cm data were added for
comparison. Some of the $\lambda$21\ cm results were re-calculated due
to the imperfect determination of the zero levels by the automatic
Gaussian fit. Some of the $\lambda$11\ cm results were obtained from
the 9$\farcm$4 resolution data (see Sect.~2) instead of the 4$\farcm$3
original data, because sometimes the original $\lambda$11\ cm data
have only fitted a part of the \ion{H}{II} region or the underlying
emission influenced the zero levels during fitting. We also list the
flux density measurements quoted in the Paladini \ion{H}{II} region
catalogue for each source in Table~\ref{pala}. However, some errors
were noticed: (1) inappropriate flux density assignment,
e.g. G24.7$-$0.2a and G24.7$-$0.2b, which should have two sets of
different values (see K97) but were given the same in the Paladini
catalogue; (2) missing flux density, e.g. G45.5+0.1, for which two
components of A78 (D80), W82, W83 should be included, but only one was
listed; (3) inappropriate estimates of flux density, especially for
all the 86~GHz measurements of W82. All these problems were reviewed
by checking the original references and fixed. There are still some
previous flux density results that cannot be understood easily, such
as those from W70. We convolved the 4$\farcm$3 angular resolution
Effelsberg $\lambda$11\ cm data to 11$\arcmin$, the same as W70 for
the Cygnus X region. We took eight isolated sources which are not
confused by the strong background emission and estimated their 2.7~GHz
flux densities and sizes by Gaussian fit. For most of the cases, the
newly derived results from the Effelsberg data are much smaller than
those reported by W70.

In Table~\ref{wise6}, it is usually the case that one $\lambda$6\ cm
radio source is found to have several counterparts in the WISE
\ion{H}{II} regions. However, due to the limit of the angular
resolution of the Urumqi data, we cannot give the exact
correspondence, and therefore list all the WISE \ion{H}{II} regions
included in one $\lambda$6\ cm source. Based on the type
classification of the WISE \ion{H}{II} regions (see Sect.~2.6), we
understand that the Q-type WISE \ion{H}{II} regions are the ones that
do not yet have radio continuum detection. Thus, Q-type WISE
\ion{H}{II} regions are not listed if there are also K-, G-, or C-type
WISE \ion{H}{II} regions for the same $\lambda$6\ cm source.

For the \ion{H}{II} regions in Table~\ref{extended}, diffuse Galactic
background emission was filtered out by the technique of
`unsharp-masking' \citep{Sofue79} prior to the flux density
integration. The shape of extended \ion{H}{II} regions, unlike the
small-diameter sources, often cannot be simply described as circular
or elliptical Gaussians. Therefore, the apparent sizes listed in
Table~\ref{extended} were all measured in the directions of Galactic
longitude and Galactic latitude. We listed the $\lambda$6\ cm extended
\ion{H}{II} regions that have WISE \ion{H}{II} region counterparts
(K-, G-, or C-type) in the upper part of Table~\ref{extended} (Rows 1
-- 74). In the lower part (Rows 75 -- 107), we show the $\lambda$6\ cm
extended \ion{H}{II} regions and \ion{H}{II} region candidates which
do not have matching WISE \ion{H}{II} regions.

\begin{figure}
\includegraphics[width=0.45\textwidth]{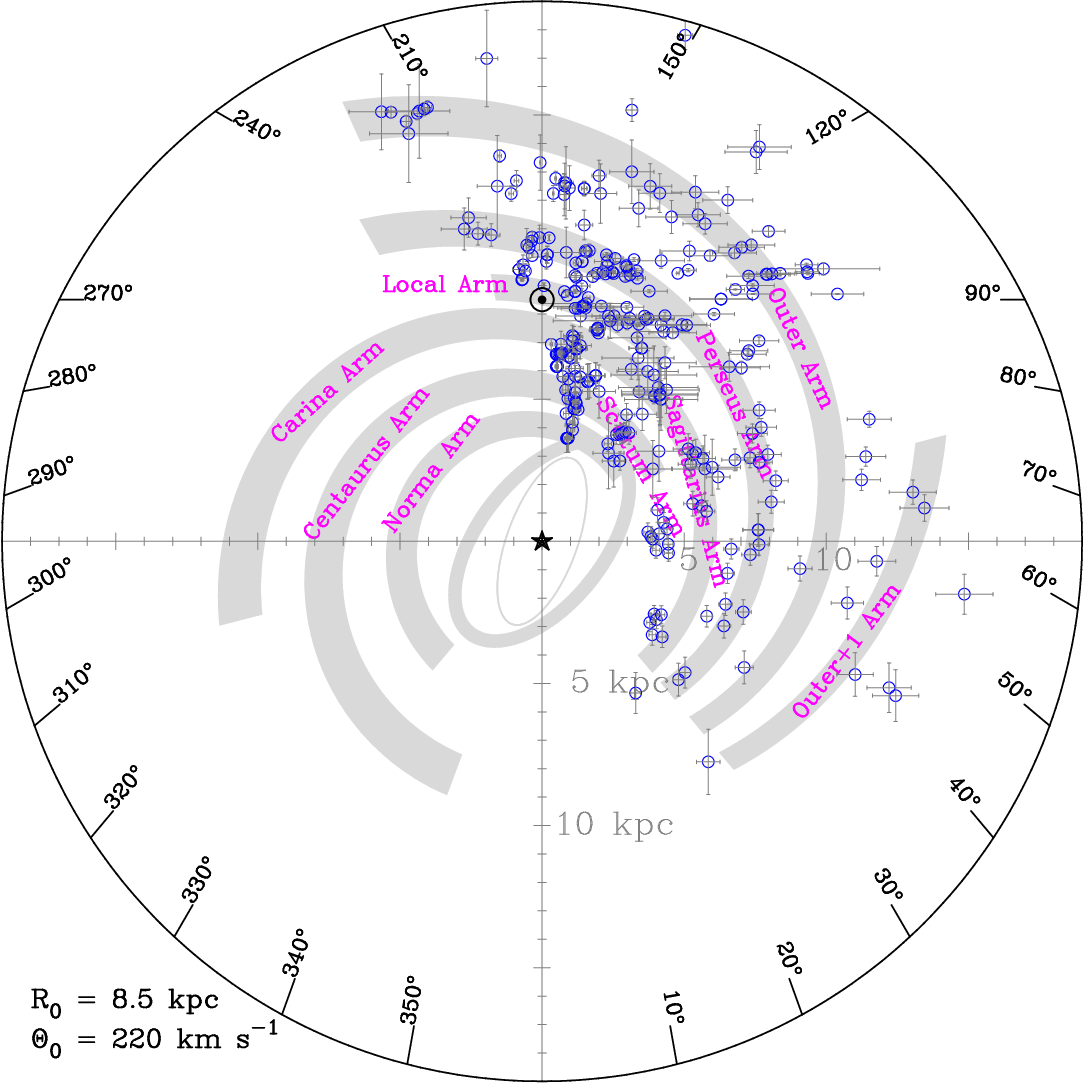}
\caption{Distribution on the Galactic plane of \ion{H}{II} regions
  identified in the Urumqi $\lambda$6\ cm survey with known
  distances.}
\label{dis}
\end{figure}

We illustrate the distribution of the identified $\lambda$6\ cm
\ion{H}{II} regions with determined distances onto a bird's eye view
of the Galactic disc in Fig.~\ref{dis}. These \ion{H}{II} regions can
be traced from the Norma Arm in the Galactic centre to the Outer+1
Arm. Most \ion{H}{II} regions are located in the spiral arms, while
some are located in the inter-arm area.

\begin{figure}
\includegraphics[angle=-90, width=0.45\textwidth]{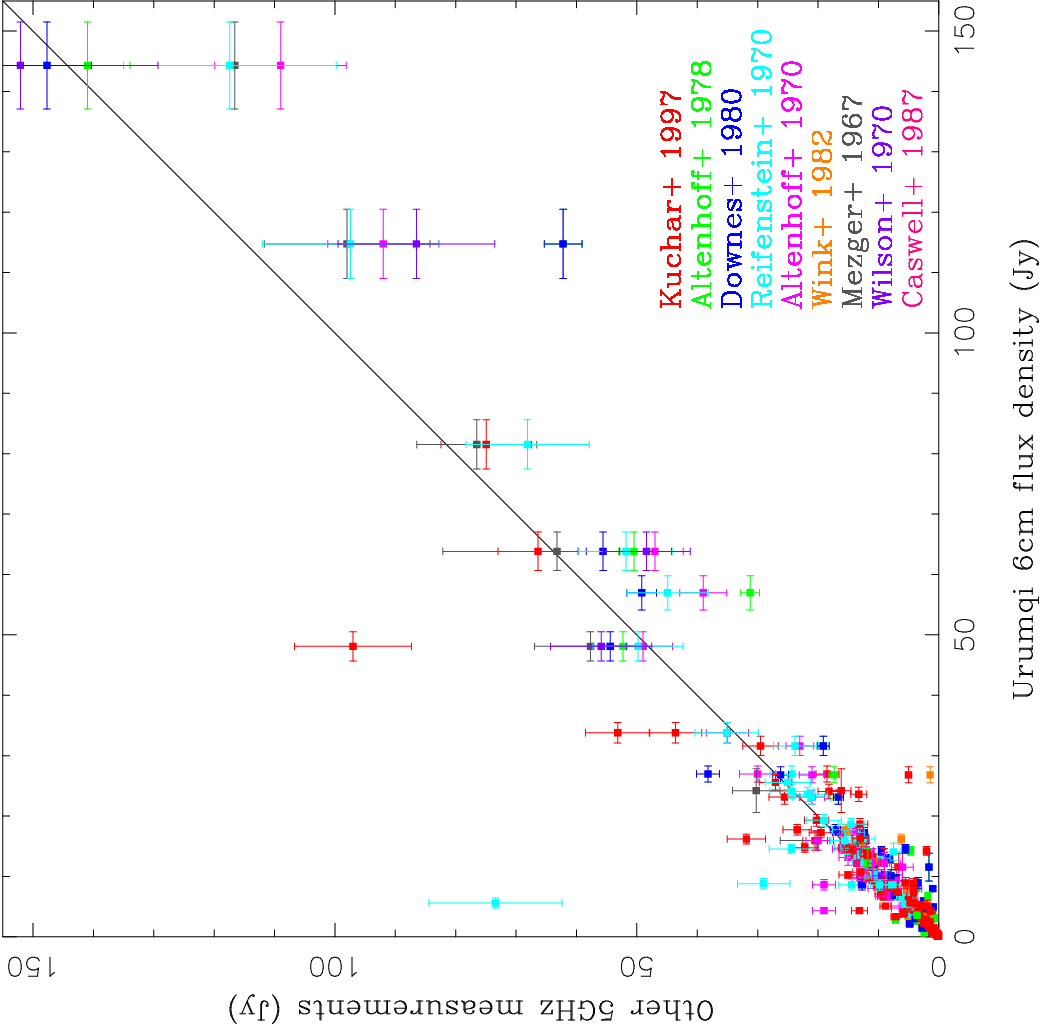}
\caption{Flux density comparison between the new $\lambda$6\ cm
  measurements and previous $\lambda$6\ cm results.}
\label{compare}
\end{figure}

\begin{figure*}
\includegraphics[angle=-90, width=0.32\textwidth]{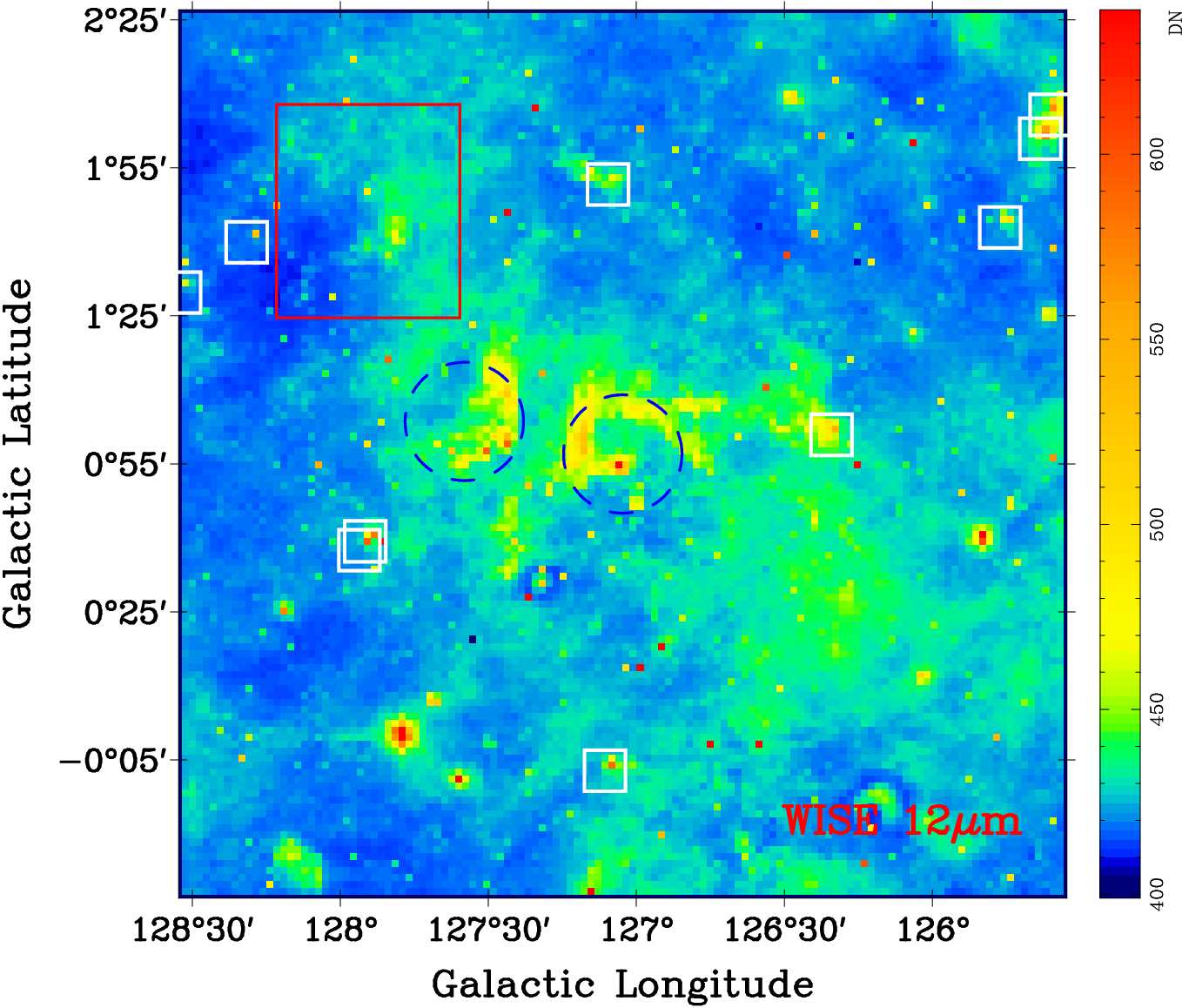}
\includegraphics[angle=-90, width=0.32\textwidth]{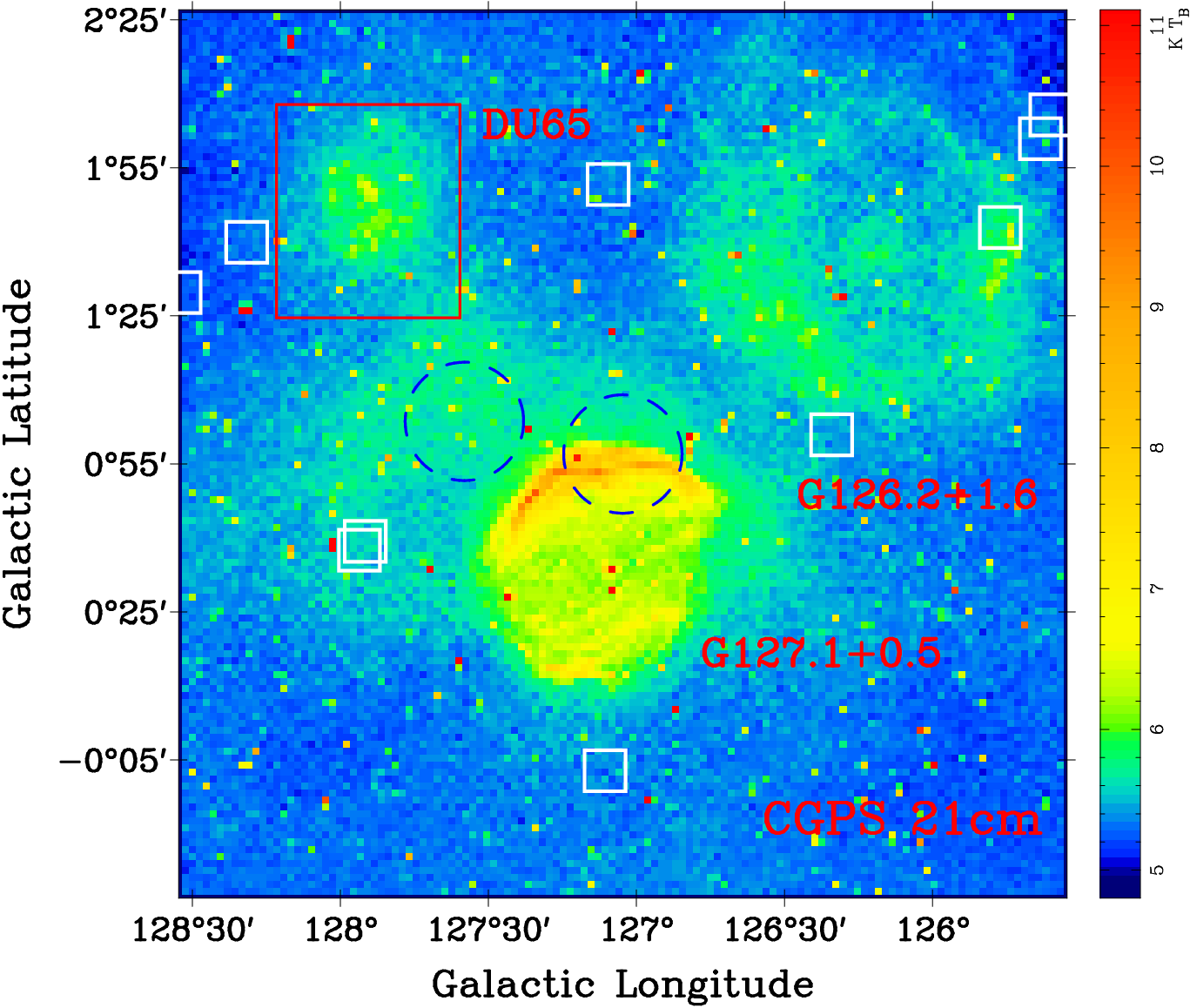}
\includegraphics[angle=-90, width=0.32\textwidth]{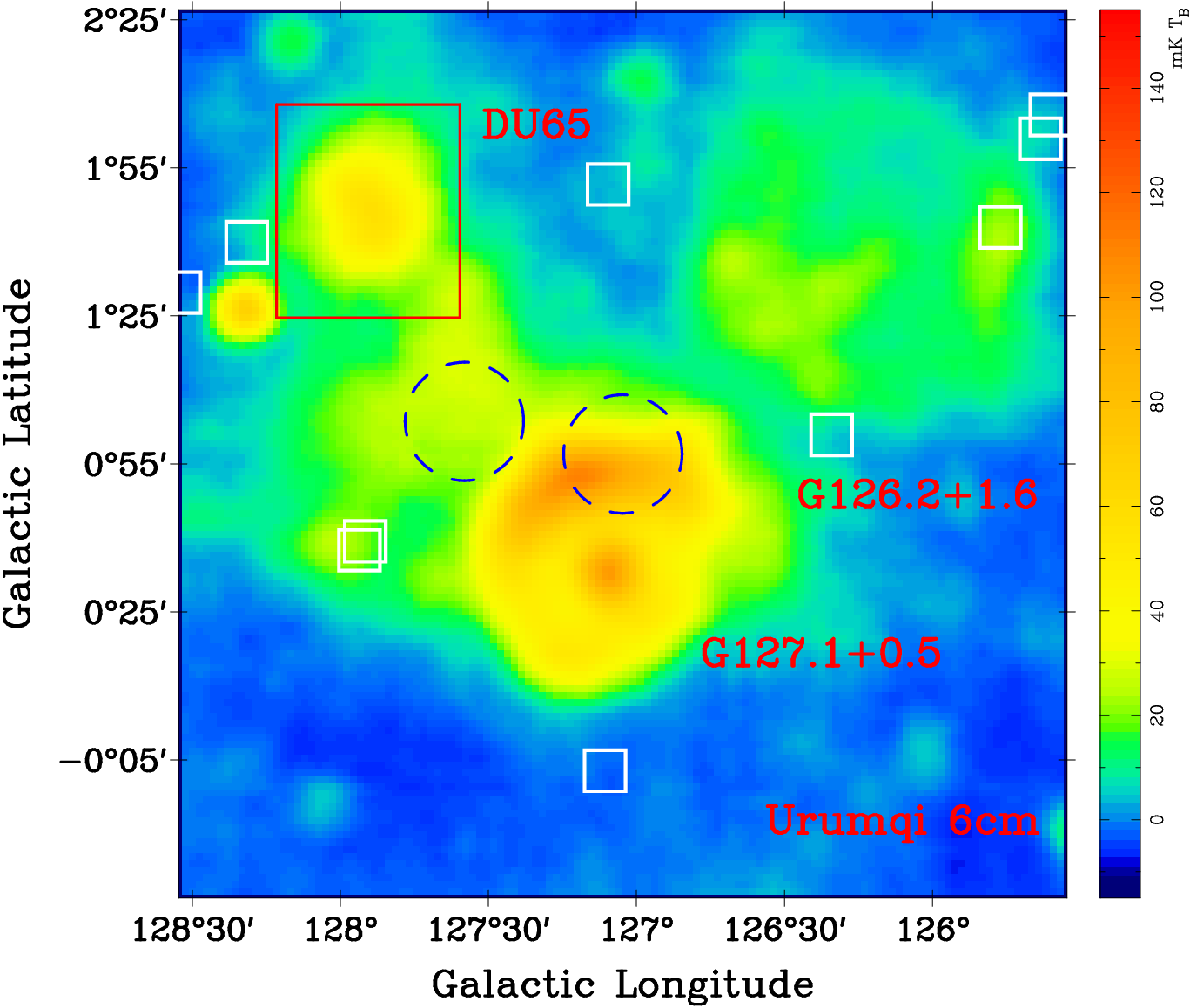}

\includegraphics[angle=-90, width=0.32\textwidth]{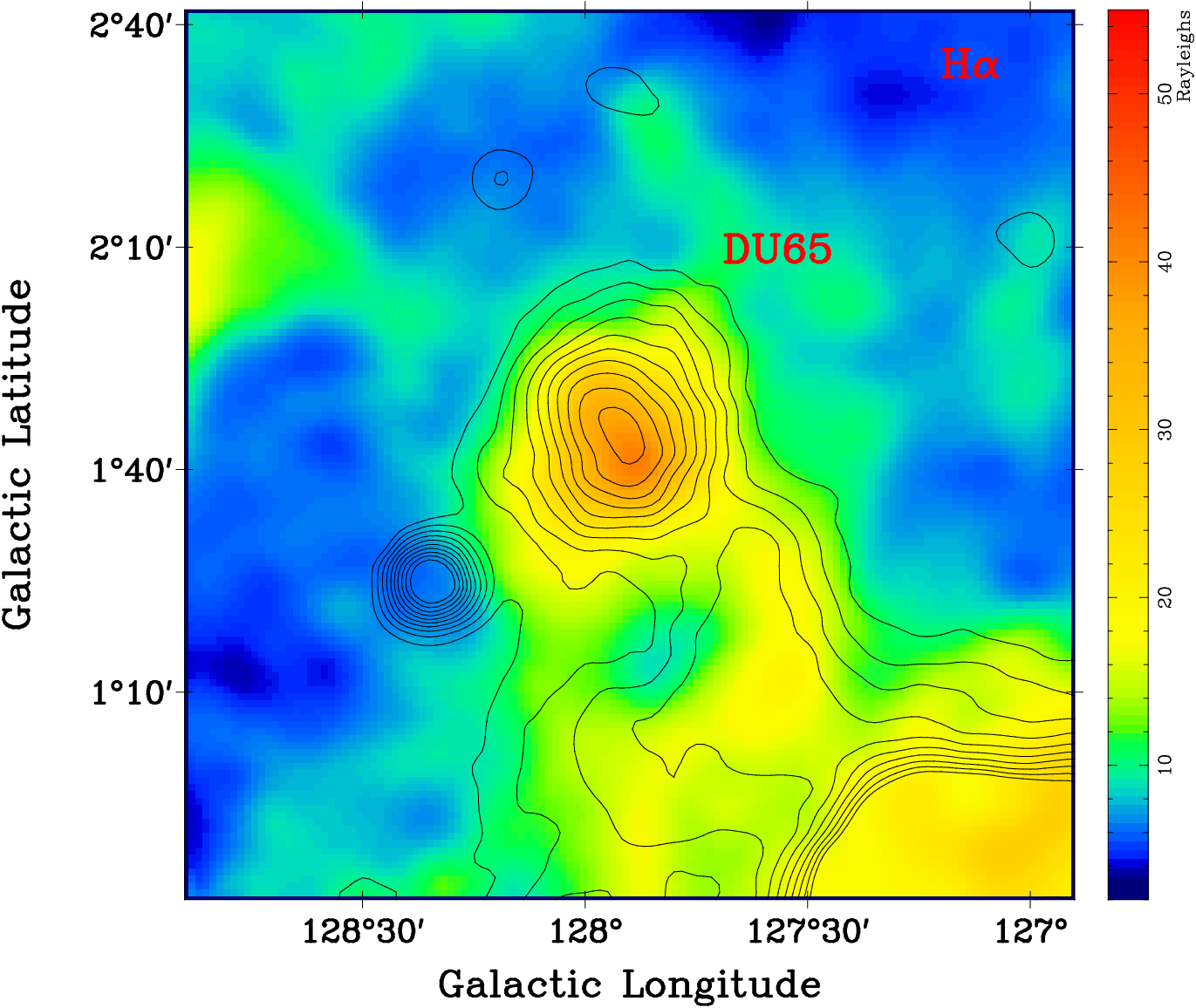}
\includegraphics[angle=-90, width=0.32\textwidth]{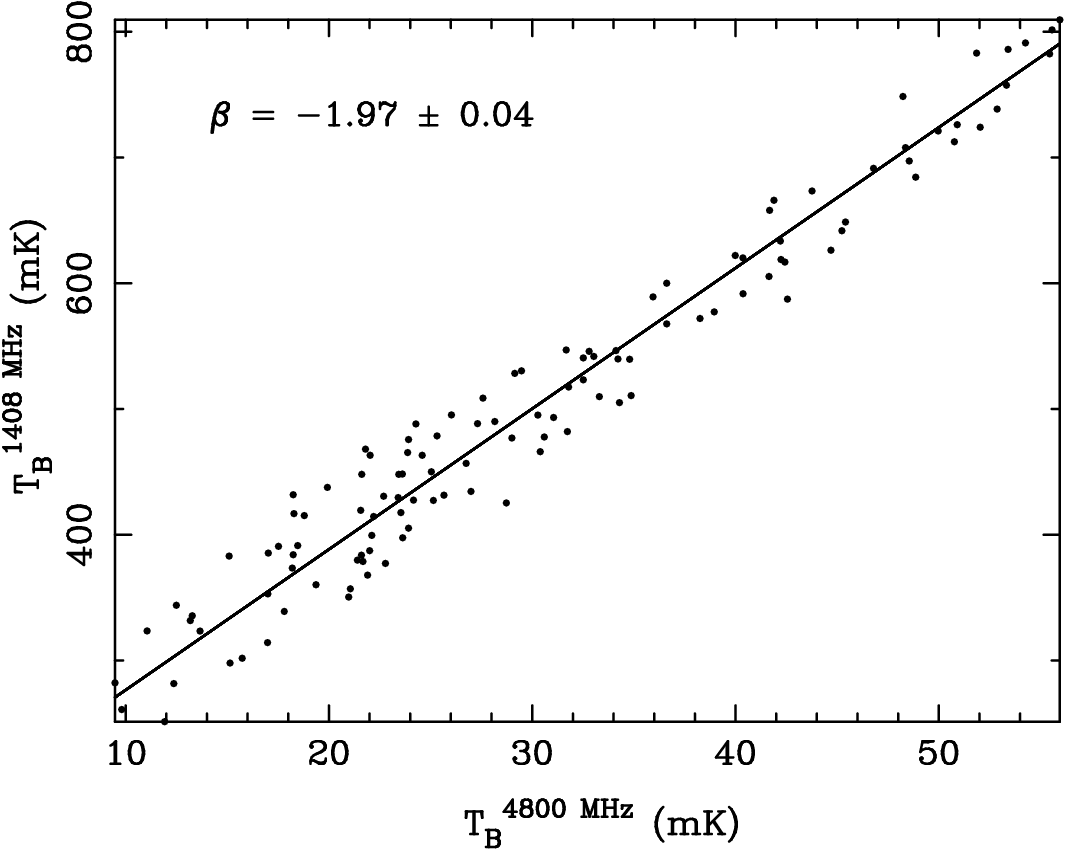}
\includegraphics[angle=-90, width=0.32\textwidth]{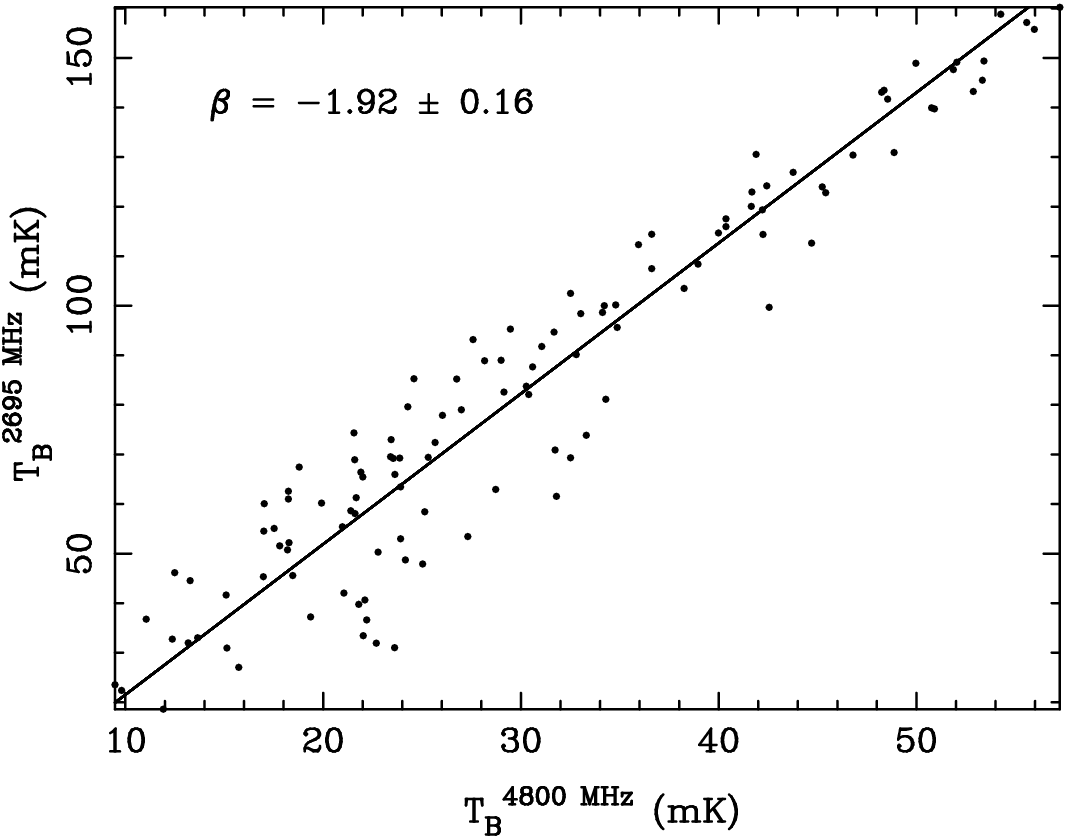}
\caption{{\it Upper panels from left to right:} WISE 12$\mu$m, CGPS
  $\lambda$21\ cm, and Urumqi $\lambda$6\ cm images centred at $\ell =
  127\fdg046, b = 0\fdg950$. Small-diameter ($<$ 7$\arcmin$) WISE
  \ion{H}{II} regions from \citet{abb+14} are marked by white squares,
  and two extended WISE \ion{H}{II} regions (diameter $>$ 20$\arcmin$)
  are indicated by blue dashed circles. Three very extended radio
  sources in both $\lambda$21\ cm and $\lambda$6\ cm images are SNR
  G126.2+1.6, SNR G127.1+0.5, and \ion{H}{II} region G127.9+1.7
  \citep[DU~65,][ red rectangle]{Dubout-Crillon76}, which is missing
  in the WISE \ion{H}{II} region catalogue. In the {\it lower left
    panel}, its H$\alpha$ emission is shown overlaid by the
  $\lambda$6\ cm total-intensity emission. The {\it lower middle and
    lower right panels} show the TT-plot results for DU~65 by using
  the Urumqi $\lambda$6\ cm against the Effelsberg $\lambda$21\ cm
  data and the Effelsberg $\lambda$11\ cm data.}
\label{G127}
\end{figure*}

As listed in Table~\ref{pala} and for some sources in
Table~\ref{extended}, \citet{Paladini03} collected many \ion{H}{II}
regions which were previously measured at $\lambda$6\ cm. We made a
comparison between the new results derived from the Urumqi data and
those flux densities quoted in \citet{Paladini03}. We show the result
in Fig.~\ref{compare}. The ratio is approximately 1 in general,
proving the consistency, but the Urumqi flux densities are somewhat
higher for those sources with higher flux densities. For some of the
sources, one \ion{H}{II} region seen in the Urumqi data can be
resolved into several components, for example in \citet{Altenhoff78}
and \citet{Kuchar97}. These components have to be added first, before
the comparison. The angular resolution of the Urumqi survey is only
slightly better than that of \citet{Altenhoff70}, but coarser than all
the other eight surveys. Individual components of \ion{H}{II} regions
or underlying emission were possibly included in $\lambda$6\ cm
sources when the Urumqi data cannot resolve them (e.g. $\lambda$6\ cm
source G192.6$-$0.0 includes five \ion{H}{II} regions: SH 2-254 to
258). The seeming overestimate of the Urumqi $\lambda$6\ cm results is
also caused by the fact that high angular resolution observations did
not cover the entire, but only a part of the \ion{H}{II} regions, for
example G119.4$-$0.8 from \citet{Kuchar97}.

There are some interesting newly identified objects in
Table~\ref{extended}, and so in the following we first discuss a
smaller number of objects in the WISE \ion{H}{II} region catalogue,
and then talk about many objects in the Paladini \ion{H}{II} region
catalogue.

\begin{figure*}[!htb]
\centering
\includegraphics[angle=-90, width=0.28\textwidth]{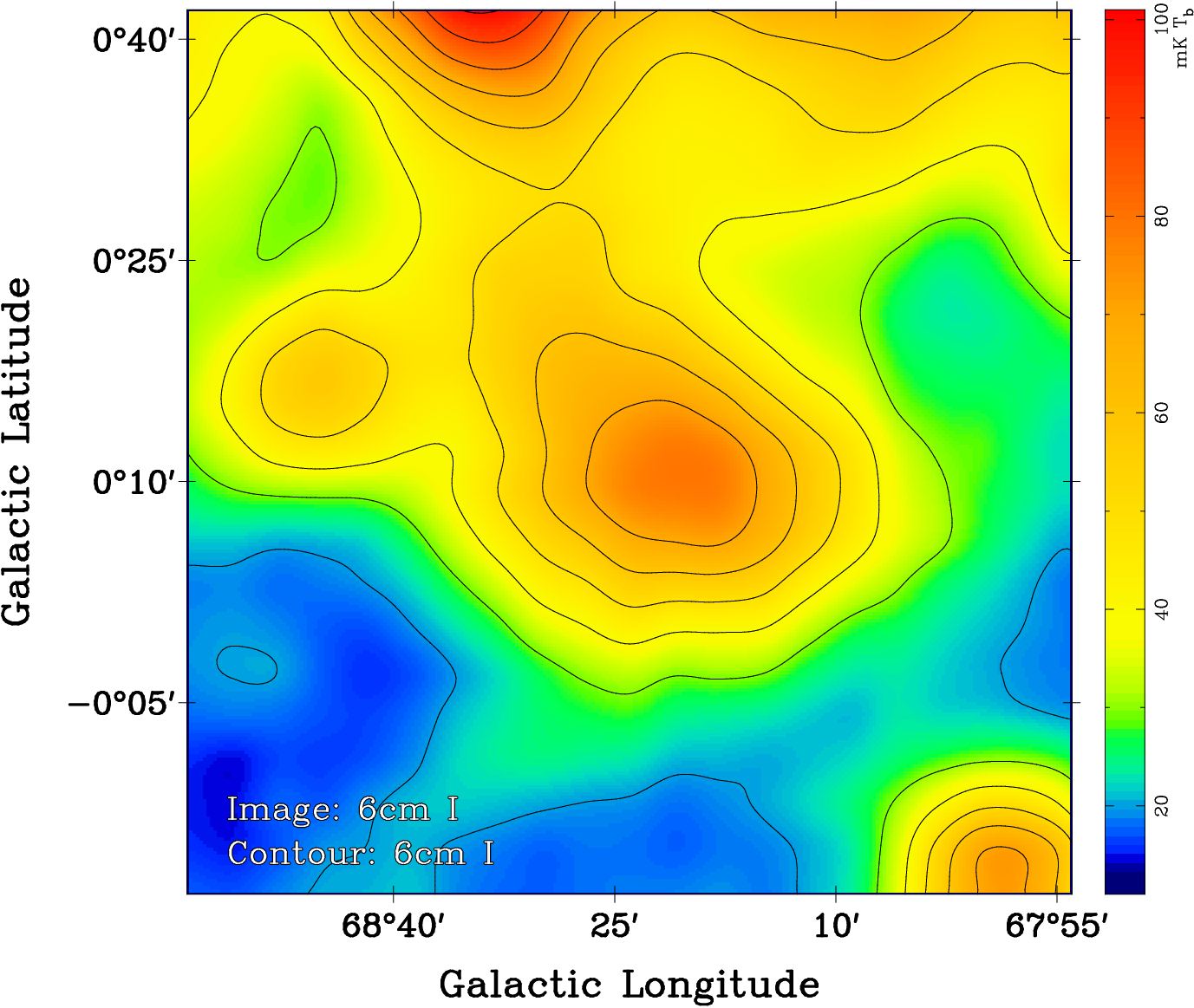}
\includegraphics[angle=-90, width=0.28\textwidth]{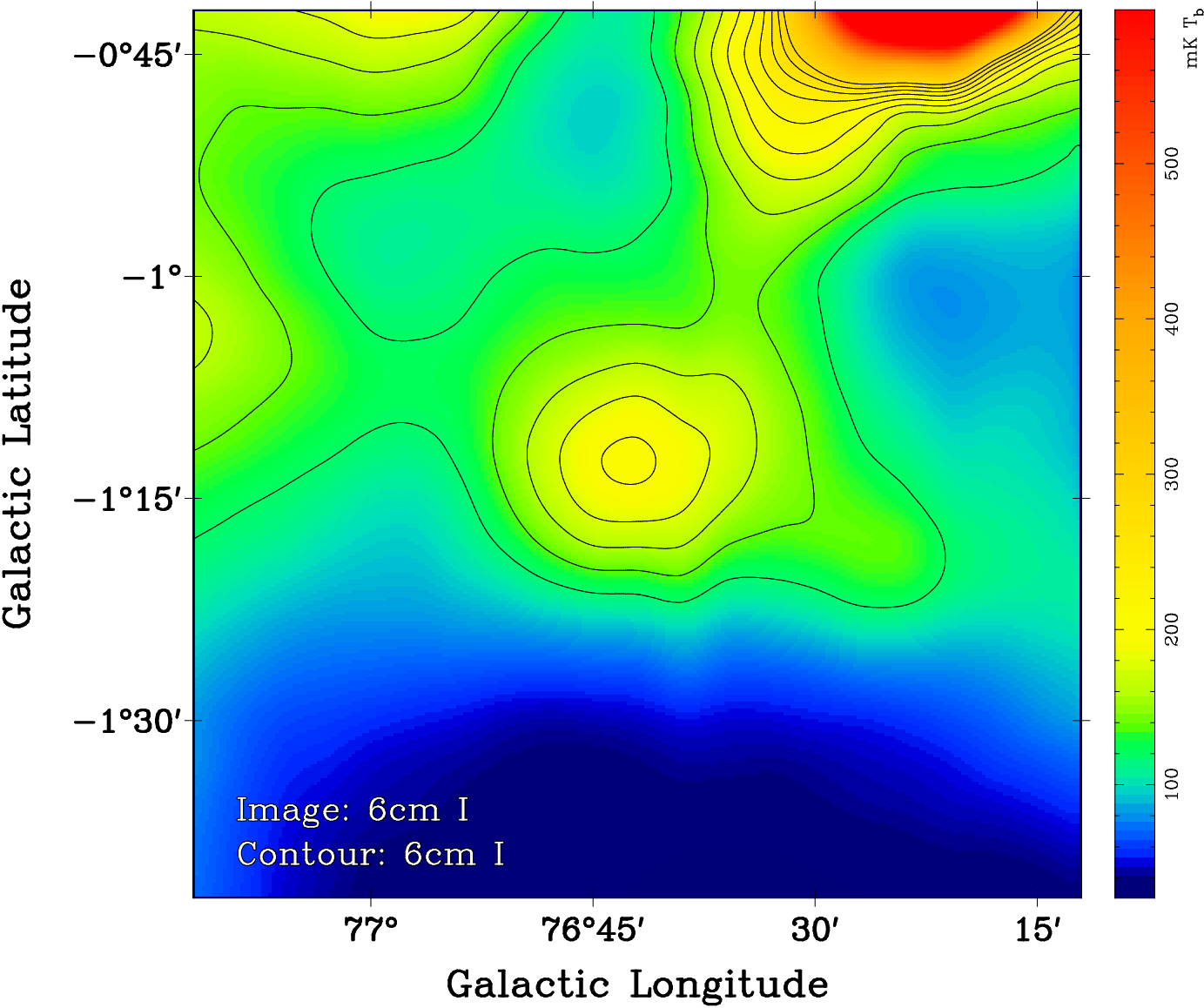}
\includegraphics[angle=-90, width=0.28\textwidth]{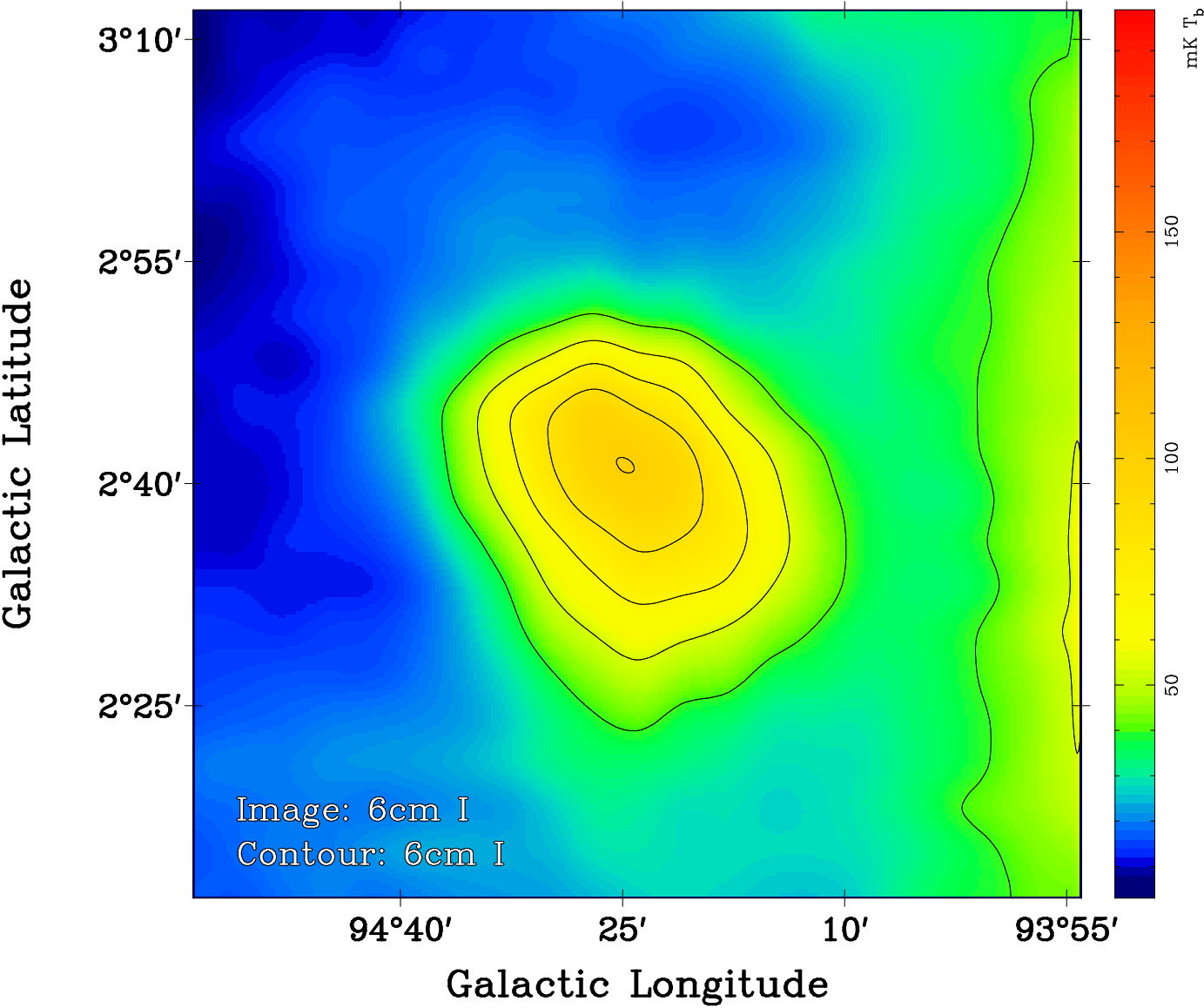}

\includegraphics[angle=-90, width=0.28\textwidth]{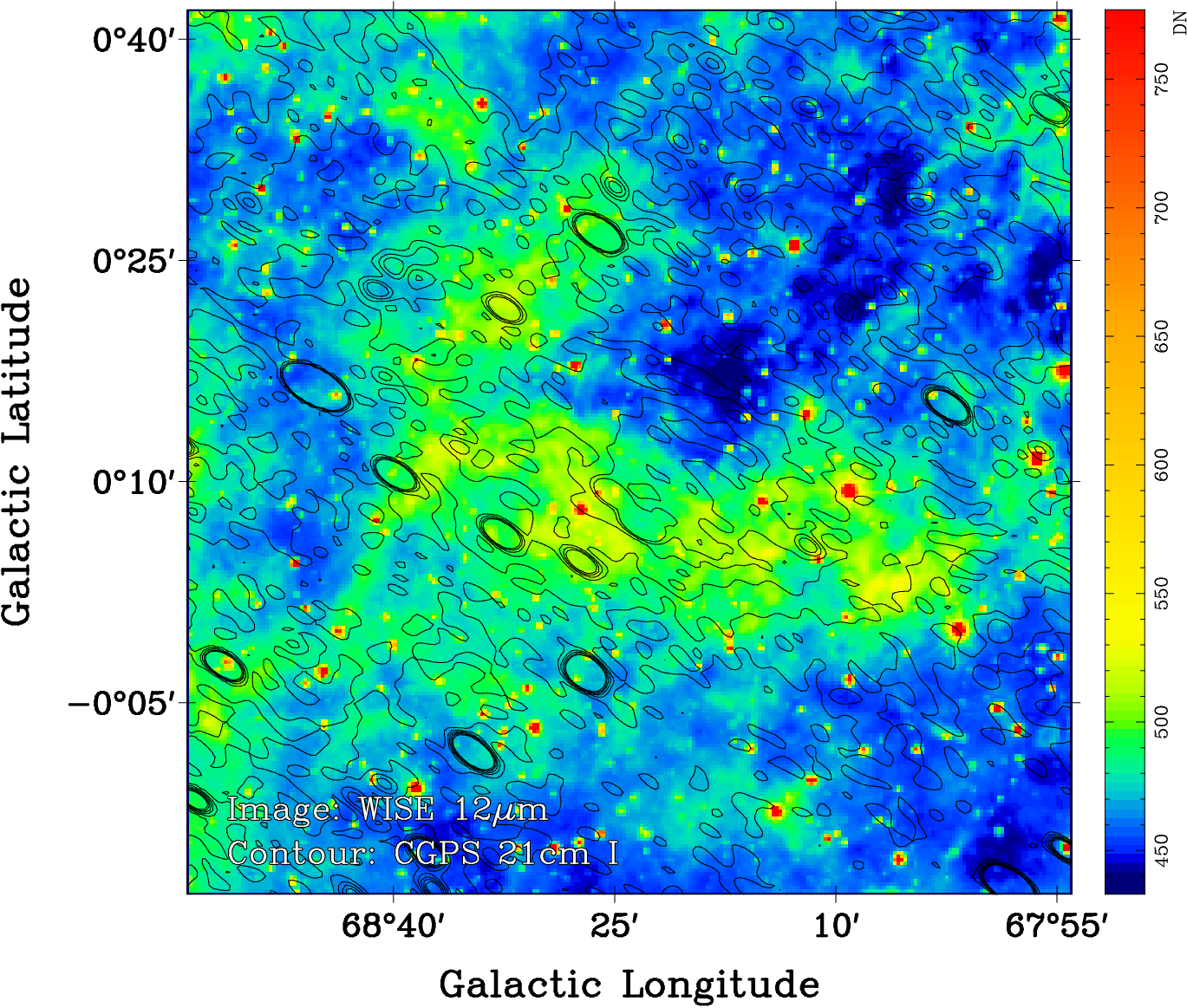}
\includegraphics[angle=-90, width=0.28\textwidth]{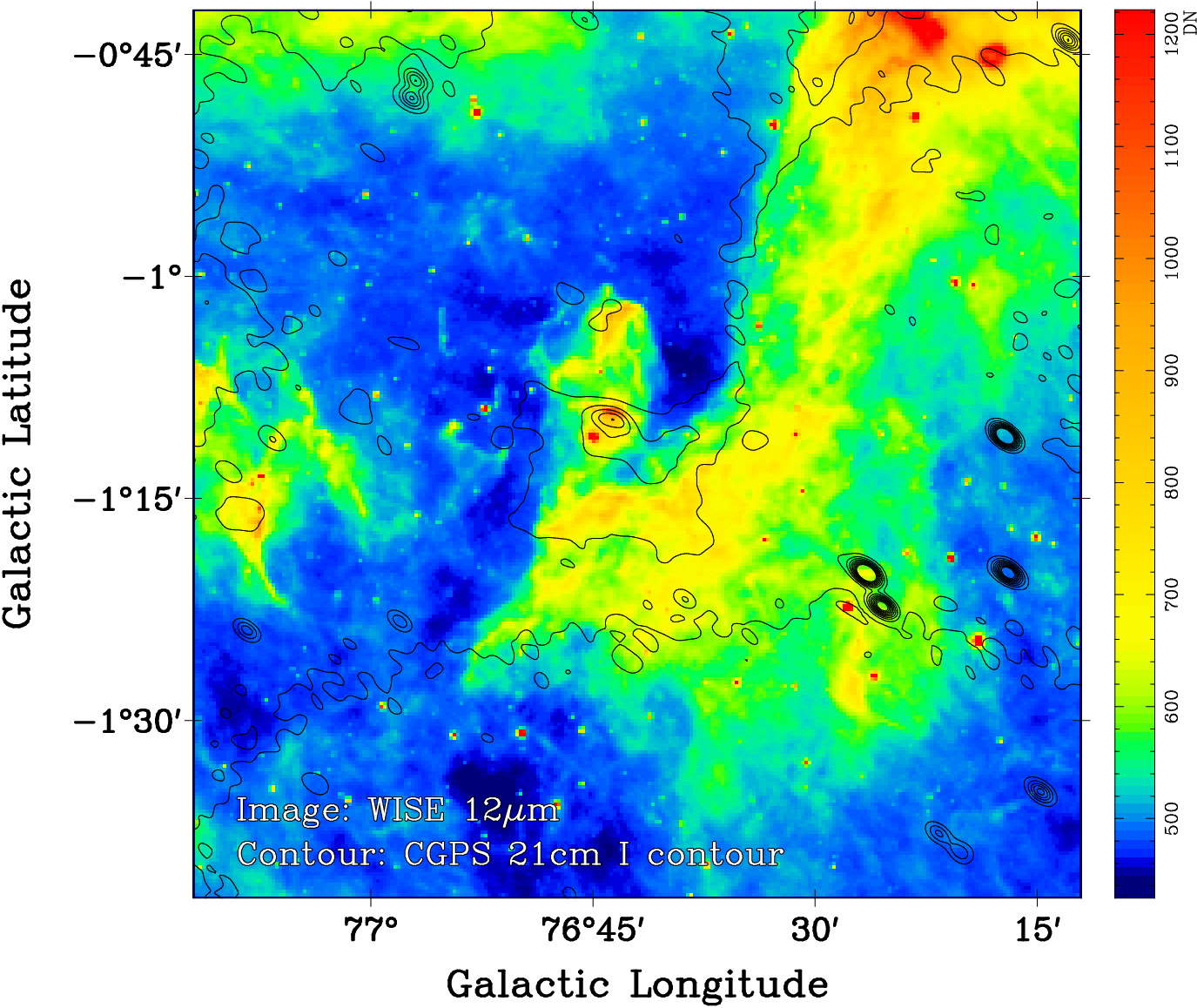}
\includegraphics[angle=-90, width=0.28\textwidth]{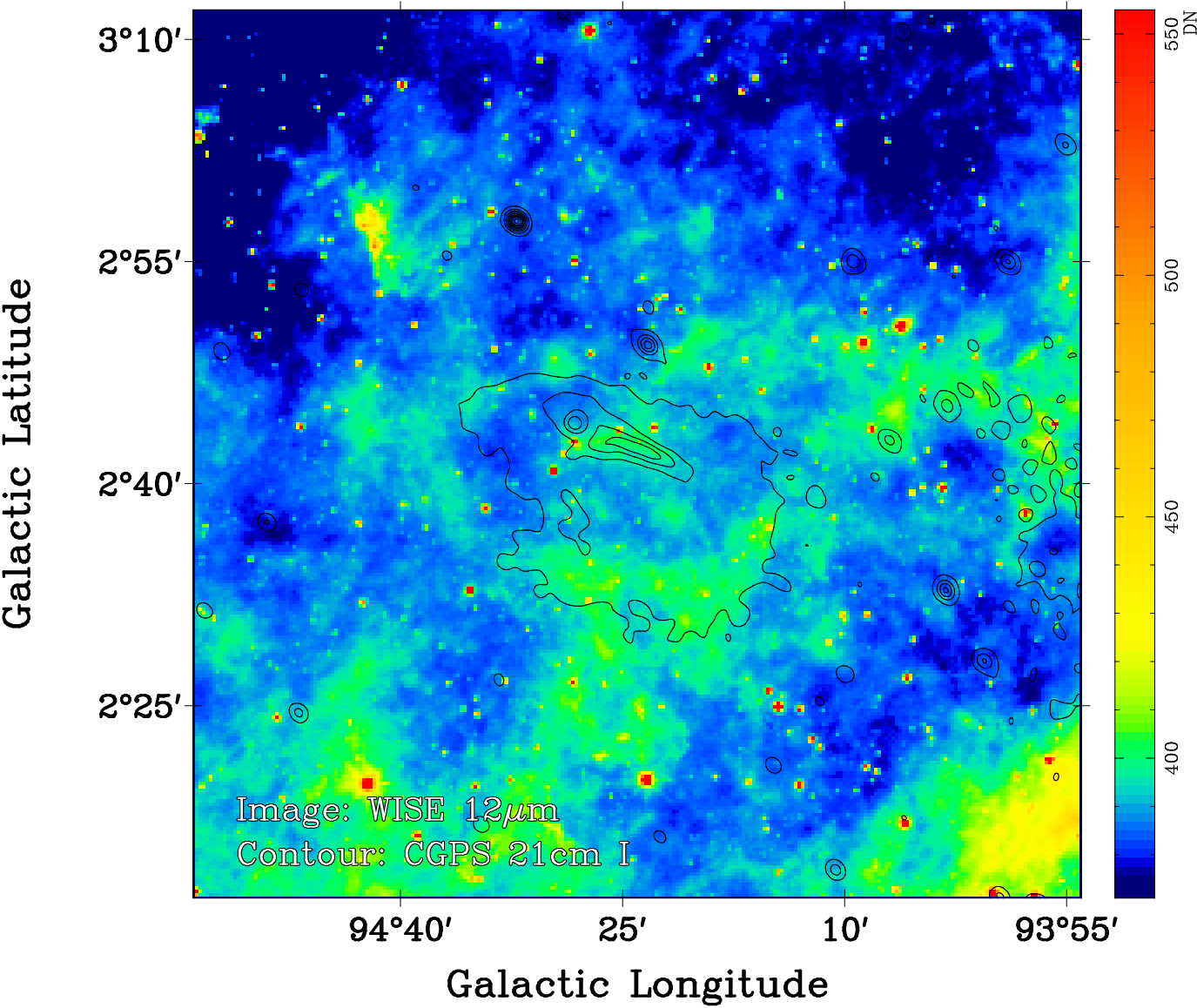}

\includegraphics[angle=-90, width=0.28\textwidth]{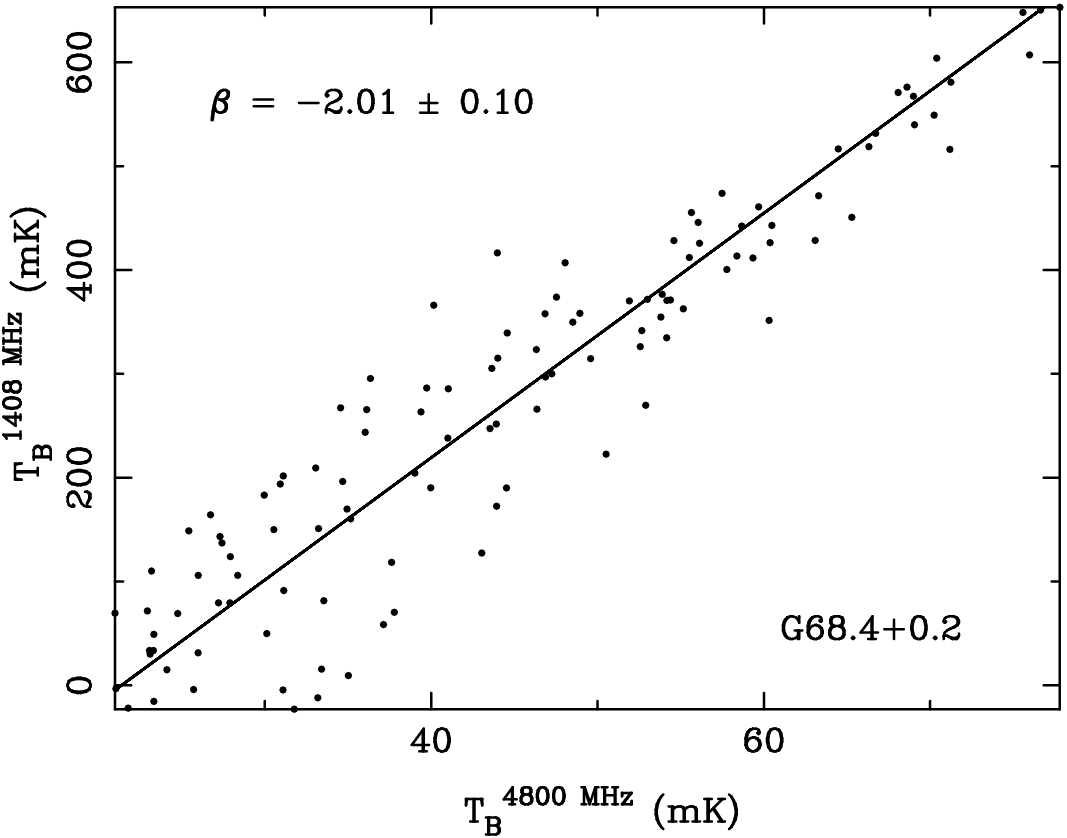}
\includegraphics[angle=-90, width=0.28\textwidth]{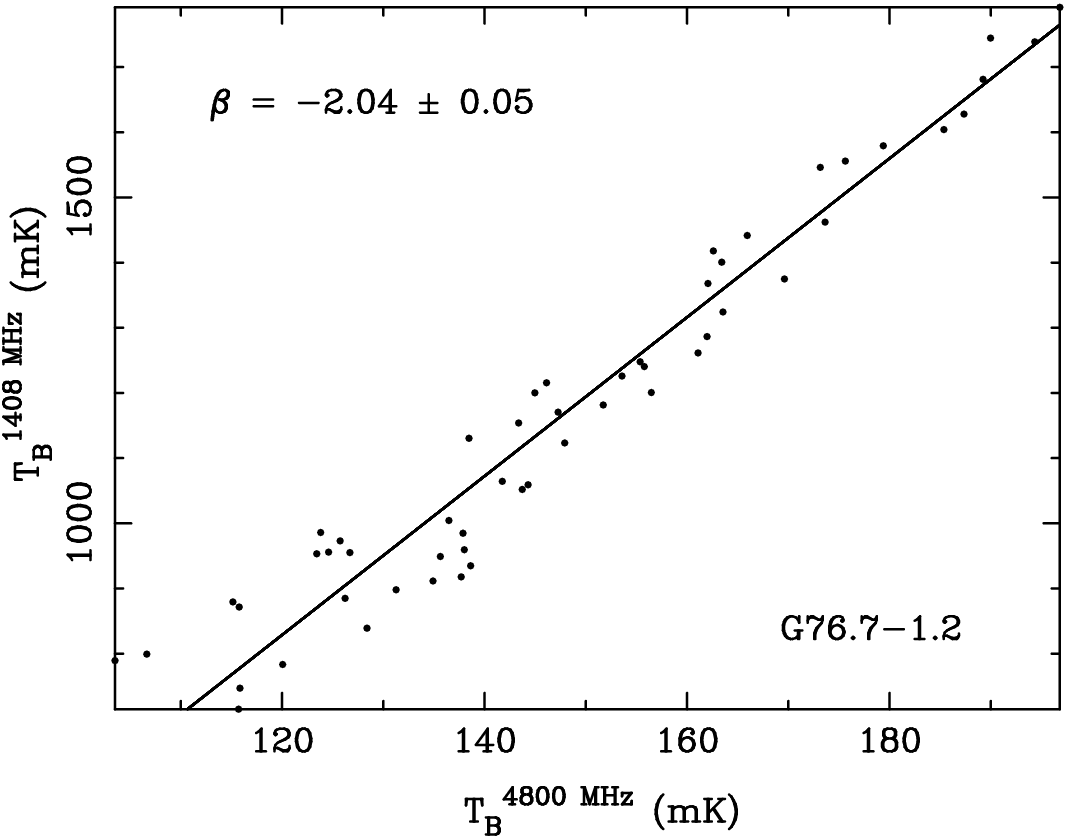}
\includegraphics[angle=-90, width=0.28\textwidth]{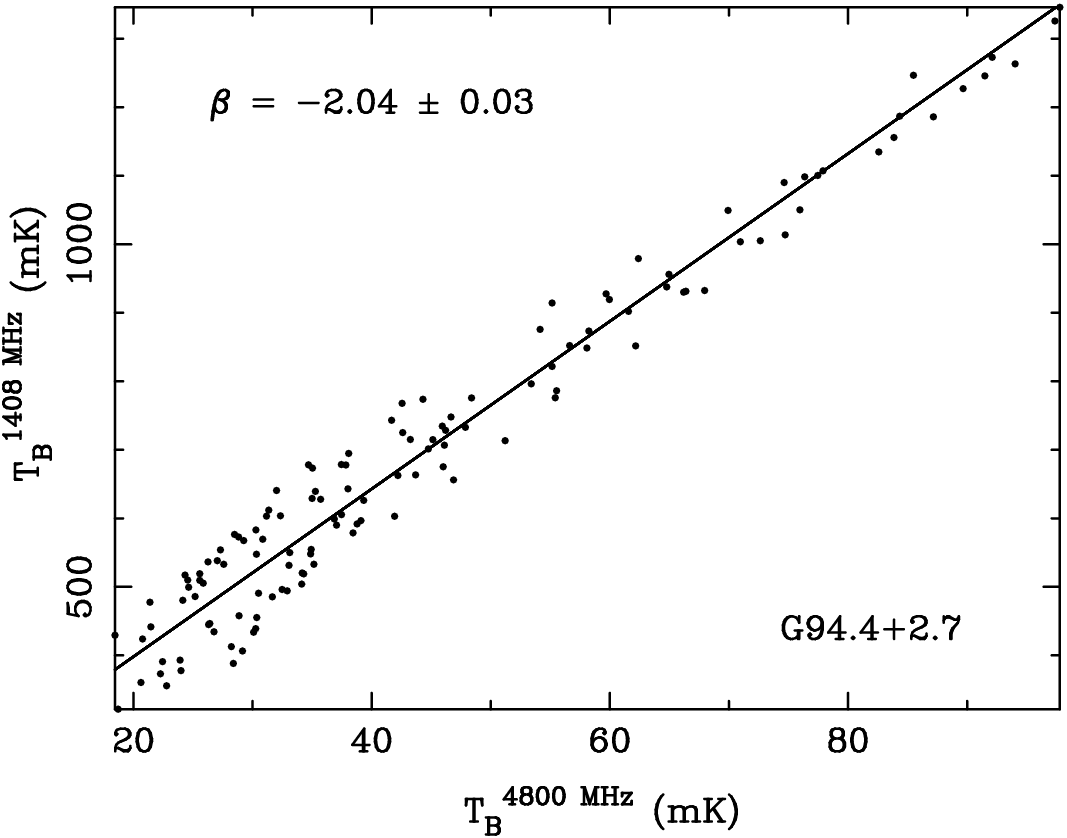}
\caption{New \ion{H}{II} regions G68.4+0.2, G76.7$-$1.2, and G94.4+2.7
  identified in the Sino-German $\lambda$6\ cm survey. We show
  $\lambda$6\ cm images and WISE 12$\mu$m emission overlaid by CGPS
  $\lambda$21\ cm total intensity in the {\it upper and middle
    panels}. We present their TT-plot results in the {\it lower
    panels}.}
\label{newHII1}
%
\centering
\includegraphics[angle=-90, width=0.28\textwidth]{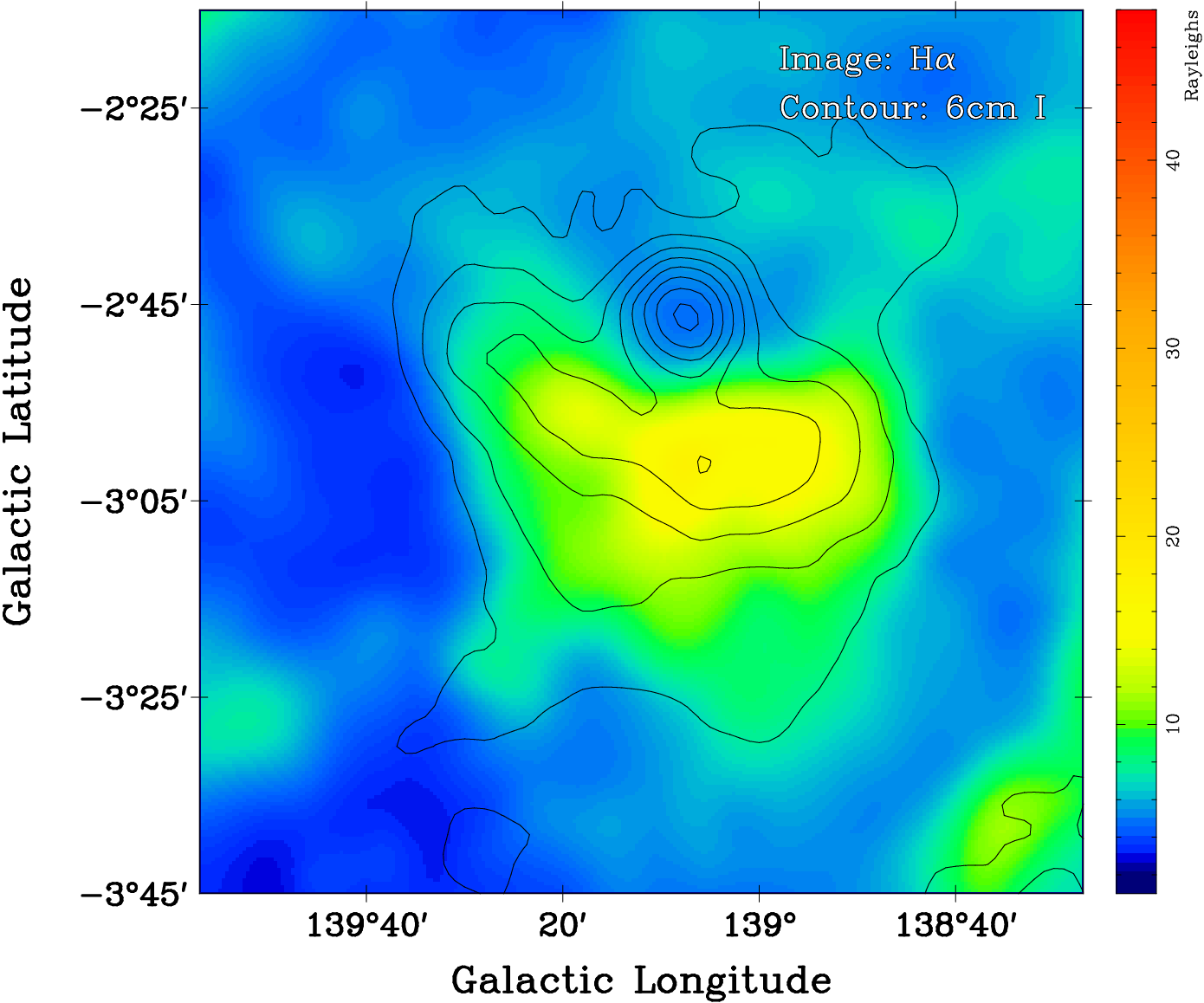}
\includegraphics[angle=-90, width=0.28\textwidth]{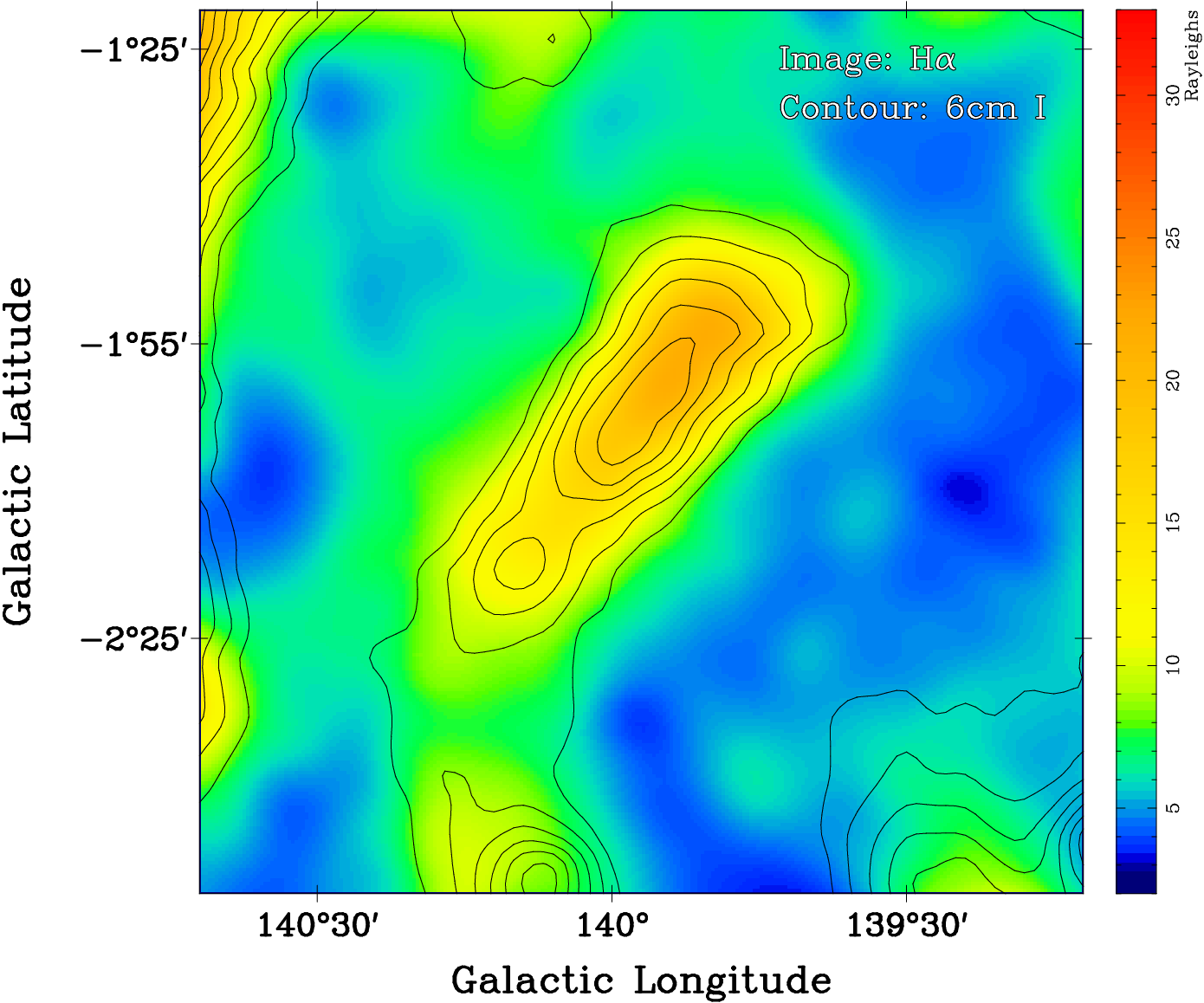}
\includegraphics[angle=-90, width=0.28\textwidth]{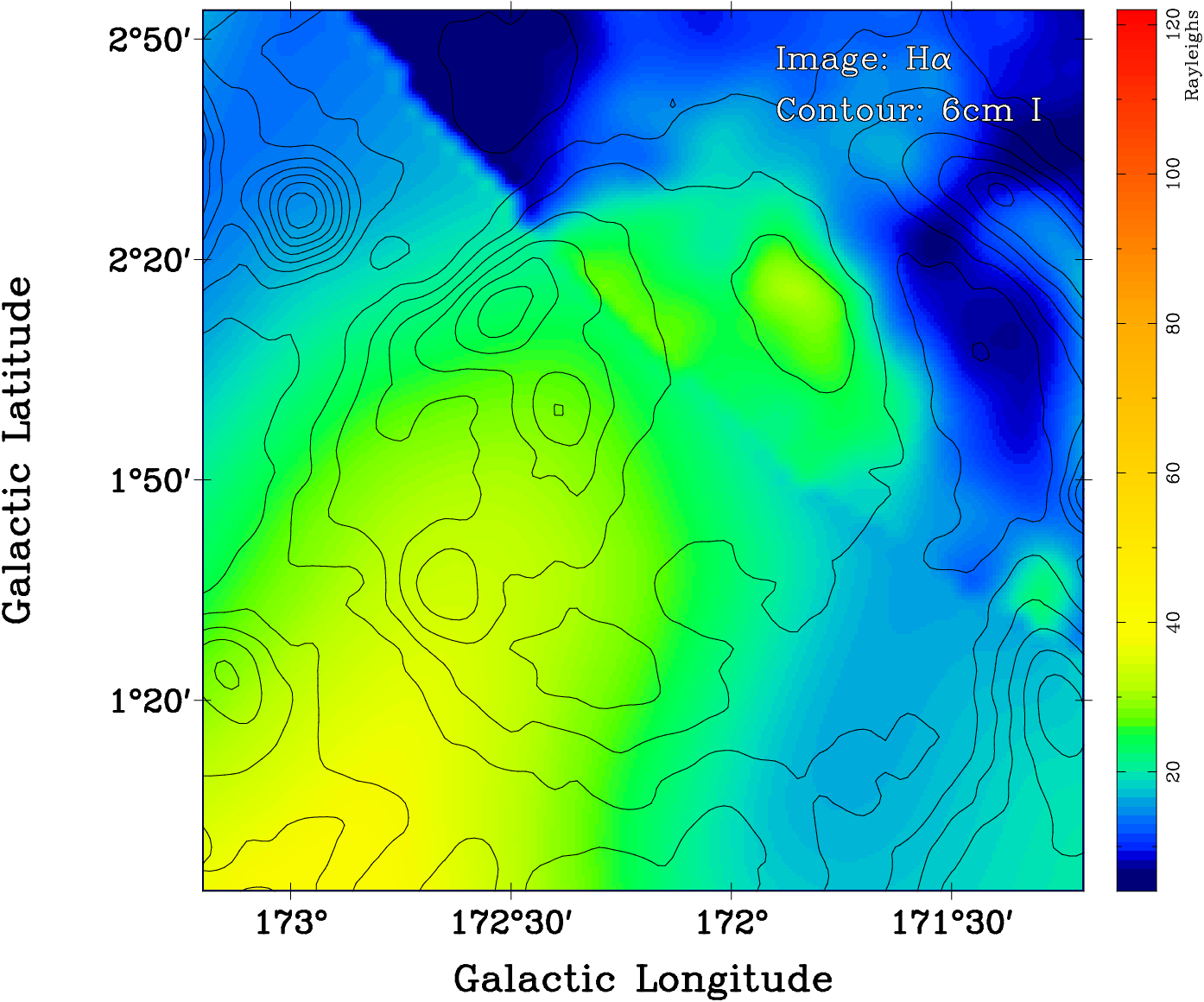}

\includegraphics[angle=-90, width=0.28\textwidth]{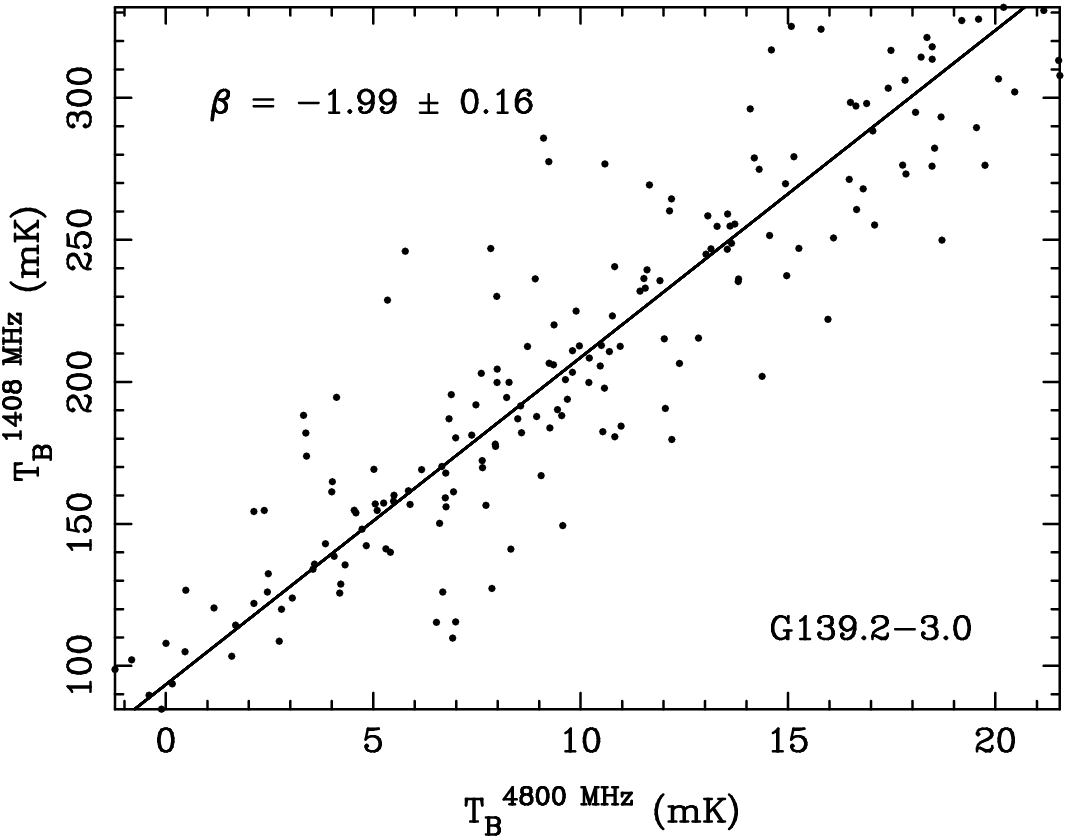}
\includegraphics[angle=-90, width=0.28\textwidth]{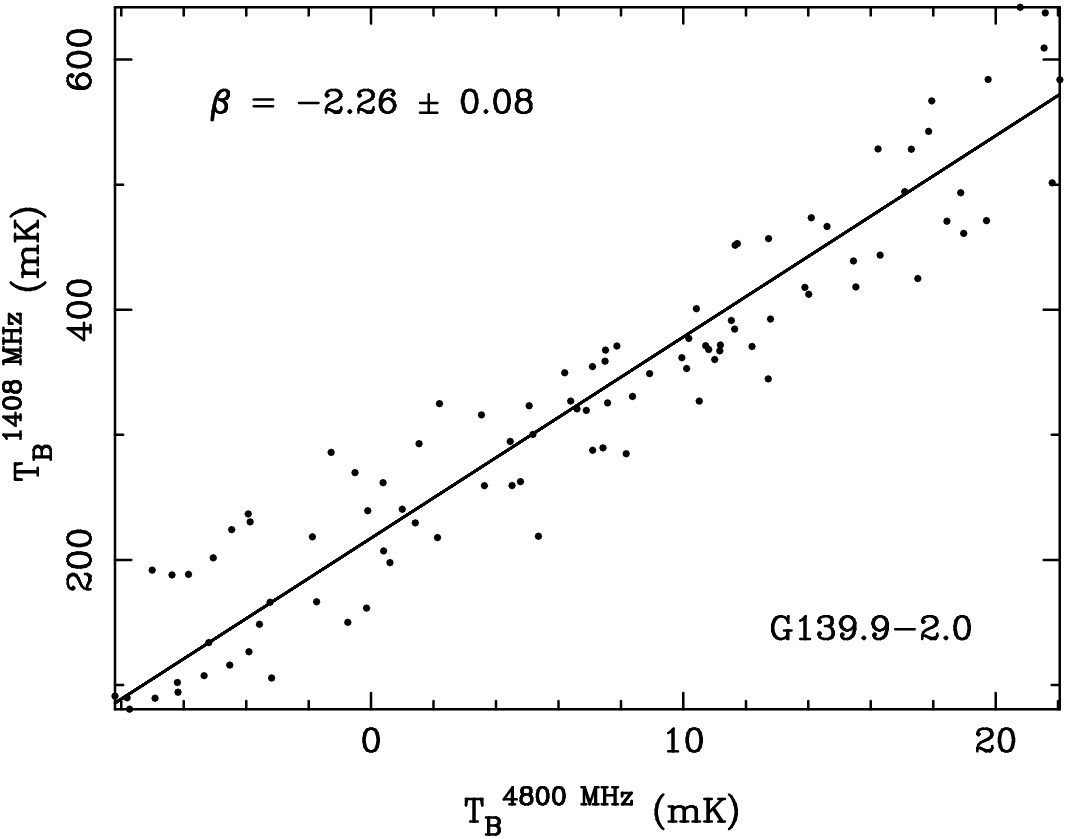}
\includegraphics[angle=-90, width=0.28\textwidth]{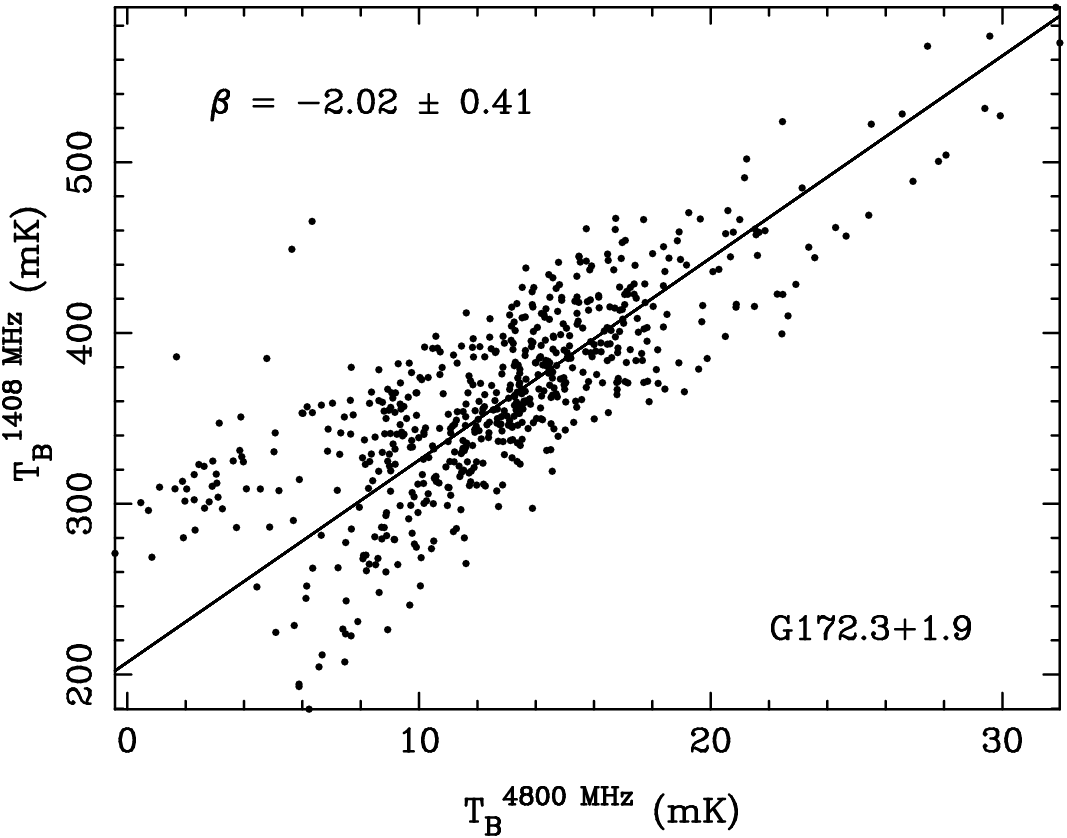}
\caption{Newly identified \ion{H}{II} regions G139.2$-$3.0,
  G139.9$-$2.0, and G172.3+1.9. H$\alpha$ emission overlaid by Urumqi
  $\lambda$6\ cm total intensity contours are shown in the {\it upper
    panel}. The image quality of the south-east part of G172.3+1.9 is
  not good; however, correlation can be clearly recognised in the
  north-west part. TT-plot results are displayed in the {\it lower
    panels}.}
\label{newHII2}
\end{figure*} 
\begin{figure*}
\begin{tabular}{ccc}
\includegraphics[angle=-90, width=0.3\textwidth]{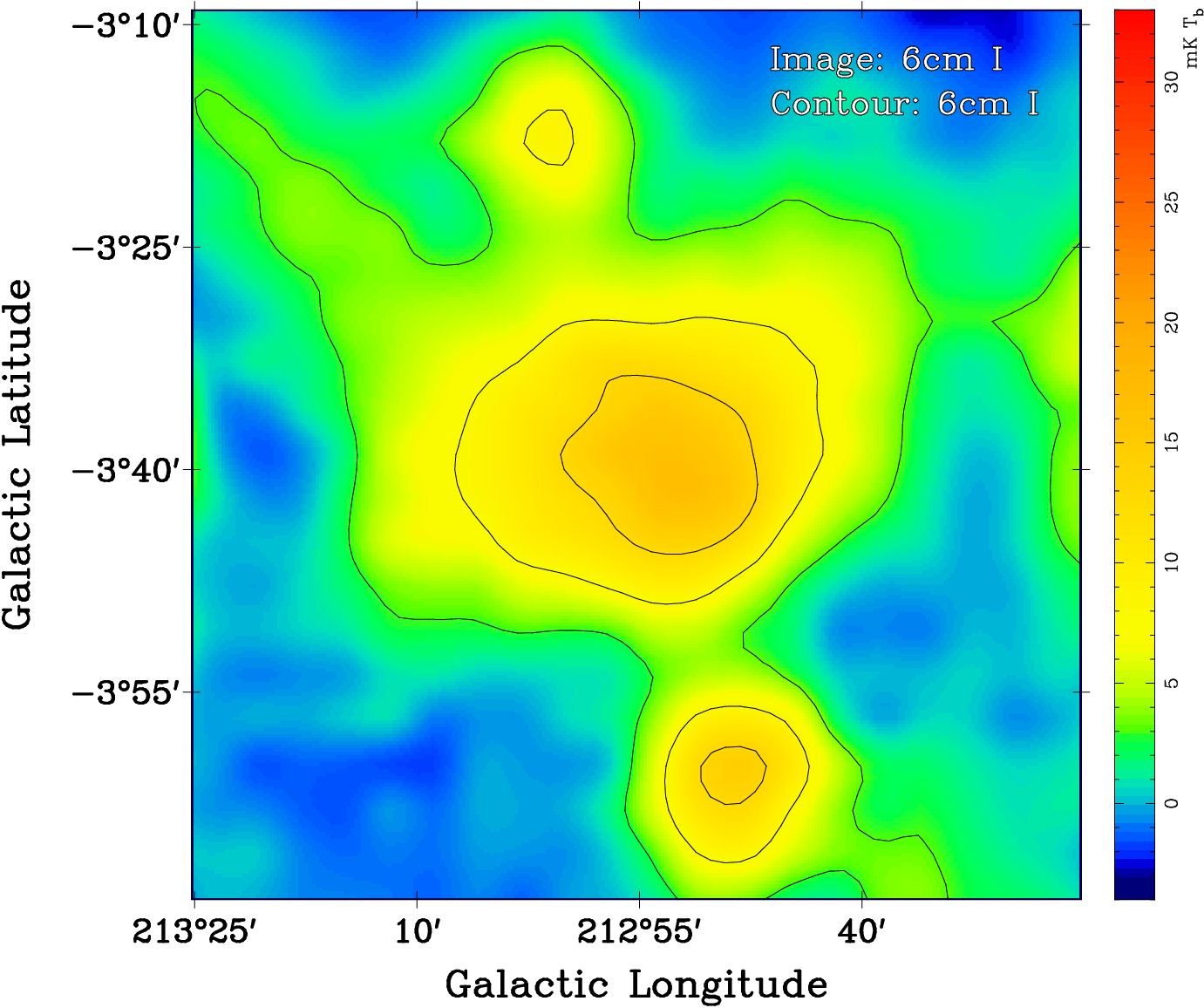} &
\includegraphics[angle=-90, width=0.3\textwidth]{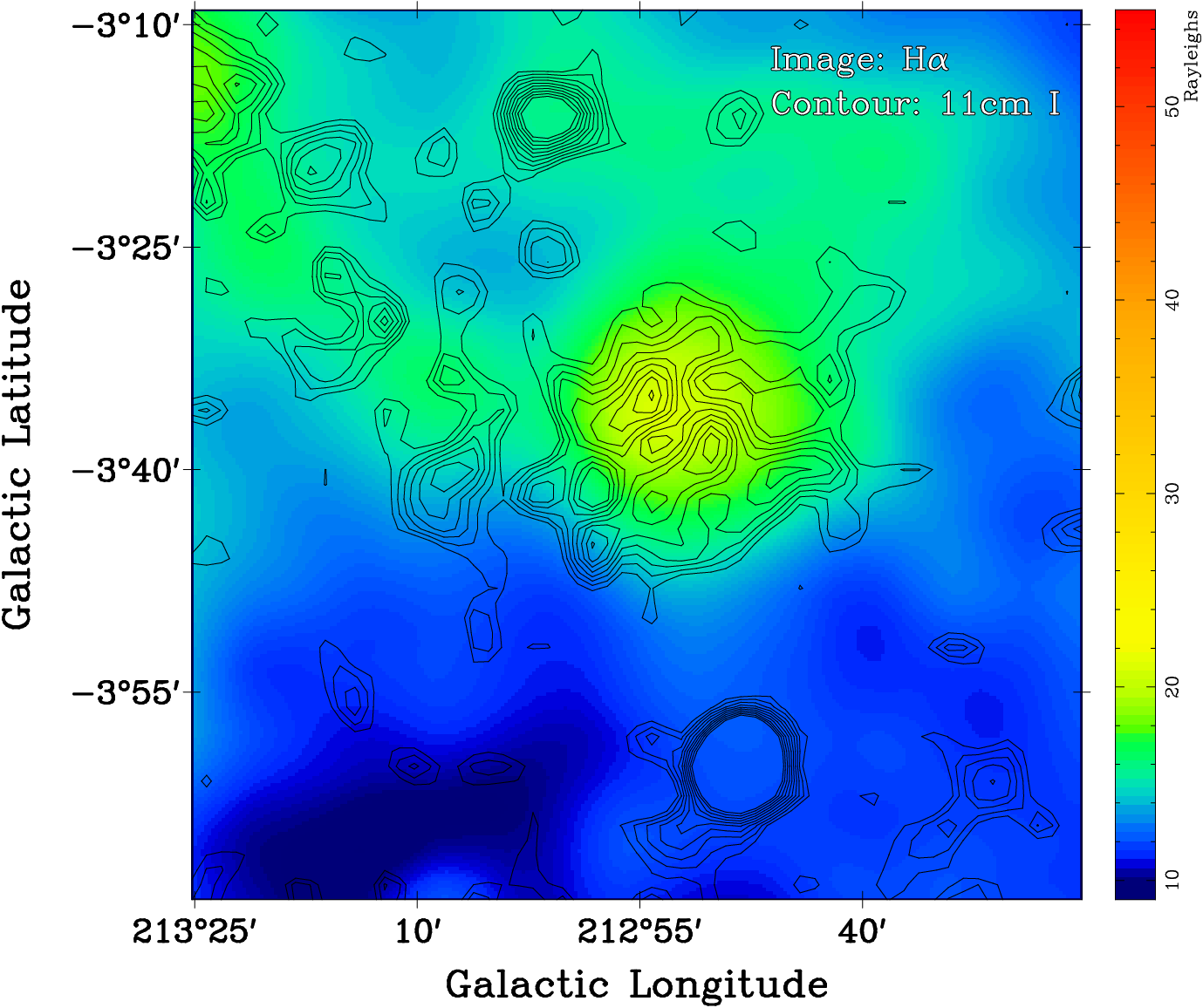} &
\includegraphics[angle=-90, width=0.3\textwidth]{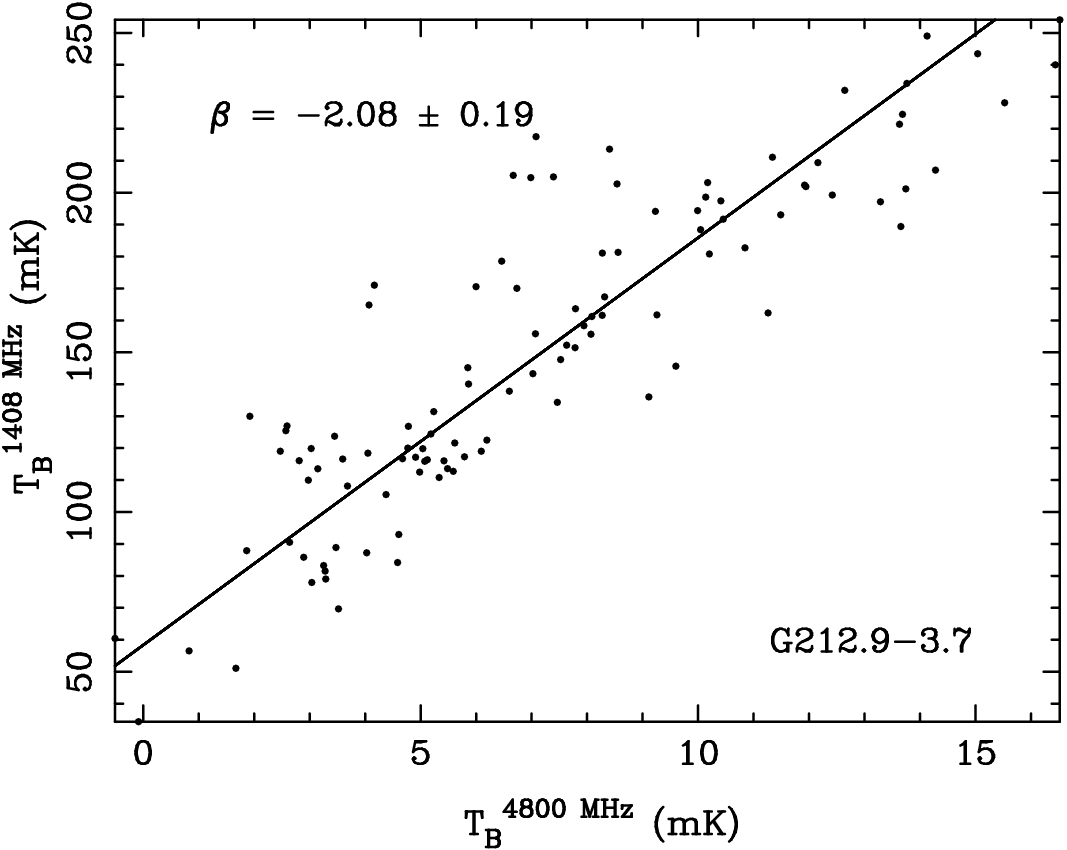} \\
\includegraphics[angle=-90, width=0.3\textwidth]{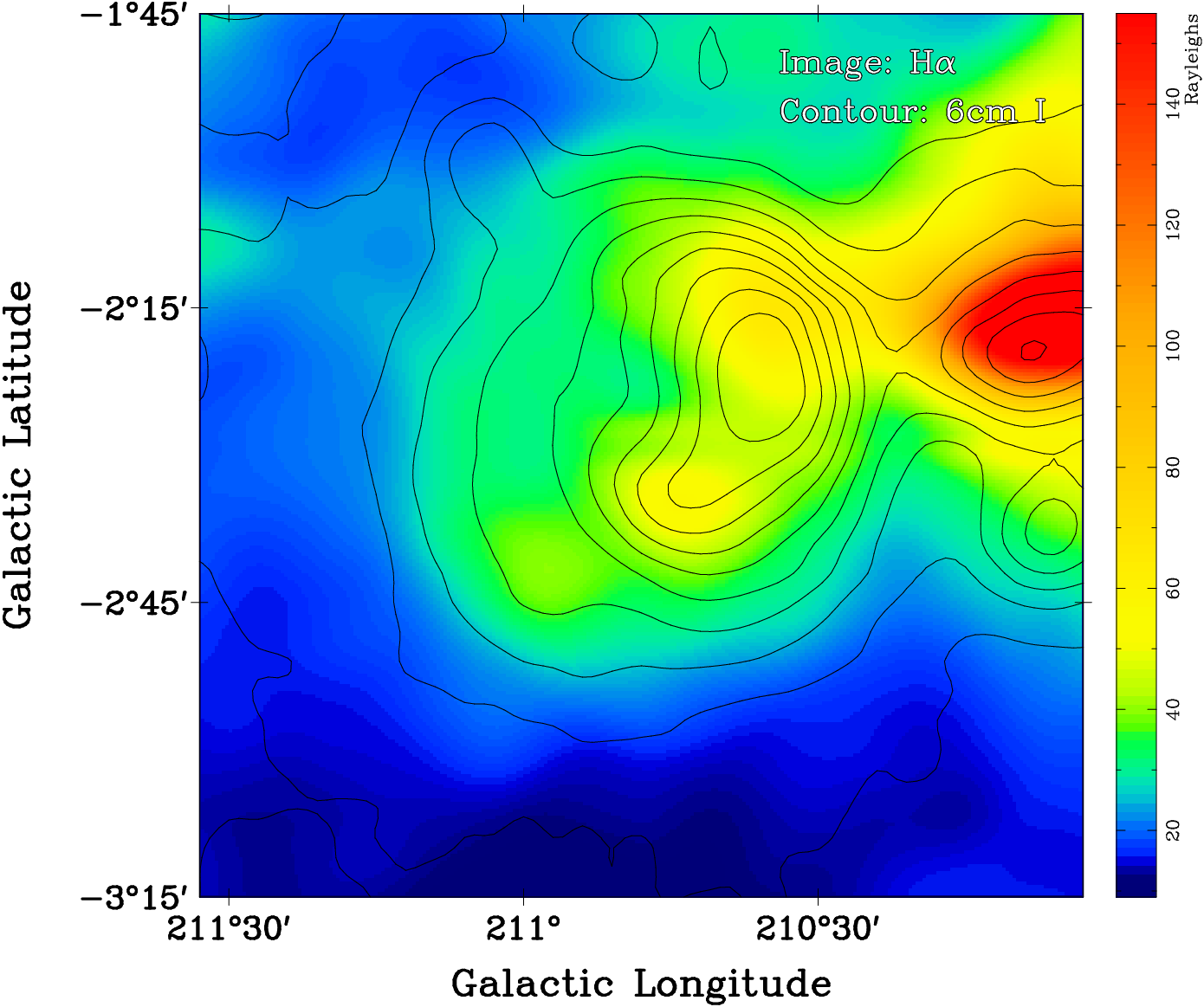} &
\includegraphics[angle=-90, width=0.3\textwidth]{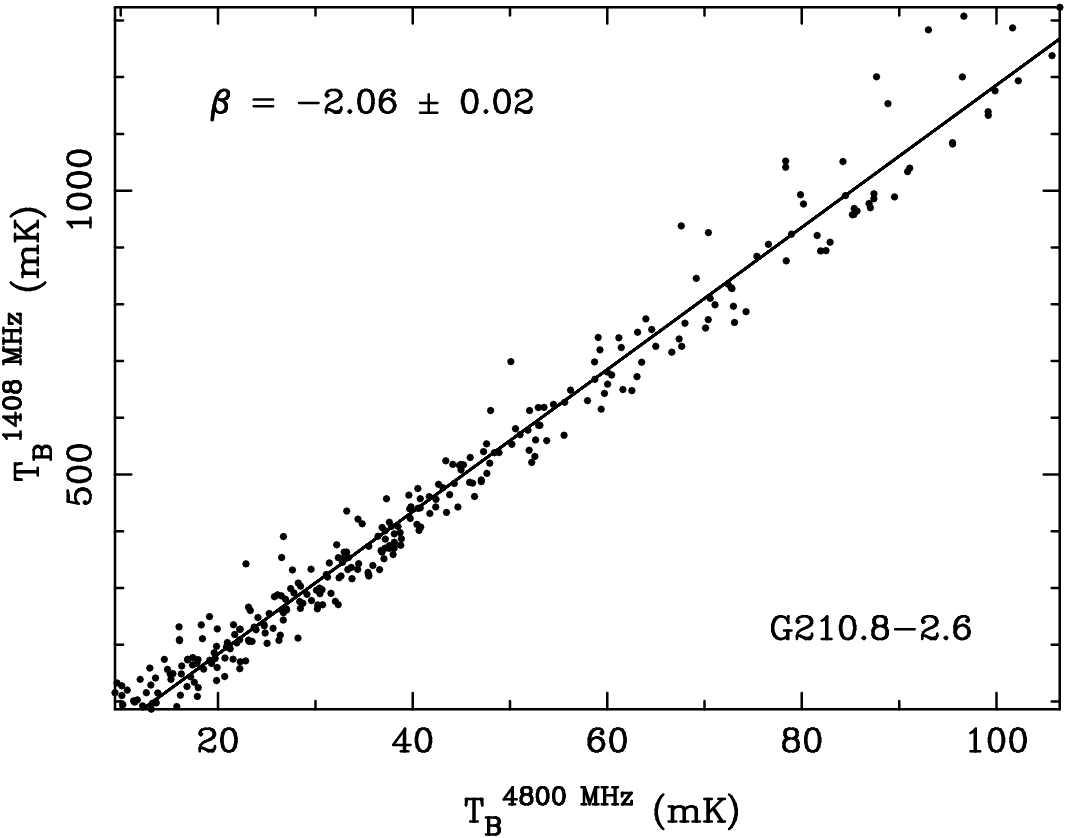} &
\includegraphics[angle=-90, width=0.28\textwidth]{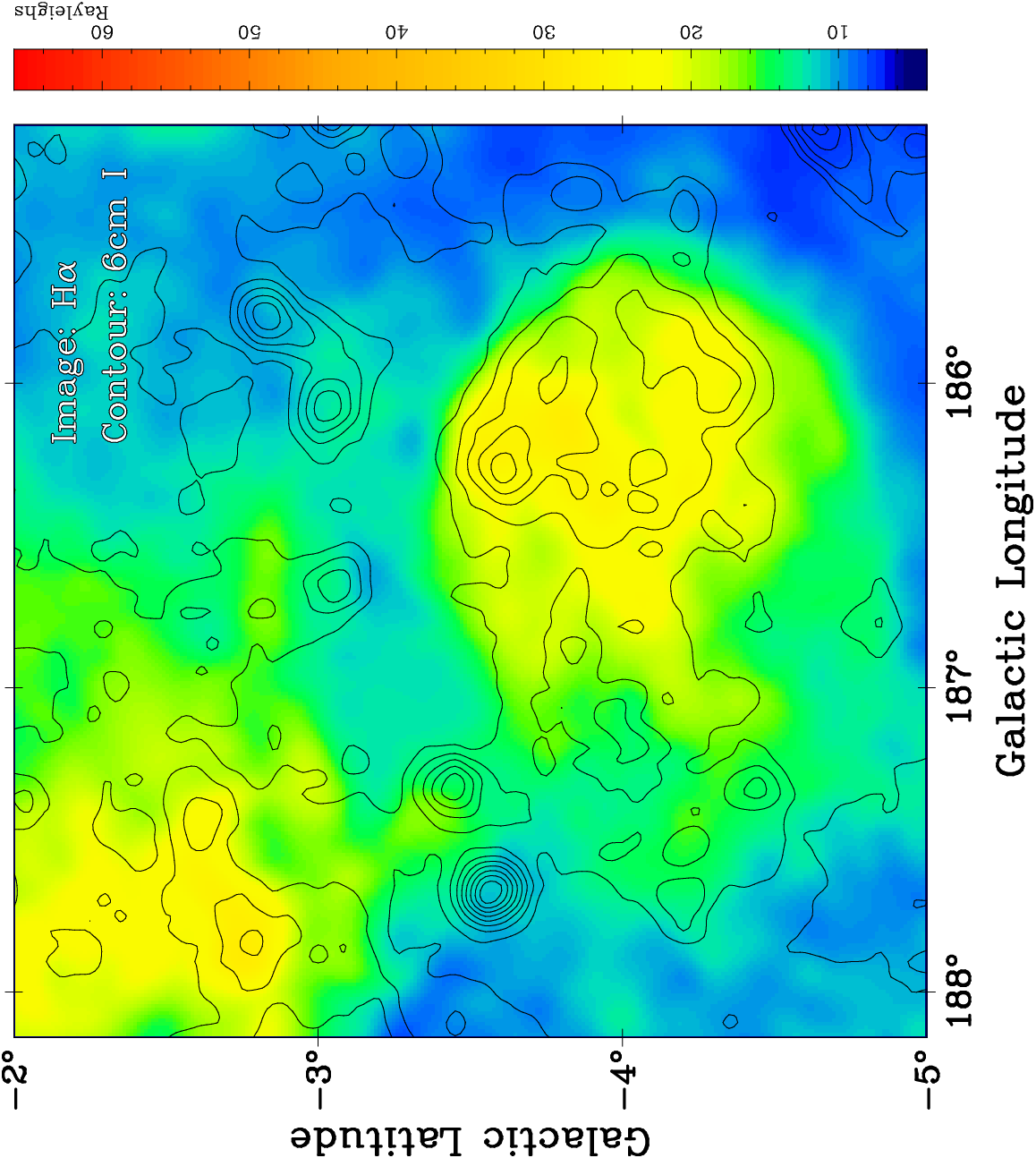} \\
\end{tabular}
\caption{{\it Upper panel from left to right:} $\lambda$6\ cm image,
  H$\alpha$ emission overlaid by Effelsberg $\lambda$11\ cm total
  intensity, and the TT-plot result for the new \ion{H}{II} region
  G212.9$-$3.7. {\it Lower panel from left to middle:} H$\alpha$
  emission overlaid by Urumqi $\lambda$6\ cm total intensity and the
  TT-plot result for the new \ion{H}{II} region G210.8$-$2.6. {\it
    Lower right panel:} Same as the {\it lower left panel}, but for
  G186.7$-$4.0.}
\label{newHII3}
\end{figure*}

\subsection{Notes on some objects in the WISE \ion{H}{II} region catalogue}

WISE \ion{H}{II} regions with sizes exceeding 8$\arcmin$ were used to
search for a sample of $\lambda$6\ cm extended \ion{H}{II}
regions. After excluding the small-diameter \ion{H}{II} regions, 107
extended \ion{H}{II} regions and \ion{H}{II} region candidates whose
flux densities can be reliably measured with the Urumqi $\lambda$6\ cm
data were catalogued and given in Table~\ref{extended}. Among these
sources, more than 30 were found without counterparts in the WISE
\ion{H}{II} region catalogue. We show an example in
Fig.~\ref{G127}. In the WISE 12$\mu$m infrared image, 12 WISE
\ion{H}{II} regions (marked with white squares and blue dashed
circles) are located in a 3$\degr \times 3\degr$ area centred at $\ell
= 127\fdg046, b=0\fdg950$. Except for G125.606+2.099 (C-type) in the
upper right corner, all the other WISE \ion{H}{II} regions are Q-type
WISE \ion{H}{II} regions. They do not show up in either the CGPS
$\lambda$21\ cm or the Urumqi $\lambda$6\ cm radio data as
expected. The radio source G127.9+1.7, indicated by the red rectangle
in Fig.~\ref{G127}, appears in both the CGPS and Urumqi images.
Strong infrared emission is seen for some parts of this source in the
WISE 12$\mu$m image. G127.9+1.7 is not recorded in either the WISE
\ion{H}{II} region catalogue or the Paladini \ion{H}{II} region
catalogue. However, it is the true Galactic \ion{H}{II} region DU~65
collected in \citet{Dubout-Crillon76}. As supporting evidence, we
displayed the H$\alpha$ emission of G127.9+1.7 and the flat radio
spectrum through TT-plot in the lower panel of Fig.~\ref{G127}
(T$_{\rm B}$ $\sim$ $\nu^{\beta}$, $\beta = \alpha -2$).

\subsubsection{New \ion{H}{II} regions}

In addition to the missing known extended \ion{H}{II} regions in the
WISE \ion{H}{II} region catalogue, a few new Galactic \ion{H}{II}
regions and \ion{H}{II} region candidates are also discovered in the
Urumqi $\lambda$6\ cm survey. They are G68.4+0.2, G76.7$-$1.2,
G94.4+2.7, G139.2$-$3.0, G139.9$-$2.0, G172.3+1.9 (\ion{H}{II} region
candidate), G186.7$-$4.0 (\ion{H}{II} region candidate), G210.8$-$2.6,
and G212.9$-$3.7. We show them in Fig.~\ref{newHII1},
Fig.~\ref{newHII2}, and Fig.~\ref{newHII3}. For G68.4+0.2,
G76.7$-$1.2, and G94.4+2.7 it is difficult to compare the
$\lambda$6\ cm radio emission with the WISE 12$\mu$m infrared emission
due to the large difference in angular resolution. We extracted the
1$\arcmin$ resolution CGPS $\lambda$21\ cm images and found that the
associated infrared emission are not for the entire radio source but
for partial shell- or ridge-like structures. The TT-plots confirmed
their thermal origin. For the six remaining newly identified
\ion{H}{II} regions and \ion{H}{II} region candidates, we found
supporting evidence both in the TT-plot results and the associated
H$\alpha$ emission, except for G172.3+1.9 and G186.7$-$4.0. For
G172.3+1.9, we did not obtain a reliable TT-plot result and for
G186.7$-$4.0, which is not fully covered by the Effelsberg
$\lambda$21\ cm survey, the sensitivity of the Effelsberg
$\lambda$11\ cm data is not enough for a clear determination. The
H$\alpha$ image quality is not high in the south-east part of
G172.3+1.9, but the data show a good correlation in the north-east
part.

\begin{figure}
\includegraphics[angle=-90, width=0.45\textwidth]{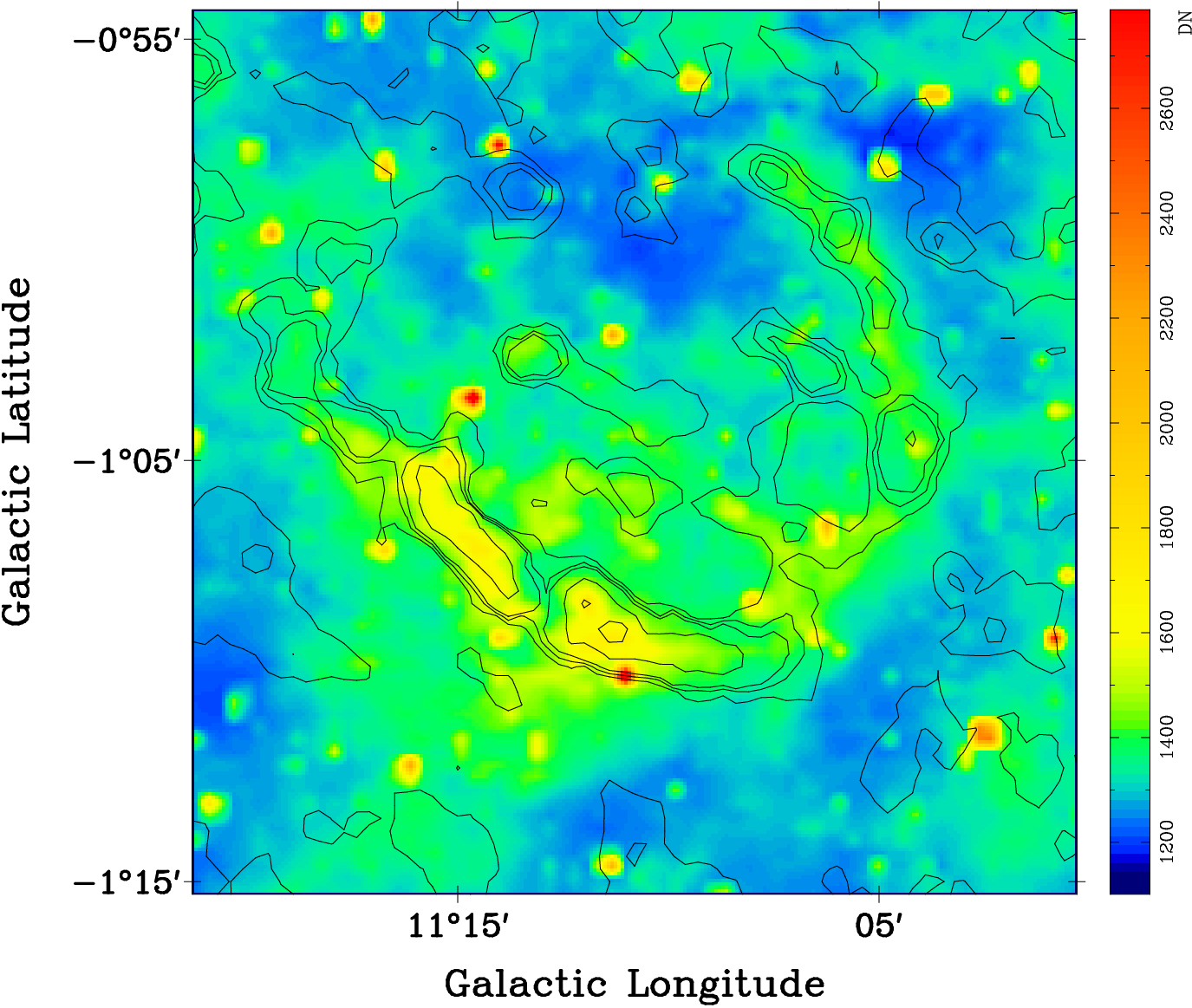}
\caption{WISE 12$\mu$m image of G11.1$-$1.0 overlaid by NVSS 1.4~GHz
  radio continuum emission.}
\label{G11}
\end{figure}

\subsubsection{\ion{H}{II} regions mistaken as SNRs}

Some of the matched WISE \ion{H}{II} regions challenge the SNR
classification collected in the Green SNR catalogue
\citep{Green17}. G11.183$-$1.063 is catalogued as a known WISE
\ion{H}{II} region, but also identified as a shell-type SNR
(G11.1$-$1.0) with a spectral index of $\alpha \sim$ $-$0.5 - $-$0.6
determined by the VLA $\lambda$90\ cm (11.0$\pm$0.3~Jy), the Southern
Galactic Plane Survey (SGPS) and VLA $\lambda$21\ cm (4.7$\pm$0.8~Jy),
and the Effelsberg $\lambda$11\ cm (4.1$\pm$0.4~Jy) data
\citep{Brogan06}. By adding the Urumqi $\lambda$6\ cm measurement
(3.40$\pm$0.25~Jy), \citet{Sun11b} found a spectral index of $\alpha =
-0.41\pm0.02$. From Fig.~1 of \citet{Sun11b}, the spectral index fit
for G11.1$-$1.0 strongly depends on the VLA $\lambda$90\ cm
result. Based on Effelsberg $\lambda$21\ cm data, \citet{Reich9021}
reported a flux intensity of S$_{\rm 21\ cm}$ = 5.17$\pm$0.52~Jy
closely agreeing with the previous SGPS/VLA result. The
$\lambda$11\ cm result of 4.082~Jy from the Effelsberg data is also
consistent with the value reported by \citet{Brogan06}. The Urumqi
$\lambda$6\ cm result from \citet{Reich14} is about 4.2$\pm$0.2~Jy, a
bit higher than that (3.40$\pm$0.25~Jy) of \citet{Sun11b}. Using all
these values except the $\lambda$90\ cm result, we fit a new spectrum
for G11.1$-$1.0 and found $\alpha = -0.16\pm0.07$, implying thermal
emission. We show the comparison of the NVSS and WISE 12$\mu$m data in
Fig.~\ref{G11}, the radio and infrared emission of G11.1$-$1.0 have a
very clear and good coincidence. Therefore, G11.1$-$1.0 seems to
favour a \ion{H}{II} region origin rather than a SNR identification.

\begin{figure}
\includegraphics[angle=-90, width=0.4\textwidth]{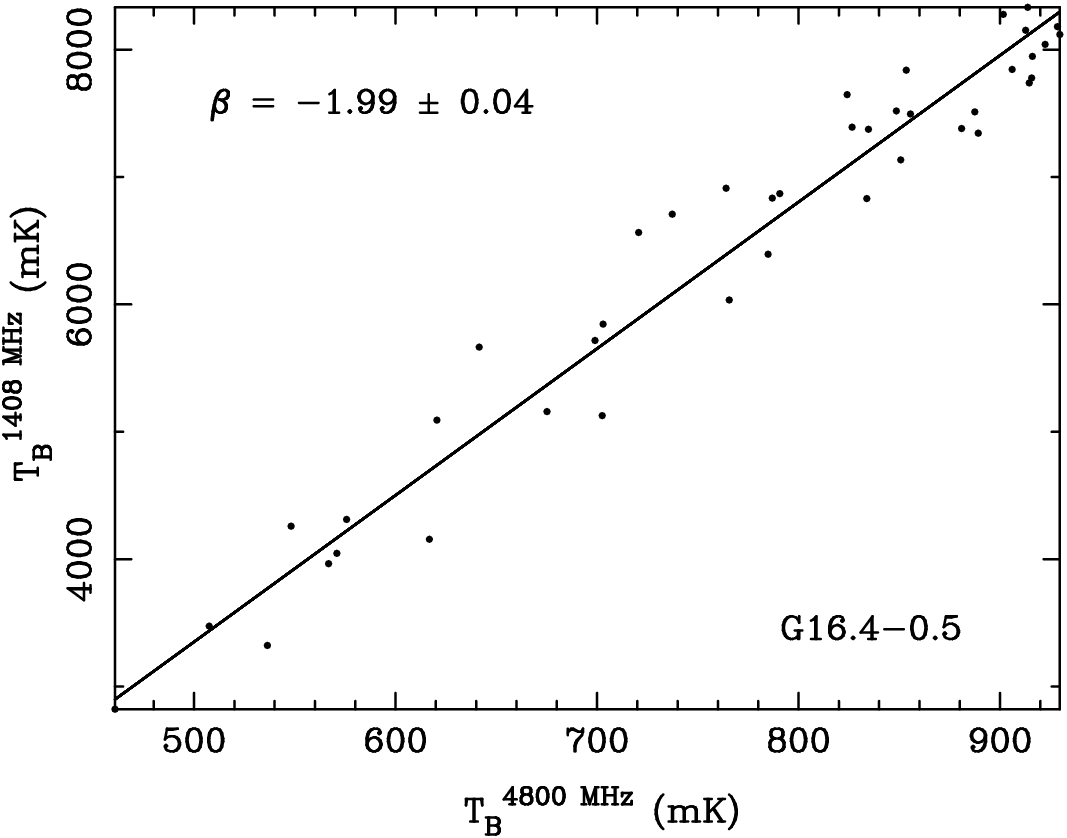}
\includegraphics[angle=-90, width=0.4\textwidth]{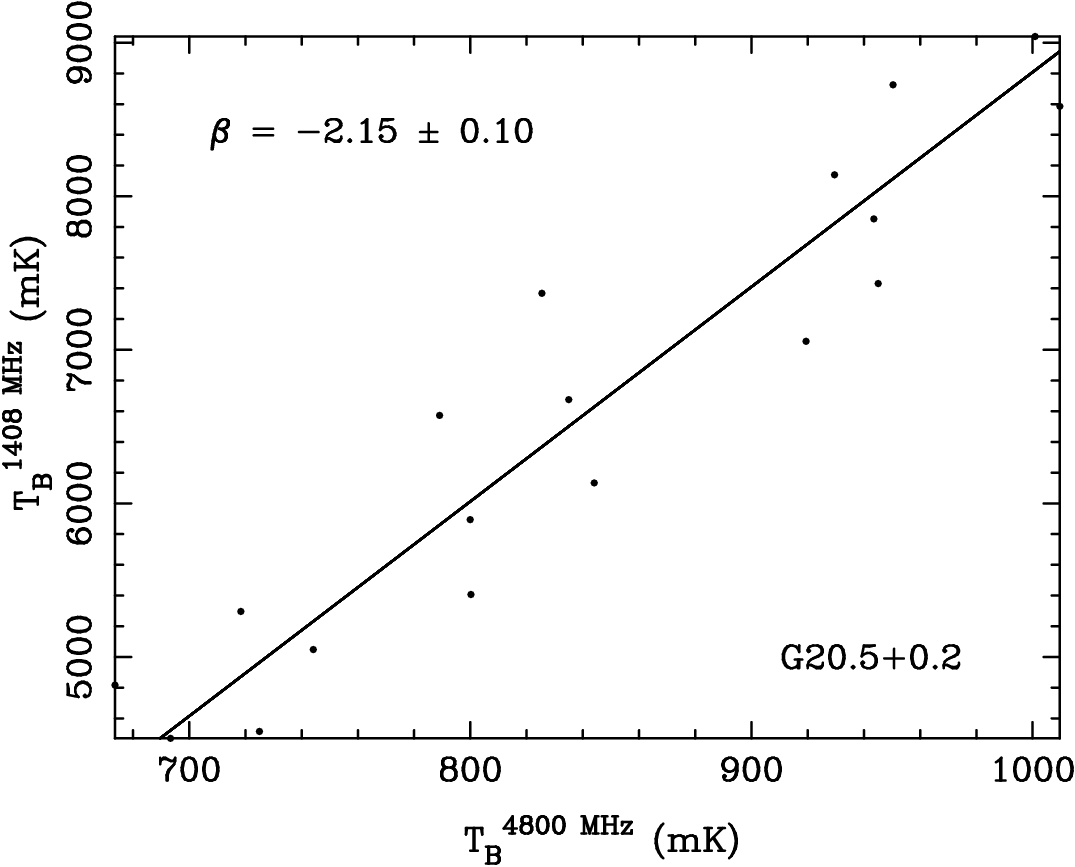}
\caption{TT-plots for G16.4$-$0.5 ({\it upper panel}) and G20.5+0.2
  ({\it lower panel}) between Effelsberg $\lambda$21\ cm data and
  Urumqi $\lambda$6\ cm data.}
\label{G20}
\end{figure}

A similar case was found for the source G20.4+0.1 ($\ell = 20\fdg45, b
=0\fdg12$) or G20.5+0.2 ($\ell = 20\fdg479, b =0\fdg165$). We found
them to be duplicated in the Paladini \ion{H}{II} region catalogue
(see Sect.~4.2). \citet{Brogan06} identified G20.47+0.16 to be a
shell-type SNR with a spectral index of about $\alpha$ $\sim$
$-$0.4. However, in the WISE \ion{H}{II} region catalogue,
G20.482+0.169 is a K-type \ion{H}{II} region with a similar size of a
few arcmins. A flat radio continuum spectrum with $\alpha =
-0.08\pm0.09$ was found in \citet{Sun11b} and confirmed in this work
(see Fig.~\ref{G20}, {\it lower panel}). These factors indicate that
G20.47+0.16 is not a shell-type SNR, which agrees with the argument in
\citet{Anderson17}.

A further investigation is also needed for the source
G16.4$-$0.5. \citet{Brogan06} identified it as a SNR by revealing a
partial shell structure and a steep spectrum with $\alpha = -0.7 \sim
-0.8$. \citet{Sun11b} derived a different spectrum with $\alpha =
-0.26\pm0.15$. Using a TT-plot between the Effelsberg $\lambda$21\ cm
data and the Urumqi $\lambda$6\ cm data, we obtained $\beta =
-1.99\pm0.04$ (see Fig.~\ref{G20}, {\it upper panel}). All these
results do not support its shell-type SNR nature. The radio emission
shown by the NVSS data is fragmented and difficult for a clear match
to the WISE infrared emission. More sensitive high angular resolution
radio observations are required to determine the shape and the nature
of G16.4$-$0.5.

By optical data, G67.8+0.5, a WISE C-type \ion{H}{II} region
(G67.816+0.511) \citep{abb+14} was identified as a SNR by
\citet{Sabin13}. The currently available radio continuum data cannot
yield a firm spectrum.

\subsection{Notes on some objects in the Paladini \ion{H}{II} region catalogue}

\begin{figure}
\includegraphics[angle=-90, width=0.45\textwidth]{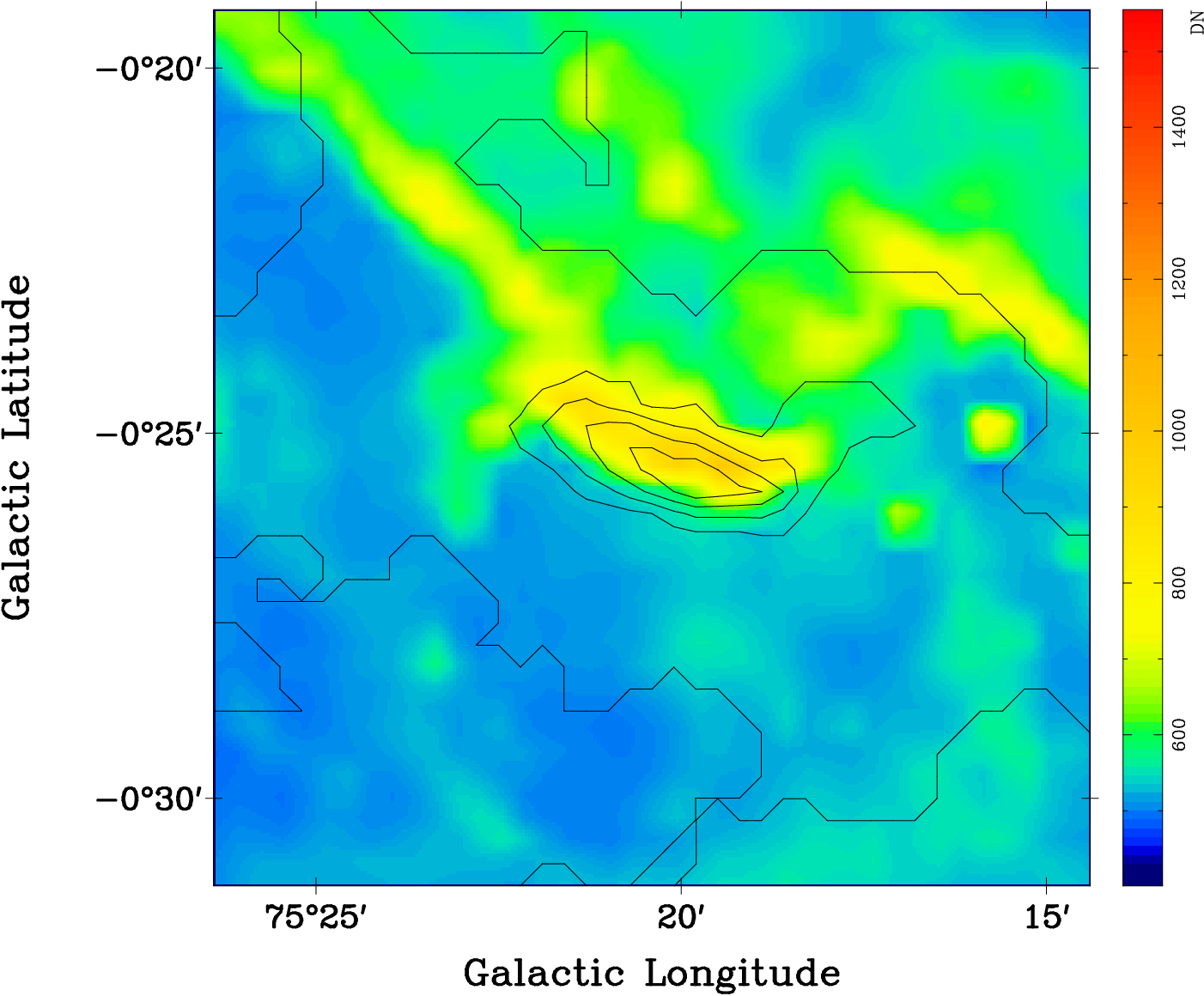}
\caption{WISE 12$\mu$m image of the \ion{H}{II} region G75.4$-$0.4
  overlaid by NVSS 1.4~GHz radio continuum emission.}
\label{G75}
\end{figure}

We cross-matched the small-diameter \ion{H}{II} regions in the
Paladini catalogue listed in Table~\ref{pala} with the WISE
\ion{H}{II} region catalogue. Only a few were found missing in the
WISE \ion{H}{II} region catalogue, for example G75.4$-$0.4 (see
Fig.~\ref{G75}) and G137.4+0.2. They both show an arc-like structure
in the WISE 12$\mu$m and the NVSS 1.4~GHz images. Their radio spectra
are flat, and the infrared and radio emission are morphologically
correlated.

Incorrect identifications are found in the Paladini \ion{H}{II} region
catalogue. Supernova remnants, planetary nebulae, and steep-spectrum
objects, likely extragalactic sources were noted. They are listed in
Table~\ref{il}.

\begin{table*}
\caption{78 misclassifications found in the Paladini \ion{H}{II}
  region catalogue with the Urumqi $\lambda$6\ cm data.}
\scriptsize
\label{il}
\begin{tabular}{cccrccccrccl}
\hline
\hline
P number  &\multicolumn{1}{c}{G-NAME} &\multicolumn{1}{c}{GLong}        &\multicolumn{1}{c}{GLat}        &\multicolumn{1}{c}{Ref} &\multicolumn{1}{c}{EB S$_{\rm 21cm}$} &\multicolumn{1}{c}{EB S$_{\rm 11cm}$} &\multicolumn{1}{c}{Urumqi S$_{\rm 6cm}$} &\multicolumn{1}{c}{S$_{\rm Ref}$} &\multicolumn{1}{c}{Freq}  &\multicolumn{1}{c}{Ref}  &\multicolumn{1}{c}{Notes} \\
          &\multicolumn{1}{c}{}     &\multicolumn{1}{c}{($\degr$)} &\multicolumn{1}{c}{($\degr$)} &                        &\multicolumn{1}{c}{(Jy)}                      &\multicolumn{1}{c}{(Jy)}                      &\multicolumn{1}{c}{(Jy)}                 &\multicolumn{1}{c}{(Jy)}               &\multicolumn{1}{c}{(GHz)}              &                         & \\
\hline

   130         &G10.1+0.7     & 10.091   & 0.727     &G70    &                    &                 &              &  0.7      & 2.7   &G70     & NGC~6537, PN \citep{Condon98p} \\
               &              &          &           &       &                    &                 &              &  0.61     &10.0   &H87     & \\
   146         &G11.2$-$0.4   & 11.172   &$-$0.354   &K97    &                    &                 &              & 10.367    & 4.9   &K97     & SNR G11.2$-$0.3 \\
               &              &          &           &       &                    &                 &              & 11        & 2.7   &A70     & \\
               &              &          &           &       &                    &                 &              & 11        & 5.0   &A70     & \\
               &              &          &           &       &                    &                 &              &  8.9      & 5.0   &R70     & \\
               &              &          &           &       &                    &                 &              &  5.43     &10.0   &H87     & \\
   147         &G11.2+0.1     & 11.207   & 0.088     &K97    &                    &  2.72           &  2.168       &  2.529    & 4.9   &K97     & SNR G11.1+0.1 \citep{Green17} \\
               &              &          &           &       &                    &                 &              &  3.20     & 4.9   &A78     & or SNR G11.18+0.11 \citep{Brogan06} \\
   157         &G11.9+2.2     & 11.914   & 2.200     &A70    &  1.74              &  0.92           &  0.404       &  1        & 1.4   &A70     & $\alpha = -1.22\pm0.09$ \\
               &              &          &           &       &                    &                 &              &  1        & 2.7   &A70     &  \\
   162         &G12.4+3.8     & 12.434   & 3.836     &A70    &  4.80              &  2.40           &  1.281       &  4        & 1.4   &A70     & $\alpha = -1.08\pm0.09$ \\
   198         &G15.0$-$1.6   & 15.041   &$-$1.569   &A70    &                    &                 &              &  5        & 1.4   &A70     & SNR G15.1$-$1.6 \\
               &              &          &           &       &                    &                 &              &  6        & 2.7   &A70     &  \\
               &$\cdots$      &          &           &       &                    &                 &              &           &       &        &  \\               
\hline\hline
\end{tabular}
\tablefoot{\scriptsize Columns 1 and 2: sequential number in the
  Paladini catalogue and G-names for sources. Columns 3--5: Galactic
  coordinates and their references. Columns 6--8: flux densities
  measured from the Effelsberg $\lambda$21\ cm, $\lambda$11\ cm, and
  the Urumqi $\lambda$6\ cm data if available. Columns 9--11: flux
  density at the corresponding frequency, and their reference
  according to \citet{Paladini03}. The flux densities are reviewed by
  comparing with those in the original literature. Column 12: remarks
  why these sources are not considered to be \ion{H}{II} regions in
  this work. Some other references in the 12th column:
  e.g. \citet{Helfand89} for P369; \citet{Vollmer10} for P863. \\ Full
  table is accessible from the CDS.}
\end{table*}

Seventeen surveys collected by \citet{Paladini03} have overlapping
regions with the Urumqi $\lambda$6\ cm survey (see
Table~\ref{survey}). These surveys were conducted from the late 1960s
to the 1990s with various radio telescopes at seven different
observing frequencies. We were aware of duplications when we noticed
that the same remarks (e.g.  Westerhout or Sharpless source names)
were given to different but very nearby Paladini sources. Therefore, a
self-intersection to the sample of the 777 Paladini objects within the
$\lambda$6\ cm survey area was made. Here we considered a coincidence
when $d \leqslant r_{p} + r_{p}'$, where $d$ is the distance of the
two centres of the Paladini sources, and $r_{p}$ and $r_{p}'$ are the
radii of the two adjacent sources. Many duplicated identifications
were found and we only discussed the ones which could be confirmed by
the Urumqi $\lambda$6\ cm data. Different angular resolutions in
different surveys accounted for many of the cases. On the one hand,
larger beam sizes include more emission and blend the peak position of
the fitted centre. On the other hand, as mentioned above, \ion{H}{II}
regions identified in low angular resolution observations can be
resolved into several discrete smaller \ion{H}{II} regions in high
angular resolution surveys. Different treatments of decimal places of
the same source in different literature cause another kind of
duplication. We discuss these duplications below. In the following, we
use the abbreviation `PXXX' (No. XXX source in the Paladini catalogue)
for a Paladini object.

\begin{enumerate}
  
\item P134 (010.3, $-$00.2) and P135 (010.3, $-$00.1):\\

 Both P134 and P135 are listed as individual point-like sources in the
 references in Table~\ref{pala}. The difference in designation very
 likely resulted from the different angular resolutions of the
 surveys. The similar flux densities achieved for P134 and P135
 support that both P134 and P135 indicate the same radio source.\\

\item P155 (011.9, +00.7) and P156 (011.9, +00.8):\\

 The 4.9~GHz source list of W82 was based on A78. The coordinates of
 P155 and P156 are so close that none of the given reference surveys
 can resolve them. It is the radio counterpart of the optical
 \ion{H}{II} region SH 2-38 \citep{Sharpless59} as indicated in G70.\\

\item P164 (012.7, +00.3) and P166 (012.8, +00.4):\\

 P166 \citep{Kuchar97, Altenhoff78} is a single isolated radio
 object. No sources as strong as $\sim$2~Jy can be found in its
 vicinity. K97 obtained a flux density of $\sim$ 2.4~Jy, which is
 comparable with the results for P164 \citep[SH
   2-40,][]{Altenhoff70}. A78 might have integrated a larger area and
 obtained a flux density exceeding 7~Jy. The optical image from the
 red plate of Digitised Sky Survey \citep{Lasker90} shows that SH 2-40
 consists of several bright knots; this may cause the shift of central
 coordinates when observed with different angular resolutions. \\

\item P170 (013.2, +00.0) and P171 (013.2, +00.1): \\

 Based on the 45$\arcsec$ resolution NVSS image, J181405-172832 with a
 flux density of S$_{\rm 1.4~GHz}$ = 2761.8$\pm$85.2~mJy is the only
 strong source in this field. The high flux density obtained in
 various references might result from the low angular resolution which
 cannot resolve the influence of the ambient emission. \\

\item P179 (013.9, +00.2) and P180 (013.9, +00.3):\\

 NVSS J181435-164530 (S$_{\rm 1.4~GHz}$ = 3307.9$\pm$120.0~mJy) is the
 only source that has a comparable flux density to that reported for
 both P179 and P180 in this area. \\

\item P189 (014.4, $-$00.7), P190 (014.4, $-$00.6) and P192 (014.5,
  $-$00.6):\\

 Based on the images from the Effelsberg $\lambda$11\ cm data, only
 one point-like source was identified. Considering that P189, P190,
 and P192 share similar central coordinates and reported flux
 densities, they should represent the same source.\\

\item P199 (015.0, $-$00.7) /  P200 (015.1, $-$00.9) / P203 (015.2,
  $-$00.8) /  P204 (015.2, $-$00.6) and P201 (015.1, $-$00.7):\\

From Table~\ref{pala}, it is clear that P199, P200, P203, and P204 are
four individual components that are resolved by A78. For the low
angular resolution observations of P199 and P201 (e.g. M67, A70), the
individual components sum up. It is not clear why D80 gave high values
for both P199 (344.5~Jy) and P201 (500~Jy). \\

\item P218 (016.9, +00.8), P219 (017.0, +00.8), P220 (017.0, +00.9),
  and P224 (017.1, +00.8):\\

P218 (G70) and P219 (e.g. A70, R70) is a complex consisting of several
components such as P218, P220, and P224 from K97. We measured the
extended complex and listed it in Table~\ref{extended}. \\

\item P221 (017.0, +01.6) and P222 (017.0, +01.7):\\

G17.0+1.6 (A70) and G17.0+1.7 (G70) were detected by observations with
angular resolutions $\geqslant$8$\arcmin$. We retrieved the 4$\farcm$3
resolution Effelsberg $\lambda$11\ cm image and cannot find point-like
sources, but we do find an extended and elongated structure. We
performed a TT-plot by using the Effelsberg $\lambda$21\ cm data and
the Urumqi $\lambda$6\ cm data. The spectral index was found to be
$\beta = -2.11\pm0.17$ ($\beta=\alpha -2$), indicating a thermal
nature. It is likely that the different designation of P221 and P222
comes from different fitted geometric centres. \\

\item P232 (018.2, $-$00.3) and P230 (018.1, $-$00.3) / P231 (018.2,
  $-$00.4) / P233 (018.2, $-$00.2) / P235 (018.3, $-$00.4) / P236
  (018.3, $-$0.3) / P239 (018.3, $-$00.4):\\

SH 2-53 is a complex consisting of many individual portions as shown
in the DSS red plate. The radio emission shown by F87 confirmed the
existence of discrete components and extended emission. It is clear
that A70, G70, and F87 measured the flux density of the entire
structure, while P230, P231, P233, P235 (P239), and P236 from A78 are
individual parts of SH 2-53. The low angular resolution flux density
measurements may be affected by the nearby SNR G18.1$-$0.1. \\

\item P255 (019.2, +02.1) and P256 (019.2, +02.2):\\

P255 is from the $\sim$11$\arcmin$ resolution data of A70. P256 comes
from the 8$\arcmin$ resolution data of G70. The 4$\farcm$3 resolution
Effelsberg $\lambda$11\ cm data only show one source. However, this
can be further resolved into two discrete sources in the NVSS data,
with flux densities of $\sim$2~Jy and $\sim$1~Jy, respectively. The
spectral index of each source cannot be found in
\citet{Vollmer10}. The overall $\alpha_{21-11-6} = -0.52\pm0.09$ was
obtained by using the Effelsberg $\lambda$21\ cm, $\lambda$11\ cm, and
the Urumqi $\lambda$6\ cm data. No WISE counterpart was found for
either of the sources. We put P255 and P256 into Table~\ref{il}, which
collects the mis-identifications of the Paladini \ion{H}{II} region
catalogue. \\

\item P261 (019.7, $-$00.1) and P258 (019.6, $-$00.2) / P259 (019.6,
  $-$00.1) / P260 (019.7, $-$00.2):\\

Based on the sources identified by A78, P258, P259, and P260 are three
individual components that can be resolved by the Effelsberg 4.9~GHz
observations. The low angular resolution measurements of G70, A70, and
R70 summed the flux densities of discrete portions. K97 listed P258
and P259 as two separated sources with flux densities of 16.6~Jy and
15.2~Jy, respectively. This would double the flux density in this area
and contradict the Effelsberg $\lambda$21\ cm and the Urumqi
$\lambda$6\ cm measurements, unless they are multi-entries by K97. \\

\item P266 (020.2, $-$00.9), P267 (020.3, $-$00.9) and P268 (020.3,
  $-$00.8): \\

These three entries share similar intrinsic source sizes of about
$\sim$10$\arcmin$ and a consistent flux density of about 3~Jy. We list
the source in Table~\ref{extended} because its apparent size slightly
exceeds 16$\arcmin$, which defines the small-diameter source of the
Urumqi $\lambda$6\ cm data. \\

\item P270 (020.4, +00.1) and P271 (020.5, +00.2):\\

Both P270 and P271 match the position of the shell-type SNR
G20.47+0.16 reported by \citet{Brogan06}. However, this may be a
\ion{H}{II} region (see detailed discussion in Sect.~4.1). \\

\item P292 (023.1, +00.5) and P293 (023.1, +00.6):\\

As pointed out by A70 for P292 and G70 for P293, they are the same
\ion{H}{II} region SH 2-58. \\

\item P303 (023.9, +00.1), P304 (023.9, +00.2), and P305 (024.0,
  +00.2):\\

The low angular resolution observations of A70 and G70 included the
contributions of P303 and P305 by K97 and A78. \\

\item P317 (024.6, +00.5) and P318 (024.6, +00.6):\\

Both P317 and P318 indicate the SNR G24.7+0.6. We listed them in
Table~\ref{il}. \\

\item P319/P320 (024.7, $-$00.2a/b) and P321 (024.7, $-$00.1):\\

P319 and P320 are two sources extracted by K97 from radio
recombination line observations of \citet{Lockman89}. They are located
at $\ell=24\fdg677, b=-0\fdg160$ and $\ell=24\fdg742, b=-0\fdg207$,
and therefore have the same abbreviation as
G24.7$-$0.2. \citet{Paladini03} gave the same flux density information
(K97, A78, D80, W82, and W83) for both G24.7$-$0.2a and G24.7$-$0.2b;
however, this seems incorrect. According to the central positions, the
peak flux densities and the measured sizes of the sources of K97 and
A78, G24.7$-$0.2a and G24.7$-$0.2b should have measured flux densities
of 9.9~Jy and 3.2~Jy in K97, and 3.19~Jy and 1.42~Jy in A78,
respectively. W82 and W83 measured the source centre at
$\ell=24\fdg676, b=-0\fdg157$ and $\ell=24\fdg68, b=-0\fdg16$, which
are closer to G24.7$-$0.2a rather than G24.7$-$0.2b. From the
Effelsberg $\lambda$11\ cm image, P321 should be the sum of P319 and
P320. However, we cannot explain the high flux densities reported by
A70 and G70.\\

\item P333 (026.1, $-$00.1) and P334 (026.1, $-$00.0):\\

By inspecting the NVSS and the Effelsberg $\lambda$11\ cm images, we
identified only one corresponding $\lambda$11\ cm radio source with
comparable flux densities. \\

\item P337 (026.5, +00.4) and P340 (026.6, +00.4):\\

The same case as P333/P334. \\

\item P343 (027.3, $-$00.2) and P344 (027.3, $-$00.1):\\

From the image of A78, G27.3$-$0.1 is a single point-like source on a
curved ridge. The SNR G27.4+0.0 is nearby. The low angular resolution
observations could possibly include the contribution of the unrelated
emission and therefore overestimated the flux density. The 5~GHz
observation of R70 measured a size of 6$\farcm$3 $\times$ 13$\farcm$7
for G27.3$-$0.2. The elongation was possibly due to the inclusion of
the point-like source itself and the underlying ridge. \\

\item P362 (028.8, +03.5) and P363 (028.9, +03.5):\\

Both P362 and P363 are marked as the \ion{H}{II} region SH
2-64/W40. The transformation from RA-DEC to L-B coordinates for P363
actually gives $\ell = 28\fdg789, b=+3\fdg488$, the same as P362. \\

\item P386 (030.7, $-$00.0) and P387 (030.8, $-$00.0):\\

Both P386 and P387 are the \ion{H}{II} region W43.\\

\item P398 (031.4, $-$00.3) and P399 (031.4, $-$00.2):\\

By inspecting the low angular resolution images, P398 from A70 and
P399 from B69 are the same point-like source. For P399, the
transformation from RA-DEC to L-B coordinates results in $\ell =
31\fdg43, b = -0\fdg25$, which should lead to the same G-name as
P398. \\

\item P403 (031.8, +01.5), P404 (031.9, +01.3), and P405 (031.8,
  +01.4):\\

G31.8+1.4 (P405) was reported to be the \ion{H}{II} region SH 2-69 in
F72 (see the Erratum of F72). P404 was marked as the \ion{H}{II}
region RCW~177 in A70. According to \citet{Dubout-Crillon76}, SH 2-69
and RCW 177 share the same coordinates, implying the same identity.
In the high angular resolution NVSS image, SH 2-69 shows a bubble-like
structure. Regarding the low flux density measured for P403, K97 might
only have measured a portion of the source.\\

\item P406 (032.1, $-$00.7) and P408 (032.2, +00.1): \\

B69 gave an incorrect G-name for P406. According to the given RA-DEC
coordinates, the centre of the source should be at $\ell = 32\fdg133,
b = 0\fdg117$, very close to P408 at $\ell = 32\fdg151, b =
0\fdg133$. According to the NVSS image, only one matched point source
is found in the field. \\

\item P419 (034.3, +00.1) and P420 (034.3, +00.2):\\

We compared the NVSS and the Effelsberg $\lambda$11\ cm
images. Considering the similar central positions and the flux
densities reported for P419 and P420, they indicate the same radio
source.  \\

\item P421 (034.5, $-$01.1) and P422 (034.6, $-$01.1):\\

The coordinates of P421 given by K97 is $\ell = 34\fdg550, b =
-1\fdg110$. The G-name should be written as G34.6$-$1.1, the same as
P422. The Effelsberg $\lambda$11\ cm data only shows one source that
is slightly extended in the field. No flux density information was
found for P422 in B69. \\

\item P431 (035.2, $-$01.8) and P432 (035.2, $-$01.7):\\

For P431 and P432, the difference in coordinates is very small (see
Table~\ref{pala}). Only one corresponding radio source is identified
in the Effelsberg $\lambda$11\ cm image. \\

\item P435 (035.3, $-$01.8) and P437 (035.4, $-$01.8):\\

The same case as P431/P432. \\

\item P440 (035.6, $-$00.0), P441 (035.6, +00.1), and P442 (035.7,
   $-$00.0):\\

P441, P442, and the high angular resolution measurements of P440
(e.g. A78, D80) are small discrete components, while the low angular
resolution results of P440 from A70 and B69 cannot resolve individual
parts but the entire structure.\\

\item P444 (036.3, $-$01.7), P445 (036.3, $-$01.6), P447 (036.4,
  $-$01.8), and P448 (036.4, $-$01.6):\\

F72 reported that the 17$\arcmin$ diameter source P448 is the radio
counterpart of the \ion{H}{II} region SH 2-72 with the optical centre
at $\ell = 36\fdg4, b =-1\fdg7$. P445 was marked by B69 as RCW~179
with a diameter of 15$\arcmin$, while P444 from A70 was also marked as
RCW~179, having a size of 15$\arcmin$ $\times$ 16$\arcmin$. SH 2-72
and RCW~179 are the same \ion{H}{II} region as indicated in
\citet{Dubout-Crillon76}. This is also supported by their similar flux
densities and sizes. K97 identified two sources P444 and P447 with
flux densities of 3.2~Jy and 0.8~Jy, respectively. They might be small
individual components which are included in this extended
structure. \\

\item P449 (036.5, $-$00.2) and P450 (036.5, $-$00.1): \\

The $\lambda$6\ cm source at $\ell = 36\fdg467, b = -0\fdg179$ is the
only source that is close to P449 and P450. As supporting evidence,
only one ring-like corresponding source is found in the NVSS data. \\

\item P466 (037.8, $-$00.3), P468 (037.9, $-$00.4), P469 (037.9,
  $-$00.3):\\
  
Based on the NVSS image, P468 (except for the result of R70) is a
single point-like component which is included in the low angular
resolution measurements of P466 and P469. \\

\item P477 (039.3, $-$00.1) and P478 (039.3, $-$00.0):\\

Both P477 (W82) and P478 (A70, D70) are marked as NRAO~591. \\

\item P480 (039.5, +00.5) and P484 (039.6, +00.6):\\

P480 and P484 have similar sizes and flux densities, as reported. Only
one extended radio structure can be identified in the Effelsberg
$\lambda$11\ cm survey data. \\

\item P487 (039.9, $-$01.4), P488 (039.9, $-$01.3), and P494 (040.0,
  $-$01.3):\\

We converted the RA-DEC coordinates reported by F72 to L-B coordinates
for P487. The centre is $\ell = 39\fdg920, b = -1\fdg355$, very close
to P488 centred at $\ell = 39\fdg904, b = -1\fdg351$. P494 was marked
as RCW~182 by A70 and P487 was recognised as SH 2-74 by F72. By
comparing the sizes and positions in \citet{Rodgers60} and
\citet{Sharpless59}, RCW~182 and SH 2-74 indicate the same \ion{H}{II}
region. \\

\item P490 (039.9, $-$00.2) and P491 (039.9, $-$00.1):\\

P490 was identified by the 8$\farcm$2 resolution Parkes data (D70),
while P491 was from the $\sim$11$\arcmin$ resolution NRAO data of
A70. From a sharper view of the 4$\farcm$3 resolution Effelsberg
$\lambda$11\ cm image, we identified only one matching source at $\ell
= 39\fdg853, b = -0\fdg180$. \\

\item P501 (041.4, +00.4) and P505 (041.5, +00.4):\\

Both P501 and P505 indicate the SNR G41.5+0.4. They are listed in
Table~\ref{il}. \\

\item P503 (041.5, +00.0) and P504 (041.5, +00.1):\\

Considering the small differences in positions and flux densities, we
identify P503 and P504 to be the same source after inspecting the NVSS
and the Effelsberg $\lambda$11\ cm image.\\

\item P511 (042.4, $-$0.3), P512 (042.5, $-$00.2), and P513 (042.6,
  $-$00.1): \\

P512 from low angular resolution observations of A70 and D70 is
reported to be an extended source. It contains small individual
components that can be resolved in higher angular resolution
images. P511 and P513 from K97 are two of the individual components
inside P512. \\

\item P519 (043.3, +00.5) and P520 (043.4, +00.5):\\

By converting RA-DEC to L-B coordinates for P519, we found that the
G-name of P519 should be written as 043.4+00.5, the same as P520. \\

\item P524 (043.9, $-$00.8) and P525 (043.9, $-$00.7):\\

We compared the NVSS and the Effelsberg $\lambda$11\ cm images and
found the $\lambda$11\ cm source located at $\ell = 43\fdg896, b =
-0\fdg801$ is the only matching one. \\

\item P529 (044.2, +00.1) and P531 (044.3, +00.1): \\

The same case as P524/P525. \\

\item P530 (044.3, $-$00.4) and P532 (044.4, $-$00.3):\\

The same case as P524/P525. \\

\item P540 (045.4, +00.1) and P541 (045.5, +00.1):\\

Both P540 from D70 and P541 from A70 were marked as the radio source
NRAO~601. \citet{Paladini03} only listed one corresponding source for
P541; we found the second component in A78/D80, W82, and W83. The sum
of the two components agreed well with the results obtained from the
low angular resolution observations (see Table~\ref{pala}). \\

\item P564 (049.5, $-$00.4) / P559 (049.3, $-$00.3), P561 (049.4,
  $-$00.5), P562 (049.4, $-$00.3), P563 (049.4, $-$00.2), P567 (049.6,
  $-$00.4): \\

The low angular resolution observations of W51A, such as A70 for P564,
smear out the individual structures (P559, P561, P562, P563, P567)
that can be resolved by high angular resolution observations. It is
not clear why the K97 measurements toward P561 and P564 both exceed
100~Jy, which result in a much larger flux density for W51A. \\

\item P570 (050.0, $-$00.1) and P571 (050.0, $-$00.0):\\

The same case as P524/P525. \\

\item P578 (051.2, $-$00.1) and P579 (051.2, +00.1):\\

P578 has a measured size of 20$\farcm$9 $\times$ 22$\farcm$4 and a
flux density of 37~Jy in R70, while A70 measured a similar size of
20$\arcmin$ $\times$ 20$\arcmin$ and comparable flux density of 35~Jy
for P579. Considering the similar size and flux density, P578 and P579
should be the same source. It is embedded in a complex spanning over
1$\degr$ (see Table~\ref{extended}). P576 and P580 are also included
in this complex. \\

\item P597 (054.1, $-$00.1) and P598 (054.1, $-$00.0):\\

A70 suspected that the source P598 might be the steep-spectrum radio
source 4C+18.57. From the NVSS image, 4C+18.57 is a single point-like
source close to P597. These two objects cannot be resolved by the
observation of A70. Thus, the measured flux density of P598 included
the contribution of the \ion{H}{II} region P597 and of 4C+18.57. It is
not easy to obtain an accurate flux density for P597 with the Urumqi
data due to the influence of the nearby SNR G54.4$-$0.3.\\

\item P605 (057.5, $-$00.3) and P606 (057.6, $-$00.3):\\

The same case as P524/P525. \\

\item P619 (063.1, +00.4), P620 (063.2, +00.4), and P621 (063.2,
  +00.5):\\

The same case as P524/P525. They are the \ion{H}{II} region SH
2-90. \\

\item P624 (064.1, $-$00.5) and P626 (064.2, $-$00.5):\\

Both P624 (F72) and P626 (A70) are marked as the \ion{H}{II} region SH
2-93. \\

\item P632 (068.1, +00.9), P633 (068.1, +01.0), and P634 (068.2,
  +01.0):\\

P633 from F72 and P634 from A70 are both reported as an extended
source and marked as the \ion{H}{II} region SH 2-98. They are
incorporated in the $\lambda$6\ cm extended \ion{H}{II} region
G68.20+1.05 in Table~\ref{extended}. From the high-resolution CGPS
image, SH 2-98 is a ring-like structure with a diameter of about
25$\arcmin$. P632 from K97 is a bright, small-diameter
($\sim$2$\arcmin$) source situated on the southern part of the
ring. It can hardly be resolved in the low angular resolution
observations such as A70 and F72. Therefore, the flux density
measurements by A70 and F72 contain the entire ring structure and
P632, while K97 only measure the point-like source. Due to influence
of the ring structure, \citet{Reich14} obtain a higher flux density
for P632 than obtained by K97. \\

\item P636 (069.9, +01.5) and P637 (069.9, +01.6):\\

In the $\lambda$6\ cm image, P636 is a point-like source which is
somehow confused by the radio emission in its north. Low angular
resolution observations might overestimate the flux density. A good
fit for the thermal radio spectrum was obtained through the Effelsberg
$\lambda$21\ cm, $\lambda$11\ cm, and the Urumqi $\lambda$6\ cm
data. However, it is strange that with the angular resolution similar
to the Effelsberg $\lambda$21\ cm data and the Urumqi $\lambda$6\ cm
data, S$_{\rm 2.7GHz}$ = 14~Jy from A70 is so high. \\

\item P651 (074.8, +00.6) and P652 (074.8, +00.7):\\

P651 from F72 and P652 from A70 were both proposed as the counterpart
of the \ion{H}{II} region SH 2-104. All the flux density estimates are
consistent, except that by F72. With the angular resolution similar to
the Urumqi data, the large flux density of 32.2~Jy is unclear.\\

\item P668 (078.0, +00.6) and P669 (078.1, +00.6):\\

P668 is reported to be a point-like source with an intrinsic size of
about 3$\arcmin$ in K97 and R70. This can be confirmed by the
Effelsberg $\lambda$11\ cm data. The source is isolated. W70 reported
that P669 has a flux density of 18.5~Jy and a size of 9$\arcmin$
$\times$ 7$\arcmin$ after deconvolution. These results are difficult
to explain. \\

\item P696 (078.9, +03.7) and P699 (079.0, +03.6):\\

P699 was measured by W70 to have an intrinsic size of 28$\arcmin$
$\times$ 20$\arcmin$, while P696 has a larger source size of
37$\farcm$2 $\times$ 37$\farcm$9 in R70. Considering the large size
and the small difference between the central coordinates, P696 and
P699 obviously indicate the same object. \\

\item P721 (080.0, +00.8) and P722 (080.0, +00.9):\\

P722 from K97 can be confirmed as a single point-like source by the
Effelsberg $\lambda$11\ cm data, which has a similar angular
resolution. W70 obtained an intrinsic source size of 16$\arcmin$
$\times$ 16$\arcmin$ for P721, which encompasses the source of
P722. The measured flux density of 17~Jy might result from the
integration of a larger area. \\

\item P723 (080.0, +01.5) and P726 (080.1, +01.5):\\

This duplication arises from different sizes found in different
literature. For the same source P723, K97 reported an intrinsic source
size of about 6$\arcmin$, but R70 reported 21$\farcm$1 $\times$
35$\farcm$5. For P726, W70 found 12$\arcmin$ $\times$ 22$\arcmin$. The
4$\farcm$3 angular resolution Effelsberg $\lambda$11\ cm image shows
an elongated structure rather than a point-like source. The different
values for P723/P726 might result from the different areas selected
for the flux density integration. \\

\item P733 (080.4, +00.4) and P734 (080.4, +00.5):\\

W70 measured an intrinsic size of 17$\arcmin$ $\times$ 22$\arcmin$ for
P733, while R70 measured 4$\farcm$4 $\times$ 32$\farcm$2 for P734. The
two extended objects overlap and should be the same source. K97
obtained a much smaller size of about 5$\arcmin$ for P733. According
to the Effelsberg $\lambda$11\ cm image, which has a similar angular
resolution to that of K97, K97 very probably measured only the compact
central part of the source. \\

\item P738 (80.8, +00.4) and P743 (080.9, +00.4):\\

P743 was identified by K97 and R70 with a size of about 4$\arcmin$. It
appears as a point-like source in the Effelsberg $\lambda$11\ cm and
Urumqi $\lambda$6\ cm images. The flux density measurements by K97 and
R70 can be clearly supported by the Effelsberg $\lambda$11\ cm and the
Urumqi $\lambda$6\ cm data. P738 was identified by W70 with a much
higher flux density of S$_{\rm 2.7~GHz}$ = 43.2~Jy and a much larger
size of 17$\arcmin$ $\times$ 16$\arcmin$. We cannot explain the W70
result with the current data. \\

\item P741 (80.9, $-$00.2) and P742 (080.9, $-$00.1):\\

High angular resolution observations of P741 and P742 region from the
CGPS data shows a circular emission region ($\sim$12$\arcmin$) with a
bright source on its north-east edge. A thin shell structure
($\sim$10$\arcmin$), which does not seem to be related, runs
underneath the circular region. By overlaying the 87GB/GB6 sources on
the image, it is possible that K97 only measured the upper bright part
of the structure. At an angular resolution of about 10$\arcmin$
(e.g. R70 and W70), these structures can no longer be resolved but
seen as a single point-like source. \\

\item P789 (097.5, +03.2) and P790 (097.6, +03.2):\\

The same case as P524/P525. \\

\item P798 (104.6, +01.3) and P799 (104.6, +01.4): \\

The \ion{H}{II} region SH 2-135 has an oval shape with a size of
22$\arcmin$ $\times$ 15$\arcmin$ \citep{Lynds65}. K97 only measured
the bright knots (P799) in its northern part, while F72 measured the
entire structure (P798). \\

\item P812 (108.8, $-$01.0) and P813 (108.8, $-$00.9): \\

Based on \citet{Sharpless59}, SH 2-152 ($\ell=108\fdg8, b=-0\fdg9$)
and SH 2-153 ($\ell=108\fdg8, b=-1\fdg0$) are two very close
\ion{H}{II} regions which the Urumqi $\lambda$6\ cm observation cannot
resolve. The flux density measurement by K97 may include both SH 2-152
and SH 2-153. The coordinate of P813 from W83 is $\ell=108\fdg76,
b=-0\fdg950$. The abbreviation should be the same as P812. W83 has a
much better angular resolution than K97, who only measured SH 2-152.\\

\item P817 (110.1, +00.0) and P818 (110.1, +00.1): \\

Both P817 from F72 and K97 and P818 from W83 are the \ion{H}{II}
region SH 2-156.\\

\item P830 (115.0, +03.1) and P831 (115.0, +03.2): \\

The same as the case of P524/P525. \\

\item P839 (119.4, $-$00.9) and P840 (119.4, $-$00.8): \\

Based on the image of Effelsberg $\lambda$11\ cm data, P839 from F72
is the \ion{H}{II} region SH 2-173 with a circular shape and a bright
western shell. From the coordinates given by K97 for P840, it is
likely that K97 measured the western shell of SH 2-173 only.\\

\item P854 (136.4, +02.5) and P856 (136.5, +02.5):\\

The same as the case of P524/P525. \\

\item P866 (151.6, $-$00.3) and P867 (151.6, $-$00.2):\\

Both P866 from F72 and P867 from W83 were claimed to be the
\ion{H}{II} region SH 2-209, which consists of a northern component SH
2-209 N ($\ell=151\fdg59, b=-0\fdg22$) and a southern component SH
2-209 S ($\ell=151\fdg64, b=-0\fdg47$). F72 reported a lower central
position, which might be influenced by SH 2-209 S.\\

\item P869 (154.6, +02.4) and P870 (154.7, +02.4):\\

The same as the case of P524/P525. \\

\item P887 (180.8, +04.0) and P888 (180.9, +04.1):\\

P888 from F72 is thought to be the radio counterpart of the
\ion{H}{II} region SH 2-241. In the NVSS image, P888 mainly includes
the source J060358+301522 ($\ell=180\fdg87, b=+4\fdg11$) and an area
of diffuse emission in its south-east. It resembles the optical image
of the \ion{H}{II} region in the DSS red plate. P887 ($\ell=180\fdg79,
b=+4\fdg03$) is catalogued by K97, but cannot be found in either the
87GB or the GB6 catalogue. We convolved the NVSS image to an angular
resolution of 3$\farcm$7, the same as K97. However, we failed to
identify any source at $\ell=180\fdg79, b=+4\fdg03$. The source
J060358+301522 has a flux density of about $S_{\rm 1.4~GHz}$ = 140~mJy
and a spectral index of $\alpha=-0.06$ \citep{Vollmer10}. It was also
identified in the GB6 catalogue with a consistent flux density of
$S_{\rm 4.85~GHz}$ = 158~mJy. Considering the similar flux density
reported by K97 (S$_{\rm K97}$ = 0.164~Jy), it might be that P887
indicates the source NVSS J060358+301522, but is listed with wrong
coordinates.\\

\item P896 (192.1, +03.6) and P897 (192.2, +03.6):\\

F72 reported that P896 is the \ion{H}{II} region SH 2-253, whose
optical centre is at $\ell=192\fdg2, b=+3\fdg6$. It appears as an
elongated diffuse structure both in the Effelsberg $\lambda$11\ cm and
in the Urumqi $\lambda$6\ cm images. A similar optical morphology was
seen in the DSS red plate. K97 gave a very small size and flux density
for P897, which likely accounts for a part of the entire \ion{H}{II}
region. \\

\item P905 (196.4, $-$01.7) and P906 (196.5, $-$01.7):\\

The same as the case of P524/P525. \\

\item P909 (197.8, $-$02.4) and P910 (197.8, $-$02.3):\\

From the NVSS image, a strong radio source is located at $\ell =
197\fdg78, b = -2\fdg32$ (P910) and a few weak sources are visible
around $\ell \sim 197\fdg76, b \sim -2\fdg45$. They show up as two
radio sources in the 4$\farcm$3 resolution Effelsberg $\lambda$11\ cm
image. They cannot be separated in the 9$\farcm$5 resolution Urumqi
$\lambda$6\ cm data. The same case is expected for P909 from F72. \\

\end{enumerate}

\section{Summary}

We identified and analysed \ion{H}{II} regions from the Sino-German
$\lambda$6\ cm polarisation survey of the $\sim$2\,200 deg$^{2}$ plane
area. The small \ion{H}{II} regions (apparent size $<$ 16$\arcmin$)
were obtained by cross-matches between the $\lambda$6\ cm
small-diameter sources of \citet{Reich14}, the Paladini radio
\ion{H}{II} region catalogue, and the WISE infrared \ion{H}{II} region
catalogue, while the extended \ion{H}{II} regions were found by
overlaying the Paladini and WISE \ion{H}{II} regions onto the Urumqi
$\lambda$6\ cm survey image and searching for coincidences by eye. The
spectra of the chosen sources were examined by using the Effelsberg
$\lambda$21\ cm, $\lambda$11\ cm, together with the Urumqi
$\lambda$6\ cm data. Finally, 401 \ion{H}{II} regions were extracted
from the $\lambda$6\ cm survey. We listed their positions,
$\lambda$6\ cm flux densities, and distances if available in
Table~\ref{pala}, Table~\ref{wise6}, and Table~\ref{extended}.

Multi-frequency and multi-domain observations are important for
\ion{H}{II} region identification. The WISE \ion{H}{II} region
catalogue, being currently the largest, provides an excellent sample
of small-diameter \ion{H}{II} regions, but misses some extended
\ion{H}{II} regions as listed in Table~\ref{extended}. Among these
$\sim$30 extended \ion{H}{II} regions that were not present in the
WISE \ion{H}{II} region catalogue, 9 are revealed in this paper for
the first time. In the Urumqi survey area, there are 78
mis-classifications and 76 pairs of duplicated identifications found
in the Paladini \ion{H}{II} region catalogue. The Mis-classifications
were mainly inherited from the literature from the 1970s. They are
chosen by spectra check and/or by comparison with the SNR or PN
catalogues. The duplications are mostly the results of the inclusion
of both high and low angular resolution observations toward the same
source.

G11.1$-$1.0, G20.4+0.1, and G16.4$-$0.5, were initially identified as
SNRs by their shell-like appearance and steep radio continuum spectra
\citep{Brogan06}. However, the newly derived radio spectra of these
three sources are all flat, which imply that their nature is
thermal. Additionally, G11.1$-$1.0 (WISE \ion{H}{II} region
G11.183$-$1.063), G20.4+0.1 (WISE \ion{H}{II} region G20.482+0.169),
and G16.4$-$0.5 (WISE \ion{H}{II} region G16.364$-$0.558) have all
been identified as K-type, namely known \ion{H}{II} regions in the
WISE \ion{H}{II} region catalogue, and G11.1$-$1.0 shows
well-correlated radio and infrared emission.

\begin{acknowledgements}
We would like to thank Dr. Thomas L. Wilson for the critical reading
of the manuscript and the anonymous referee for the helpful
comments. The Chinese authors are supported by the National Key R\&D
Program of China (NO. 2017YFA0402701), the Open Project Program of the
Key Laboratory of FAST, NAOC, the Chinese Academy of Sciences, the
National Natural Science foundation of China (11303035, 11503033), and
the Partner group of the MPIfR at NAOC in the framework of the
exchange programme between MPG and CAS for many bilateral visits. XYG
acknowledges financial support from the CAS-NWO cooperation programme
and Young Researcher Grant of National Astronomical Observatories,
Chinese Academy of Sciences, and the FAST FELLOWSHIP. The FAST
FELLOWSHIP is supported by Special Funding for Advanced Users,
budgeted and administrated by the Center for Astronomical
Mega-Science, Chinese Academy of Sciences (CAMS). LGH is grateful for
the support by the Youth Innovation Promotion Association CAS. The
Digitized Sky Surveys were produced at the Space Telescope Science
Institute under U.S. Government grant NAG W-2166. The images of these
surveys are based on photographic data obtained using the Oschin
Schmidt Telescope on Palomar Mountain and the UK Schmidt
Telescope. The plates were processed into the present compressed
digital form with the permission of these institutions. The National
Geographic Society - Palomar Observatory Sky Atlas (POSS-I) was made
by the California Institute of Technology with grants from the
National Geographic Society. The Wisconsin H$\alpha$ Mapper and its
H$\alpha$ Sky Survey have been funded primarily by the National
Science Foundation. The facility was designed and built with the help
of the University of Wisconsin Graduate School, Physical Sciences Lab,
and Space Astronomy Lab. NOAO staff at Kitt Peak and Cerro Tololo
provided on-site support for its remote operation. This publication
makes use of data products from the Wide-field Infrared Survey
Explorer, which is a joint project of the University of California,
Los Angeles, and the Jet Propulsion Laboratory/California Institute of
Technology, funded by the National Aeronautics and Space
Administration.

\end{acknowledgements}

\bibliographystyle{aa}
\bibliography{bbfile}

\end{document}